\documentclass[11pt,a4paper]{article}
\pdfoutput=1
\usepackage{jheppub}
\usepackage[all]{xy}
\usepackage{amsfonts}
\usepackage{amscd}
\usepackage{amssymb}
\usepackage{amsmath,bbm}
\usepackage{graphicx}
\usepackage{epsfig}
\usepackage{latexsym}
\usepackage{mathtools}
\usepackage{hyperref}
\usepackage[vcentermath]{youngtab}
\usepackage{accents}
    
\def \be  {\begin{equation}}
\def \ee  {\end{equation}}
\def \bea {\begin{equation}\begin{aligned}}
\def \eea {\end{aligned}\end{equation}}
\def \ba  {\begin{eqnarray}}
\def \ea  {\end{eqnarray}}
\def \bb  {\begin{thebibliography}}
\def \eb  {\end{thebibliography}}
\def \lab #1 {\label{#1}}
 \def\Im{{\rm Im}}

\newcommand\cH{\mathcal{H}}

\newcommand\cK{\mathcal{K}}
\newcommand\cL{\mathcal{L}}
\newcommand\cM{\mathcal{M}}
\newcommand\cN{\mathcal{N}}
\newcommand\cO{\mathcal{O}}
\newcommand\cP{\mathcal{P}}

\newcommand\cT{\mathcal{T}}

\newcommand\cV{\mathcal{V}}
\newcommand\cW{\mathcal{W}}

\newcommand\al{\alpha}

\newcommand\tf{\mathfrak{t}}

\newcommand\m{\mathfrak{m}}
\newcommand\C{\mathbb{C}}
\newcommand\CP{\mathbb{CP}}
\newcommand\ep{\epsilon}

\newcommand\fM{\mathfrak{M}}

\newcommand\fL{\mathfrak{L}}

\newcommand\R{\mathbb{R}}

\renewcommand{\t}{\widetilde }

\newcommand\la{\langle}
\newcommand\ra{\rangle}

\newcommand\tr{\mathrm{Tr}}

\newcommand\Hom{\mathrm{Hom}}
\newcommand\vir{\mathrm{vir}}

\definecolor{cardinal}{rgb}{0.6,0,0}
\definecolor{darkgreen}{rgb}{0,0.5,0}
\definecolor{golden}{rgb}{0.92, 0.7, 0}
\definecolor{midnight}{rgb}{0, 0, 0.5}
\definecolor{darkblue}{rgb}{0.2, 0, 0.8}

\title{Twisted Indices of 3d $\cN=4$ Gauge Theories and Enumerative Geometry of Quasi-Maps}
\author[1]{Mathew Bullimore,}
\author[1,2]{Andrea Ferrari,}
\author[2]{Heeyeon Kim}
\affiliation[1]{Department of Mathematical Sciences, Durham University, Lower Mountjoy, Stockton Road, Durham, DH1 3LE, UK}
\affiliation[2]{Mathematical Institute, University of Oxford, Woodstock Road, Oxford, OX2 6GG, UK}

\abstract{
We explore the geometric interpretation of the twisted index of 3d $\cN=4$ gauge theories on $S^1 \times \Sigma$ where $\Sigma$ is a closed Riemann surface. We focus on a rich class of supersymmetric quiver gauge theories that have isolated vacua under generic mass and FI parameter deformations. We show that the path integral localises to a moduli space of solutions to generalised vortex equations on $\Sigma$, which can be understood algebraically as quasi-maps to the Higgs branch. We show that the twisted index reproduces the virtual Euler characteristic of the moduli spaces of twisted quasi-maps and demonstrate that this agrees with the contour integral representation introduced in previous work. Finally, we investigate 3d $\cN=4$ mirror symmetry in this context, which implies an equality of enumerative invariants associated to mirror pairs of Higgs branches under the exchange of equivariant and degree counting parameters. 
}

\begin{document}
\today
\maketitle


\section{Introduction and Summary}
\label{sec:intro}

The Witten index \cite{Witten:1982df} of supersymmetric quantum mechanics,
\be
I = \text{Tr}_{\cH} (-1)^F\ ,
\ee 
is a powerful tool to study geometric aspects of supersymmetric theories. For example, in a 1d $\cN=(0,2)$ sigma model to a compact target $M$ endowed with a holomorphic vector bundle $E$, the Witten index can be identified with the holomorphic Euler characteristic
\be\label{holomorphic euler intro}
\chi\left(\cM,K_M^{1/2} \otimes E\right) = \int_{M} \hat A(T M)~\text{ch}(E) \, .
\ee
In the presence of flavour symmetry, this can be promoted to a flavoured Witten index that computes the equivariant holomorphic Euler characteristic.

In this paper, we study the twisted indices of 3d $\cN=4$ supersymmetric gauge theories on $S^1 \times \Sigma$. This can be regarded as the flavoured Witten index of the effective supersymmetric quantum mechanics on $S^1$ obtained by performing a topological twist on a genus $g$ Riemann surface $\Sigma$. There are two distinct twists that utilise a $U(1)$ subgroup from each factor of the R-symmetry $SU(2)_H \times SU(2)_C$. We refer to them as the `H-twist' and `C-twist' respectively. 

The twisted index can be defined by
\be
I^{H,C}(a,q, t) = \text{Tr}_{\cH_{H,C}} (-1)^F a^{J_H}q^{J_C}  t^{J_t}\ ,
\ee
where $\cH_{H,C}$ is the Hilbert space of supersymmetric ground states on $S^1 \times \Sigma$ and
\begin{itemize}
\item $J_H$ is the generator of the Cartan subalgebra of the Higgs branch flavour symmetry $G_H$ acting on the hypermultiplets. The associated fugacity is $a = e^{2\pi i m}$ where $m$ are real mass parameters.
\item $J_C$ is the generator  of the Cartan subalgebra of the Coulomb branch flavour symmetry $G_C$, which is realised as a topological symmetry in the UV. The associated fugacity is $q = e^{2\pi i \zeta}$ where $\zeta$ is the real Fayet-Iliopoulos parameter.   
\item $J_t$ is the generator of the combination $U(1)_t = U(1)_H-U(1)_C$ of R-symmetries, that commutes with the two supercharges preserved in both the H-twist and the C-twist. 
\end{itemize}

The twisted indices of 3d supersymmetric gauge theories were first computed by Nekrasov and Shatashvili~\cite{Nekrasov:2014xaa} using the topological A-model on $\Sigma$ in the context of the Bethe/Gauge correspondence. More recently, the twisted indices of 3d $\cN=2$ supersymmetric gauge theores have also been derived from the UV Coulomb branch localisation~\cite{Benini:2015noa,Gukov:2015sna,Benini:2016hjo,Closset:2016arn}. The result can be expressed as a sum of contour integrals over the complexified maximal torus of the gauge group $G$. When $G$ is a product of unitary groups, as we consider in this paper, we have
\be\label{twisted index residue intro}
I^{H,C}(a,q,t) = \frac{1}{|W_{G}|} \sum_{\underline{\m} \in \Lambda_{G}}  (-q)^{\text{tr}(\underline{\m})}~\underset{u=u*}{\text{JK-Res}}~du~ Z_{\underline{\m}}^{\text{1-loop}}(u,a,t) H^g(u,a,t)\ ,
\ee
where the summation is over GNO quantised flux on $\Sigma$, or co-character lattice $\Lambda_G$ of the gauge group $G$. The contribution from each flux sector is given by a Jeffrey-Kirwan residue that specifies the choice of contour.

The main purpose of this paper is to provide a geometric interpretation of this contour integral as a holomorphic Euler characteristic, as in equation~\eqref{holomorphic euler intro}. We focus on 3d $\cN=4$ superconformal quiver theories that have isolated massive vacua in the presence of generic mass and FI parameters. By introducing an alternative localising action, we show that the path integral can be localised to solutions of the generalised vortex equations on $\Sigma$, which take the schematic form
\begin{gather}
*F_A+ e^2 \left(\mu_\R -2[\varphi^\dagger , \varphi ] -\tau  \right) = 0 \nonumber \\
 \bar\partial_A X= 0 \quad \bar\partial_AY= 0 \quad \bar\partial_A\varphi = 0  \label{eq:vortex intro}
 \\
\varphi \cdot X = 0 \quad \varphi \cdot Y = 0 \quad X \cdot Y = 0 \ , \nonumber
\end{gather}
where $(X,Y)$ are the hypermultiplet scalar fields transforming in a quaternionic representation of $G$ and $\varphi$ is the vector multiplet complex scalar field in the adjoint representation. The solutions to these equations form a moduli space $\fM$, which is a disjoint union of topologically distinct sectors labelled by the degree of the gauge bundle 
\be\label{decomposition intro}
\fM = \bigcup_{\m\in \Lambda_{C}^\vee}\fM_\m\ ,
\ee
where $\Lambda_C^\vee$ is the character lattice of the Coulomb branch flavour symmetry $G_C$. 

The description of the moduli space depends on a parameter $\tau$, valued in a Cartan subalgebra $\mathfrak{t}_C$ of the Coulomb branch flavour symmetry. Although this parameter appears in an exact deformation of the action, we expect intricate wall-crossing behaviour in this parameter space.
In this paper, 
we formally take the parameter $\tau\rightarrow \infty$ in a given chamber, which is relevant for three-dimensional mirror symmetry.  
In this limit, $\fM_\m$ has an algebraic description as the moduli space of quasi-maps to the Higgs branch, $\Sigma \rightarrow \cM_H$, of degree $\m$~\cite{CIOCANFONTANINE201417}. \footnote{The moduli space of quasi-maps and their enumerative geometry have been discussed in various contexts, e.g., \cite{Okounkov:2016sya,Pushkar:2016qvw,Aganagic:2017gsx,Koroteev:2017nab,Jockers:2018sfl,Bonelli:2013mma}.} More precisely, in the H-twist we recover the twisted quasi-maps to holomorphic symplectic quotients introduced in \cite{kim:2016}, while in the C-twist we find a generalisation to arbitrary genus of a construction of \cite{Okounkov:2015spn}.

In order to provide a concrete interpretation of the contour integral representation of the twisted index~\eqref{twisted index residue intro} in terms of the enumerative geometry of the moduli space $\fM$, we carefully study the massless fluctuations of the bosonic and fermionic fields around a point $p\in \fM$. 
From a mathematical viewpoint, these massless fluctuations can be identified with the virtual tangent bundle to the moduli space $\fM$ and gives rise to perfect obstruction theory, which coincides with those  considered in \cite{kim:2016,Okounkov:2015spn}. We remark that related constructions have also been extensively studied in \cite{Nekrasov:2014nea,Okounkov:2015spn} in the context of the K-theoretic Donaldson-Thomas invariants of Calabi-Yau three-folds.
From this discussion, we argue that the localised path integral for the twisted index reproduces a generating function of virtual Euler characteristics of $\fM_\m$ defined by
\be 
I^{H,C} = \sum_{\m \in \Lambda_C^\vee}(-q)^\m\int_{\fM_\m} \hat A(T^{\text{vir}})\ .
\ee

In general, the moduli spaces $\fM_\m$ are non-compact and these integrals are not well-defined. However, by turning on a real mass parameter with associated fugacity $t$, we can localise further to the compact fixed locus of the $U(1)_t$ symmetry. This fixed locus $\fL \subset \mathfrak{M}$ coincides with the moduli space of quasi-maps to a holomorphic Lagrangian $\cL_H \subset \cM_H$ known as the compact core. The virtual tangent bundle then decomposes on the fixed locus as
\be
T^{\text{vir}}|_{\fL_\m} = T\fL_\m+ N_\m\ ,
\ee
where $T\fL_\m$ is the virtual tangent bundle to the fixed locus and $N$ is the virtual normal bundle. The path integral then reproduces the virtual Euler characteristic defined by localisation with respect to the $U(1)_t$ action,
\be
I^{H,C} = \sum_{\m \in \Lambda_C^\vee}(-q)^\m\int_{\fL_\m} \frac{ \hat  A\left(T\fL_\m\right) }{ \text{ch}\left( \widehat \wedge^\bullet N_\m^{\vee}\right) }\, .
\ee
where the notation $\widehat\wedge^\bullet$ indicates the exterior algebra normalised by the square root of the determinant bundle. This gives a concrete geometric interpretation to the twisted index.



In order to perform explicit calculations, we can localise further to the fixed locus of the maximal torus $T_H$ of the flavour symmetry $G_H$ by turning on mass parameters with associated fugacity $a$, which play the role of equivariant parameters. Under our assumptions, we show that the fixed locus is a disjoint union of smooth compact spaces $\fM_{\underline{\m},I}$, where $I$ labels the fixed points on $\cM_H$ and $\underline{\m} \in \Lambda_G$ is a GNO quantised flux with $\mathrm{tr}(\underline{\m}) = \m$. Each component is given by a product of the symmetric product of the curve $\Sigma$,
\be
\fM_{\underline{\m},I} = \prod_{a=1}^{\text{rk}(G)}\text{Sym}^{{\frak n}_{I_a}}\Sigma\ ,
\ee
where ${\frak n}_{I_a}$'s are non-negative integers which depend on the twist and a component of the magnetic flux $\underline{\m}$. On the fixed locus, the virtual tangent space decomposes as
\be
T^{\text{vir}}\big|_{\cM_{\underline{\m},I}} = T\fM_{\underline{\m},I} + N_{\underline{\m},I}\ ,
\ee 
where $N_{\m,I}$ are the virtual normal bundles and non-zero weights under the $U(1)_t \times T_H$ action.
The path integral then reproduces the equivariant virtual Euler characteristic via virtual localisation,
\bea\label{virtual localisation intro}
I^{H,C}
& = \sum_{\underline{\m} \in \Lambda_G}(-q)^{\m} \sum_{I}\int_{\fM_{\underline{\m},I}} \frac{\hat A (T\fM_{\underline{\m},I})}{\text{ch}(\widehat \wedge^\bullet N_{\underline{\m},I}^\vee)}\ .
\eea
The intersection theory on the symmetric product of a curve is well-known \cite{macdonald1962symmetric,thaddeus1994stable,arbarello1985geometry} allowing us to convert the expression \eqref{virtual localisation intro} into a sum of the residue integrals. We show explicitly that this reproduces the contour integral representation of the twisted index \eqref{twisted index residue intro}. In particular the fixed loci of $U(1)_t \times T_H$ are in one-to-one correspondence with the poles selected by the Jeffrey-Kirwan residue integral. \footnote{The geometric interpretation of the twisted index for an $\cN=2$ supersymmetric Chern-Simons theory with an adjoint chiral multiplet has been studied in the references \cite{Andersen:2016hoj,Hausel:2017iel}.}

Sending $t\rightarrow 1$, the twisted index preserves four supercharges that generate a 1d $\cN=(2,2)$ and $\cN=(0,4)$ supersymmetric quantum mechanics in the H-twist and C-twist respectively. This enables us to add further exact terms to the localising action to further constrain the moduli space. In particular, the C-twisted index can be localised to the space of constant maps to the Higgs branch $\cM_H$. In this limit, the virtual Euler characteristic is independent of $q$ and reduces to the equivariant Rozansky-Witten invariants \cite{Rozansky:1996bq} of $\cM_H$, associated with the three-manifold $S^1 \times \Sigma$,
\be\label{RW intro}
I^C \big|_{t\rightarrow 1} = \int_{\cM_H} \hat A(T\cM_H)~\text{ch}\left( \widehat{\wedge}^{\bullet} T^* \mathcal{M}_H \right)^g\ .
\ee
On the other hand, the H-twisted index reduces to a generating function of the Euler classes of the $G_H$-fixed loci,
\be\label{euler intro}
I^H \big|_{t\rightarrow 1} = \sum_{\underline{\m} \in \Lambda_G} (-q)^\m \sum_{I}(-1)^{\text{dim}_{\mathbb{C}}{(\fM_{\underline{\m}, I})}}\int_{\fM_{\underline{\m},I}} e(\fM_{\underline{\m},I})\ ,
\ee
which is independent of the fugacity $a$.

An important feature of the class of 3d $\cN=4$ supersymmetric gauge theories we consider is the existence of mirror symmetry, which exchanges the H-twist and the C-twist of a dual pair of theories $\cT$ and $\cT^\vee$. This implies the following relation between the twisted indices of these theories,
\be
I_H[\cT](q,a, t) = I_C[\cT^\vee](a,q,t^{-1})\ ,
\ee
This provides extremely non-trivial relationship between enumerative invariants of quasi-maps to pairs of Higgs branches $\cM_H$ and $\cM_H^\vee$ under the exchange of the degree counting parameters $q$ and equivariant parameters $a$. In the limit $t\rightarrow 1$, we explicitly prove this relation for $T[SU(N)]$ theories, which are self-dual under mirror symmetry.

The paper is organised as follows. In section \ref{sec:background}, we describe the class of 3d $\cN=4$ theories we consider in this paper. In particular, we summarise the construction of the Higgs branch $\cM_H$, which plays an important role for later discussions. In section \ref{sec:twist}, we explain the procedure of the topological reduction of 3d $\cN=4$ theories on $\Sigma$. For this purpose, we review the localisation process which gives rise to the moduli space $\fM$ and study the algebraic description in terms of twisted quasi-maps to $\cM_H$. Then we study the massless fluctuations of the bosonic and fermionic fields at a point on the moduli space $\fM$, from which we construct the virtual tangent bundle $T^{\text{vir}}$ over $\fM$. From this discussion, we provide a geometric interpretation of the contour integral formula as the virtual Euler characteristics constructed from $T^{\text{vir}}$. In section \ref{sec:t1limit}, we study the reduced moduli space that preserves four supercharges and discuss the relation to the twisted indices evaluated in the limit $t\rightarrow 1$. In section \ref{sec:examples}, we explore the geometric interpretations of the twisted indices through various concrete examples. Finally, in section \ref{sec:mirror-symmetry}, we study the implications of mirror symmetry in this context, and explicitly check the proposed dualities for the $T[SU(N)]$ theories in the limit $t\rightarrow 1$.

\section{Background and Notation}
\label{sec:background}


\subsection{Quiver Gauge Theories}
\label{sec:quivers}

A renormalisable 3d $\cN=4$ supersymmetric gauge theory is specified by a compact group $G$ and a linear quaternionic representation $Q$ - we refer the reader to~\cite{Bullimore:2015lsa,Bullimore:2016nji} for a summary and further background. In this paper, we will focus on unitary quiver gauge theories. Introducing an index $I = 1,\ldots,L$ labelling the nodes of the quiver, this corresponds to the choice
\be
G = \prod_I U(V_I) \qquad Q =T^* M
\ee
where
\be
M = \bigoplus_I \Hom(W_I,V_I) \oplus \bigoplus_{I \leq J} \Hom(V_I,V_J) \otimes Q_{IJ}    \, .
\ee 
is a unitary representation of $G$. Here $V_I$, $W_I$ denote complex vector spaces while $Q_{IJ}$ are multiplicities. In physical parlance, there is a dynamical vectormultiplet for the gauge group $G$ and
\begin{itemize}
\item $Q_{II}$ hypermultiplets in the adjoint representation of $U(V_I)$,
\item $Q_{IJ}$ hypermultiplets in the bifundamental representation of $U(V_I) \times U(V_J)$ for $I < J$,
\item and $\dim_\C W_I$ hypermultiplets in the fundamental representation of $U(V_J)$.
\end{itemize}

\begin{figure}[htp]
\centering
\includegraphics[height=1.25cm]{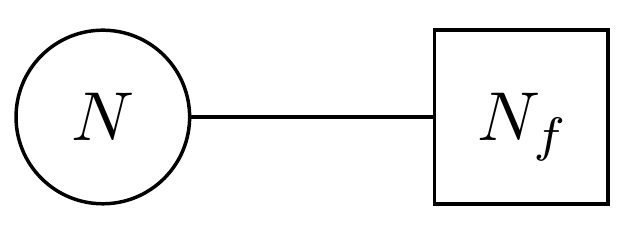}
\caption{Quiver for $U(N_c)$ supersymmetric QCD with $N_f$ fundamental hypermultiplets}
\label{fig:sqcd}
\end{figure}

An example is the single node quiver with $V = \C^{N_c}$, $W = \C^{N_f}$ and unitary representation $M = \Hom(W,V)$. This is supersymmetric QCD with $G=U(N_c)$ and $N_f$ fundamental hypermultiplets, as illustrated in figure~\ref{fig:sqcd}. In the following sections~\ref{sec:twist} and \ref{sec:t1limit}, we will formulate our constructions for a general unitary quiver (subject to an assumption explained in section~\ref{sec:assumption}) but our explicit examples in section~\ref{sec:examples} will be almost exclusively supersymmetric QCD.

In what follows, we use euclidean $SU(2)$ spinor indices $\al$ in addition to spinor indices $A$, $\dot A$ for the $SU(2)_H \times SU(2)_C$ R-symmetry, with uniform conventions summarized in Appendix~\ref{app:susy algebra}. With this notation, the vectormultiplet includes a gauge connection $A_{\al\beta}$, scalar fields $\phi^{\dot A\dot B}$, and gauginos $\lambda_\al^{A\dot A}$ transforming in the adjoint representation of $G$. The hypermultiplets contains complex scalars $X_A$ and fermionic spinors $\Psi_\al^{\dot A}$ transforming in the unitary representation $M$.

It will be convenient to decompose the supermultiplets under a fixed maximal torus $U(1)_H \times U(1)_C$ of the R-symmetry. The vectormultiplet scalars decompose into real and complex components $\sigma$, $\varphi$, $\varphi^\dagger$ transforming with $U(1)_C$ charge $0,+1,-1$ respectively, while the hypermultiplet scalars decompose into a pair of complex scalars $X$, $Y$ transforming with $U(1)_H$ charge $+\frac{1}{2}$. The charges of these fields are shown in table~\ref{tab:charges}.

\begin{table}[htp]
\centering
\begin{tabular}{ c | c | c c | c }
&  $G$ & $U(1)_H$ & $U(1)_C$  & $U(1)_t$ \\ \hline
$\sigma$ & Adj & $0$ & $0$  & $0$ \\
$\varphi$ & Adj & $0$ & $+1$  & $-1$ \\
$X$ & $M$ & $+\frac12$ & $0$ &  $+\frac{1}{2}$ \\  
$Y$ & $M^*$ & $+\frac12$ & $0$  & $+\frac{1}{2}$    
\end{tabular}
\caption{Summary of gauge and R-symmetry representations.}
\label{tab:charges}
\end{table}

The flavour symmetry is a product $G_H \times G_C$ where:
\begin{itemize}
\item $G_H$ acts on the hypermultiplets. It is given by the normalizer of $G$ inside $USp(M)$ modulo the gauge group $G$:
\be
 G_H = N_{USp(M)}(G)/G \, .
\ee
\item $G_C$ contains topological symmetry $U(1)^{L}$ under which monopole operators are charged. This may be enhanced in the IR to a non-abelian group with maximal torus $U(1)^L$. 
\end{itemize}
  We turn on associated real mass deformations valued in the Cartan subalgebras $\tf_H$, $\mathfrak{t}_C$ of the flavour symmetry factors:
\begin{itemize}
\item Real mass parameters $m \in \mathfrak{t}_H$ are vacuum expectation values for the real scalar in a background vectormultiplet for $G_H$.
\item Real FI parameters $\zeta \in \mathfrak{t}_C$ are vacuum expectation values for the real scalar in a background twisted vectormultiplet for $G_C$.
\end{itemize}
We do not consider complex mass and FI parameters in this paper.

In supersymmetric QCD, $G_H = PSU(N_f)$ and $G_C = U(1)$, enhanced to $G_C = SU(2)$ when $N_f = 2N_c$. Correspondingly, we introduce real mass parameters $m = (m_1,\ldots,m_{N_f}) \in \R^{N_f-1}$ satisfying $\sum_j m_j = 0$ and a single FI parameter $\zeta \in \R$.

It will also be important to introduce a real mass parameter that breaks $\cN=4$ to $\cN=2$ supersymmetry. Given the maximal torus $U(1)_H \times U(1)_C$ with generators $T_H$, $T_C$, we may decompose the supermultiplets under the $\cN=2$ supersymmetry commuting with the $U(1)_t$ generated by
\be
T_t  = T_H - T_C \, .
\ee
From this perspective, $U(1)_t$ is a distinguished flavour symmetry. We can then choose an integer $R$-symmetry for the $\cN=2$ supersymmetry algebra generated by $R_H = 2 T_H$ or $R_C = 2 T_C$. This choice is important when performing a topological twist on $S^1 \times \Sigma$.

From the perspective of $\cN = 2$ supersymmetry $\sigma$ transforms in a vectormultiplet, while $\varphi$, $X$, $Y$ transform in chiral multiplets whose charges are summarised in table~\ref{tab:charges}. There are also superpotentials  
\be
W_I = \tr_{V_I}(\varphi X Y)
\label{eq:superpotential}
\ee
at each node whose $R$-charges are always $+2$. The real mass parameters $m$ are now obtained by coupling to a background $\cN=2$ vectormultiplet for the flavour symmetry $G_H$ while $\zeta$ is an FI parameter for the dynamical $\cN=2$ vectormultiplet. 

We can now explicitly break to $\cN=2$ supersymmetry by introducing a real mass parameter $m_t$ for the distinguished $U(1)_t$ flavour symmetry. This is the mass deformation mentioned in the introduction and, as anticipated there, it will play an important role in this paper as a localisation parameter.

\subsection{Moduli Spaces of Vacua}
\label{sec:assumption}

The moduli space of vacua of 3d $\cN=4$ supersymmetric gauge theory includes a Higgs branch and a Coulomb branch, denote by $\cM_H$ and $\cM_C$ respectively. They are both hyper-K\"ahler,  such that the R-symmetries $SU(2)_H$, $SU(2)_C$ rotate the complex structure on $\cM_H$, $\cM_C$. Furthermore, the flavour symmetries $G_H$, $G_C$ act by tri-hamiltonian isometries of $\cM_H$, $\cM_C$.

The choice of maximal torus $U(1)_H \times U(1)_C$ selects a complex structure on $\cM_H$ and $\cM_C$. From this point of view, they are K\"ahler manifolds equipped with holomorphic symplectic forms of weight $+1$ under K\"ahler isometries $U(1)_H$, $U(1)_C$. The flavour symmetries $G_H$, $G_C$ act by Hamiltonian isometries of $\cM_H$, $\cM_C$ that leave invariant the holomorphic symplectic form.

In this paper, we make a crucial assumption that the supersymmetric quiver gauge theory flows to a superconformal fixed point and has isolated massive vacua when generic real mass and FI parameters are turned on. This translates into the assumption that $\cM_H$, $\cM_C$ are conical symplectic resolutions with isolated fixed points under infinitesimal $T_H$, $T_C$ transformations. Furthermore, $\mathfrak{t}_H$, $\mathfrak{t}_C$ describe K\"ahler resolution parameters for $\cM_C$, $\cM_H$ under the identifications
\be
\mathfrak{t}_H = H^2(\cM_C,\R) \, , \qquad \mathfrak{t}_C = H^2(\cM_H,\R) \, .
\label{eq:equiv-resol}
\ee
In more physical terms:
\begin{itemize}
\item The mass parameters $m \in \tf_H$ are resolution parameters for $\cM_C$ and generate an infinitesimal Hamiltonian isometry of $\cM_H$,
\item The FI parameters $\zeta \in \tf_C$ are resolution parameters for $\cM_H$ and generate an infinitesimal Hamiltonian isometry of $\cM_C$.
\end{itemize}

This assumption will permeate our considerations on $S^1\times \Sigma$, allowing explicit computations to be performed while encompassing an infinite and rich class of examples. Further motivation comes from the fact that such theories transform straightforwardly under 3d mirror symmetry and play an important role in connection with symplectic duality~\cite{Braden:2014iea,BPW-I}. For further motivation and background we refer the reader to reference~\cite{Bullimore:2016nji}. We will return to this connection in section~\ref{sec:mirror-symmetry}.

\subsection{Higgs Branch Geometry}
\label{sec:hb}

The Higgs branch is particularly important for consideration of the twisted index on $S^1 \times \Sigma$. We therefore explain its construction in more detail now. We first set the mass parameters $m = 0$. The classical vacuum configurations are solutions to
\bea
\mu_\R - \zeta & = 0 \qquad  \mu_\C  = 0   \\
\sigma \cdot X &= 0 \qquad \varphi \cdot X = 0 \qquad \varphi^\dagger \cdot X = 0    \\
\sigma \cdot Y & = 0 \qquad \varphi \cdot Y  = 0 \qquad \varphi^\dagger \cdot Y  = 0 \\
[\sigma, \varphi] &= 0 \qquad [\varphi,\varphi^\dagger] = 0 \, ,
\label{eq:vacuum_higgs}
\eea
modulo gauge transformations. Here it is understood that vectormultiplet scalars act on $(X,Y)$ in the representation $T^*M$. Finally, 
\be
\mu_\R = X \cdot X^\dagger- Y^\dagger \cdot Y  \qquad
\mu_\C =  X \cdot Y 
\ee
are the real and complex moment maps for the $G$ action on $T^*M$. 

Equations~\eqref{eq:vacuum_higgs} may be decomposed into contributions from each node labelled by an index $I = 1,\ldots,L$. Here we are employing shorthand notation such as $\zeta = \{\zeta_1,\ldots,\zeta_L\}$ and $\mu_\R = \{ \mu_{\R,1} , \ldots, \mu_{\R,L}\}$ to express the contributions from all of the nodes simultaneously.

For future applications, it is useful to reconsider the vacuum equations in the language of $\cN=2$ supersymmetry. From this perspective the vacuum equations are 
\bea
&\mu_\R -  2[\varphi^\dagger,\varphi] - \zeta  = 0 \\
&\varphi \cdot X  = 0 \qquad \varphi \cdot Y = 0 \qquad \mu_\C  = 0   \\
& \sigma \cdot X  = 0 \qquad \sigma \cdot Y  = 0  \qquad [\sigma, \varphi] = 0  \, ,
\label{eq:hbN=2}
\eea
where the first line contains the $D$-term equations and the second line the $F$-term equations associated to the superpotential $W = \tr_{V}(\varphi X Y)$.
Note that the $D$-term equation involves an additional commutator compared to~\eqref{eq:vacuum_higgs}. However, by squaring the D-term equation and imposing the $F$-term equations, 
\be
\| \mu_\R - 2 [\varphi^\dagger,\varphi] - \zeta \|^2 = \| \mu_\R  - \zeta \|^2 + 4 \| [\varphi^\dagger,\varphi] \|^2 + 2 \| \varphi \cdot X^\dagger \|^2 + 2 \| \varphi \cdot Y^\dagger \|^2 \, ,
\ee
which requires $[\varphi^\dagger, \varphi] = 0$ separately and recovers the remaining equations in~\eqref{eq:vacuum_higgs}. 

\subsubsection{Hyper-K\"ahler Quotient}

Under our assumption of section \ref{sec:assumption}, the FI parameter $\zeta$ can be chosen such that $G$ acts freely on solutions to the vacuum equations~\eqref{eq:vacuum_higgs}. This typically requires that the FI parameter lies in the complement of hyperplanes
\be 
\zeta \in \mathbb{R}^L \backslash \cup_\al H_\al \, ,
\ee
which split the parameter space $\mathfrak{t}_C = \mathbb{R}^L$ into chambers. In supersymmetric QCD, this means we assume that $N_f \geq N_c$ and $\zeta >0$ or $\zeta <0$.

The implies $\sigma = \varphi = 0$ on solutions of the vacuum equations, which would otherwise generate unbroken gauge transformations. 
The remaining equations then describe the Higgs branch as a smooth hyper-K\"ahler quotient
\be
\mathcal{M}_{\zeta,H} := T^*M/\!/\!/_\zeta \, G  \, ,
\ee
is a Nakajima quiver variety~\cite{nakajima1998,Nakajima:1994nid}. We note that the holomorphic symplectic form on the Higgs branch is independent of $\zeta$ within each chamber, while the real symplectic form or K\"ahler structure depends explicitly on $\zeta$. 

The assumption of section~\ref{sec:assumption} requires that
\be
\nu : \cM_{H,\zeta} \to \cM_{H,0}
\label{eq:res}
\ee
is a conical symplectic resolution. The inverse image 
\be
\cL_{H,\zeta} := \nu^{-1}(0)
\ee
is then a compact holomorphic lagrangian known as the `compact core'. This has a convenient K\"ahler quotient description as follows. The choice of chamber selects a holomorphic Lagrangian splitting $T^*M = L \oplus L^*$, corresponding to a decomposition of the hypermultiplet fields $(X_L,Y_L)$ where $Y_L = 0$ on the compact core. We then have 
\be
\cL_{H,\zeta} = L /\!/_\zeta \, G =  \{ \mu_\R |_L = 0 \} / G \, .
\ee
We frequently fix a chamber and omit the dependence on $\zeta$, writing $\cM_H$ and $\cL_H$ respectively for the Higgs branch and its compact core.

In supersymmetric QCD, this assumption requires that $N_f \geq 2N_c$. In this case, the Higgs branch is a cotangent bundle to the grassmannian of $N_c$-planes in $N_f$ complex dimensions, $\cM_{H} = T^*G(N_c,N_f)$. The map~\eqref{eq:res} is the Springer resolution of the nilpotent cone closure $\bar \cN_\rho \subset \mathfrak{sl}(N_f,\C)$ labelled by $\rho^T = (N_c,N_f-N_c)$.  The compact core $\cL_{H} = G(N_c,N_f)$ is the grassmannian base, where:
\begin{itemize}
\item In the chamber $\zeta > 0$, $\cL_H$ is characterised by the decomposition $(X_L,Y_L) = (X,Y)$ and corresponds to configurations with $Y = 0$. 
\item In the chamber $\zeta < 0$, $\cL_H$ is characterised by the decomposition $(X_L,Y_L) = (Y,-X)$ and corresponds to  configurations with $X = 0$.
\end{itemize}

\subsubsection{Algebraic Description}
\label{sec:higgs-algebraic}

The Higgs branch has an algebraic description as a holomorphic symplectic quotient by omitting the D-term equation in favour of an appropriate stability condition and dividing by complex gauge transformations.

Starting from $(X,Y) \in T^*M$, solutions of the F-term equation cut out the subspace $\mu_\C^{-1}(0) \subset T^*M$. We then impose a stability condition depending on the chamber of $\zeta \in \R^L \backslash \cup_\al H_\al$ and quotient by complex gauge transformations $G_\C$. Under our assumptions, stability coincides with semi-stability and we obtain a smooth quotient,
\be
\cM_H = \mu_\C^{-1}(0)^s / G_\mathbb{C} \, .
\ee
We do not describe the stability condition for a general quiver, instead focussing later on the example of supersymmetric QCD~\footnote{An account of the appropriate stability condition for a general quiver that is close to the perspective taken here can be found in~\cite{1611.10000}.}. 

This provides an algebraic description of the tangent bundle to $\cM_H$, which will reappear in section~\ref{sec:algebraic}. Considering small fluctuations of the hypermultiplets $(\delta X, \delta Y)$ compatible with the F-term equation, modulo infinitesimal complex gauge transformations, generates the following complex 
\be
0 \longrightarrow  \mathfrak{g}_\C \overset{\alpha}{\longrightarrow} T^*M \overset{\beta}{\longrightarrow}  \mathfrak{g}_\C^* \longrightarrow 0 
\label{eq:higgs-complex}
\ee
of trivial $G_\C$-equivariant vector bundles on $T^*M$. The maps
\be
\alpha  : \delta g \mapsto (\delta g \cdot X, \delta g \cdot Y) \qquad
\beta  : (\delta X, \delta Y) \mapsto X \cdot \delta Y + \delta X \cdot Y
\ee
at a point $(X,Y) \in T^*M$ correspond to infinitesimal complex gauge transformations and the differential of the complex moment map respectively. On restriction to the stable locus $\mu^{-1}_\C(0)^s$, $\al$ is injective and $\beta$ surjective, and equation~\eqref{eq:higgs-complex} descends to a complex of vector bundles on $\cM_H$ whose cohomology is the tangent bundle,
\be
T \cM_H = \mathrm{Ker}(\beta) / \mathrm{Im}(\alpha) \, .
\ee

In supersymmetric QCD in the chamber $\zeta >0$, the stable locus consists of solutions where $X$ has maximal rank and defines a complex $N_c$-plane in $W = \C^{N_f}$. The holomorphic symplectic quotient
\be
\left\{ \, X,Y \, | \, X \cdot Y = 0 , \, \mathrm{rk}(X) = N_c \, \right\} / GL(N_c,\C)
\ee
provides an algebraic description of $\cM_H = T^*G(N_c,N_f)$.  
The tangent bundle is the cohomology of the complex
\be
0 \longrightarrow \Hom(\cV,\cV) \overset{\alpha}{\longrightarrow} T^*\Hom(\cW,\cV) \overset{\beta}{\longrightarrow} \Hom(\cV,\cV) \longrightarrow 0  \, ,
\ee
where $\cV$ is the tautological complex vector bundle with fiber $V = \C^{N_c}$ and $\cW$ is the trivial complex vector bundle with fiber $W$. The maps are the infinitesimal complex gauge transformation $\al : \delta g \mapsto (\delta g X,-Y \delta g)$ and the differential of the complex moment map $\beta : (\delta X,\delta Y) \mapsto \delta X  Y + X \delta Y$. 

\subsection{Mass Parameters and Fixed Loci}
\label{sec:masses}

We now consider the fate of the Higgs branch vacua in the presence of real mass parameters $m_t$ and $m$ associated to flavour symmetries $U(1)_t$ and $G_H$ respectively.

\subsubsection{$U(1)_t$ Mass Parameter}

The mass parameter $m_t$ is a vacuum expectation value for a background $\cN=2$ vectormultiplet for the flavour symmetry $U(1)_t$. Accordingly, the supersymmetric vacuum equations~\eqref{eq:vacuum_higgs} are modified by replacing $\sigma \to \sigma + m_t$ (acting in the appropriate representation). More precisely,
\be
 \sigma \cdot X + \frac{m_t}{2} X  = 0 \qquad \sigma \cdot Y + \frac{m_t}{2} Y  = 0  \qquad [\sigma, \varphi] - m_t \varphi = 0 
\ee
in view of the charges presented in table~\ref{tab:charges}. The remaining supersymmetric vacua correspond to configurations $(X,Y,\varphi)$ solving the modified vacuum equations, for which there exists a $\sigma$ such that the combined infinitesimal gauge and $U(1)_t$ transformation generated by $\sigma$ and $m_t$ leaves the configuration invariant. 

Such configurations are found by setting $Y_{L} = 0$ where $T^*M = L \oplus L^*$ is the Lagrangian splitting introduced above. It is useful to note that under the combined gauge and $U(1)_t$ transformation that leaves this configuration invariant, the hypermultiplet fields $(X_L,Y_L)$ transform with weight $(0,1)$. This property could be used to characterise the holomorphic Lagrangian splitting.

Geometrically, the remaining supersymmetric vacua correspond to the fixed locus of the $U(1)_t$ K\"ahler isometry of $\cM_H$ generated by the mass parameter $m_t$. From the discussion above, this coincides with the compact core,
\be
\cM_H^{U(1)_t} = \cL_H \, .
\ee
In the algebraic description, the $U(1)_t$ isometry becomes a $\C^*$ action that transforms the holomorphic symplectic form with weight $+1$. This will play an important role in the definition of the enumerative invariants to be considered in section~\ref{sec:twist}.

For example, in supersymmetric QCD with $N_f \geq 2N_c$ in the chamber $\zeta >0$, the mass deformation requires $\sigma = - \frac{m_t}{2} {1}_{N_c}$ and $Y = 0$. Indeed, $U(1)_t$ acts on the fibres of  $\cM_H = T^*G(N_c,N_f)$ with weight $+1$ such that the remaining supersymmetric vacua coincide with the compact core, $\cM_H^{U(1)_t} = G(N_c,N_f)$.

\subsubsection{$G_H$ Mass Parameters}
\label{section:Mass parameter}

Let us now add real mass parameters $m \in \mathfrak{t}_H$ by turning on a vacuum expectation value for a background $\cN=2$ vectormultiplet for the $G_H$ flavour symmetry. The vacuum equations~\eqref{eq:vacuum_higgs} are modified by 
\be
\sigma \to \sigma + m + m_t \, ,
\ee
where again it is understood that the mass parameters act in the appropriate representation of $U(1)_t \times G_H$. 
The remaining vacua now correspond to configurations $(X,Y,\varphi)$ solving the modified vacuum equations, for which there exists a $\sigma$ such that the combined infinitesimal gauge and $G_H \times U(1)_t$ transformation generated by $\sigma$ and $m+ m_t$ leaves the configuration invariant.

Geometrically, the remaining vacua correspond to the fixed locus of the $T_H \times U(1)_t$ isometry of $\cM_{H}$ generated $m +m_t$.
The assumption of section~\ref{sec:assumption} requires that for generic mass parameters $m$, the fixed locus is a set of isolated points
\be
\cM_H^{T_H \times U(1)_t} = \{ v_I \} \, .
\ee
The fixed points necessarily lie in the compact core. Each massive vacuum corresponds to a configuration of $\mathrm{rk}(G)$ non-vanishing hypermultiplet fields chosen from $X_L$, which we denote collectively by $\{ Z_a \}$. We note that in the algebraic description, $T_H$ is promoted to a $ (\C^*)^{\mathrm{rk}(G)}$ action leaving the holomorphic symplectic form invariant.

In supersymmetric QCD the flavour symmetry $G_H = PSU(N_f)$ acts by K\"ahler isometries on $\cM_H = T^*G(N_c,N_f)$. Turning on generic mass parameters $m = \{m_1,\ldots,m_{N_f}\}$ obeying $\sum_{i=1}^{N_f} m_i = 0$, there are ${N_f}\choose{N_c}$ massive supersymmetric vacua labelled by distinct subsets $I = \{i_1,\ldots,i_{N_c}\} \subset \{1,\ldots,N_f\}$ where
\be\label{isolated vacua 2}
v_I \quad : \quad \sigma_a = m_{i_a} \qquad \varphi_a = 0 \qquad Z_a = X^a{}_{i_a} \, .
\ee
They are the fixed points of a generic $T_H \times U(1)_t$ isometry of $\cM_H$ and coincide with the coordinate hyperplanes in the grassmannian base $\cL_H = G(N_c,N_f)$.

\section{Twisted Theories on $S^1 \times \Sigma$}
\label{sec:twist}

In this section, we consider $\cN=4$ supersymmetric quiver gauge theories on $S^1 \times \Sigma$. The construction is a special case of $\cN=2$ supersymmetric gauge theories on $S^1 \times \Sigma$, which have been extensively covered starting with~\cite{Benini:2015noa} and continuing in~\cite{Benini:2016hjo,Closset:2016arn}. These works considered a localisation action where the partition function is expressed as a contour integral in the complexified maximal torus of the gauge group $G$, with the contour specified by a Jeffrey-Kirwan residue prescription. An important motivation for this paper is to understand the geometric origin of this contour prescription, as in the original mathematical constructions~\cite{JK1995}.

We consider here an alternative localising action akin to that introduced in~\cite{Closset:2015rna}, which localises the path integral to solutions of generalised vortex equations on $\Sigma$. In this section, we briefly review the process of the twisted reduction of the gauge theories on $\Sigma$ and study the massless  fluctuations around a point on the moduli space of solutions to these equations. From this we will provide a general relation between the twisted indices and enumerative invariants of the moduli space.

\subsection{Topological Twists}
\label{sec:twisting}

We consider an $\cN=4$ supersymmetric quiver gauge theory on $S^1 \times \Sigma$ with a topological twist along a closed orientable Riemann surface $\Sigma$ of genus $g$. The topological twist can be performed using either $U(1)_H$ or $U(1)_C$, leading to a pair of supersymmetric quantum mechanics on $S^1$ with four supercharges, which we refer to as the $H$-twist and $C$-twist respectively. Their properties are summarised as follows:
\begin{itemize}
\item The $H$-twist preserves the same supersymmetry as a 1d $\cN=(2,2)$ supersymmetric quantum mechanics on $S^1$ with R-symmetry $U(1)_H \times SU(2)_C$.
\item The $C$-twist preserves the same supersymmetry as a 1d $\cN=(0,4)$ supersymmetric quantum mechanics on $S^1$ with R-symmetry $U(1)_C \times SU(2)_H$.
\end{itemize}

Turning on the real mass parameter $m_t$, both twists preserve a common 1d $\cN=(0,2)$ subalgebra that commutes with the $U(1)_t$ symmetry. From the perspective of 3d $\cN=2$ supersymmetry, we are performing topological twists on $\Sigma$ using the integer valued R-symmetries generated by $R_H$ and $R_C$. Both topological twists are compatible with real mass parameters $m$ and FI parameters $\zeta$.

\subsection{Localising Actions}
\label{sec:lagrangians}

We first consider an $\cN=2$ supersymmetric gauge theory with gauge group $G$ and chiral multiplets of $R$-charge $r$ transforming in a unitary representation $R$. After performing the topological twist, we denote the fields in the $\cN=2$ vector multiplet by
\be
V = (\sigma, A_\mu, \lambda, \bar\lambda,\Lambda,\bar\Lambda, D) \, .
\ee 
This can be regarded as a 1d $\cN=2$ vectormultiplet $(\sigma+iA_0, \lambda, \bar\lambda, D)$ for the group of gauge transformations $g : \Sigma \to G$ and a chiral multiplet $(\bar\partial_A, \bar\Lambda)$ transforming in the adjoint representation where $\bar \partial_A$ denotes the anti-holomorphic covariant derivative on $\Sigma$.
The standard Yang-Mills lagrangian for the vectormultiplet is \footnote{Here we used frame indices
$$
e^0 = dt\ ,~~e^{1} = \sqrt{\bar2g_{z\bar z}}dz\ ,~~e^{\bar 1} = \sqrt{2g_{z\bar z}} d\bar z\ ,
$$
so that the metric on the Riemann surface is $ds^2 = e^1 e^{\bar 1} = 2g_{z\bar z} dz d\bar z\ .$
We also defined $F_{\mu\nu} = \partial_\mu A_\nu - \partial_{\nu}A_\mu -i[A_\mu,A_\nu]$, where $*F = -2i F_{1\bar 1}$ is hermitian. (Throughout the paper, we will use $*$ to denote the hodge dual on the 2d Riemann surface $\Sigma$.) The holomorphic derivatives and the gauginos are $(\bar\partial_A, \bar\Lambda) = (D_{\bar 1} e^{\bar 1}, \Lambda_{\bar 1}e^{\bar 1} ) \text{ and }(\partial_A, \Lambda) = (D_{ 1} e^{ 1}, \Lambda_{ 1}e^{ 1})\ .$} 
\bea\label{n=2 ym}
\cL_{\text{YM}} 
& = \text{tr}\left[\frac{1}{2}F_{01}F_{0\bar 1} + \frac12 (-2i F_{1\bar 1})^2 + \frac12 D^2 + \frac12 |D_\mu\sigma|^2 - i \bar\lambda D_0\lambda - i\bar\Lambda_{\bar 1} D_0 \Lambda_{1}\right. \\
&~~~\left.+ 2i\bar\Lambda_{\bar 1} D_1 \lambda -2i\Lambda_1 D_{\bar 1} \bar\lambda -i\bar\Lambda_{\bar 1} [\sigma, \Lambda_1] + i \bar\lambda[\sigma,\lambda] \right]\ .
\eea
This action is exact with respect to the two supercharges $\delta$, $\t\delta$ preserved by the topological twist on $\Sigma$. In addition, we can introduce a real FI parameter for each $U(1)$ factor in the gauge group. For example, when $G=U(N)$ the FI parameter gives a contribution
\be\label{FI term}
\cL_{\mathrm{FI}}[V] = -\frac{i\zeta}{2\pi}~ \text{tr}(D)\ .
\ee
Similarly, we denote the fields of the $\cN=2$ chiral multiplet transforming in a representation $R$ by
\be
\Phi= (\phi,\psi,\eta, F) \, .
\label{N=2chiralfields}
\ee
On a curve $\Sigma$, this reduces to a 1d $\cN=(0,2)$ chiral multiplet $(\phi, \psi)$ transforming as a smooth section $\Omega_\Sigma^{0,0}(P_\Phi)$ and a fermi multiplet transforming as a smooth section $\Omega_\Sigma^{0,1}(P_\Phi)$. Here we define the associated vector bundle $P_\Phi := K_\Sigma^{r/2} \otimes ( P \times_G R)$ where $P$ is the principal gauge bundle on $\Sigma$.
We use the following lagrangian for the chiral multiplet, 
\bea\label{n=2 chiral}
\cL_{\Phi} 
& = \phi^\dagger (- D_0^2 - 4D_{1}D_{\bar 1} + \sigma^2 + iD-2iF_{1\bar 1})\phi -  F^\dagger F \\
&~~~ -\frac{i}{2}\bar\psi (D_0 + \sigma) \psi - 2i \bar \eta ( D_0 - \sigma)\eta + 2i\bar\psi D_1 \eta - 2i \bar\eta D_{\bar 1} \psi \\
&~~~ -i \bar\psi \bar\lambda \phi + i \phi^\dagger \lambda \psi - 2i \phi^\dagger \Lambda_1 \eta + 2i \bar\eta\bar\Lambda_{\bar 1}\phi\ ,
\eea
which is also $\delta , \widetilde\delta$-exact.
Finally the superpotential term is given by
\be
\cL_W[\Phi] + \cL_{\bar W}[\Phi^\dagger] = \int d^2\theta ~W(\Phi) + \text{h.c.}
\ee
where $W$ is a holomorphic function of chiral multiplets with total $U(1)_R$-charge 2.

We now consider a $\cN=4$ supersymmetric quiver gauge theory as described in section~\ref{sec:quivers}. We can regard this as an $\cN=2$ supersymmetric quiver theory with R-symmetry $U(1)_H$ or $U(1)_C$. In particular,
\begin{itemize}
\item The $\cN=4$ vectormultiplet decomposes into an $\cN=2$ vectormultiplet $V$ and an $\cN=2$ chiral multiplet $\Phi_\varphi = (\varphi,\psi_{\varphi},\eta_{\varphi},F_{\varphi})$ in the adjoint representation. 
\item The $\cN=4$ hypermultiplet decomposes into a pair of $\cN=2$ chiral multiplets denoted by $\Phi_X = (X,\psi_X,\eta_X, F_X)$ and $\Phi_Y = ( Y,\psi_Y,\eta_Y, F_Y)$ transforming in the unitary representations $M$ and $M^*$ respectively.
\end{itemize}
Accounting for the $R$-charges summarised in table~\ref{tab:charges}, the chiral multiplets mentioned above transform as sections of the associated bundles
\bea
\label{associated bundle 1}
P_{\varphi}& :=\left(P\times_G \mathfrak{g}\right) \otimes K_\Sigma^{1-r} \\
P_X & :=\left(P \times_G M\right) \otimes K^{r/2}_\Sigma \\
P_Y & :=\left(P\times_G M^*\right) \otimes K_\Sigma^{r/2} \, ,
\eea
where
\be
r: = \begin{cases}
1 & \text{H-twist} \\
0 & \text{C-twist} \, .
\end{cases}
\ee
In addition there is an $\cN=2$ superpotential $W = \langle Y  , \Phi \cdot X \ra$ of $R$-charge $+2$ and transforms as a section of $K_\Sigma$. \footnote{By $\langle \cdot , \cdot \rangle$ we always denote the natural pairing between a space and its dual.}  These supermultiplets further decompose into 1d $\cN=(0,2)$ supermultiplets as detailed above.

Summing the above lagrangian contributions from these multiplets gives
\be \label{localizing action 1} 
\cL = \frac{1}{e^2} \cL_{\mathrm{YM}} + \frac{1}{g^2} \left(  \cL_X + \cL_Y + \cL_\Phi \right)+ \frac{1}{g_W^2} \cL_W +\cL_{\mathrm{FI}}
\ee
where we have inserted parameters $e^2$, $g^2$ and $g_W^2$ in front of the exact contributions.
By taking the limit as these parameters tend to zero, we can localise the path integral to the critical loci of the combinations $\cL_{\mathrm{YM}}$, $\cL_{X} + \cL_Y +\cL_\Phi$ and $\cL_W$ separately.
By imposing a suitable reality condition for all the fields except for the auxiliary field, the path integral localises to solutions of the following equations
\bea
~&*F = -iD\ ,~~\bar\partial_A X = 0 \quad \bar\partial_A Y = 0 \quad \bar\partial_A \varphi = 0 \ ,  \\
&d_A\sigma =0\ ,~~ F_{0 1} = F_{0\bar 1}=0\ ,~~ \sigma \cdot X=0 \quad \sigma \cdot Y=0 \quad \sigma \cdot \varphi=0 \ , \\
&\varphi \cdot X = 0 \quad \varphi \cdot Y = 0 \quad X \cdot Y = 0
\eea
as found in \cite{Closset:2016arn,Benini:2016hjo,Benini:2015noa}. This leads to a contour integral representation of the twisted index.

We consider here an alternative localising action akin to the one introduced in section 9 of reference~\cite{Closset:2015rna} in the context of the twisted partition function of 2d $\cN=(2,2)$ theories on $S^2$. In particular, we add a $(\delta +\t\delta)$-exact term,
\be\label{Q exact 1}
\cL_{\text {H}} = \frac{1}{2i}\left(\delta+\t\delta\right)\left[ \left(\lambda+\bar\lambda\right)\left(\mu_\R  -2 [\varphi^\dagger, \varphi] -\tau \right)\right] \, ,
\ee
whose bosonic part is
\be
\cL^{\text{bos}}_{\text{H}}= i\left(D-2F_{1\bar 1}\right) \left(\mu_\R -2 [\varphi^\dagger, \varphi]  -\tau \right)\ .
\ee
We emphasise that the parameter $\tau \in \mathfrak{t}^*$ is distinct from the physical FI parameter $\zeta$ introduced in equation~\eqref{FI term}. We then replace the vectormultiplet action by
\be\label{localizing action 2}
\frac{1}{{ e}^2}\cL_{\mathrm{YM}}~\rightarrow~ \frac{1}{t^2}\left(\frac{1}{e^2}\cL_{\mathrm{YM}} + \cL_{\text{H}}\right)
\ee
and consider the limit as $ t,g\rightarrow 0$ such that $t/g\rightarrow 0$, while keeping $e$ finite. After integrating out the auxiliary field, the path integral localises to configurations solving the following set of `generalised vortex equations' on $\Sigma$, 
\begin{gather}
*F_A+ e^2 \left(\mu_\R -2[\varphi^\dagger , \varphi ] -\tau  \right) = 0  \quad d_{A} \sigma = 0 \nonumber \\
 \bar\partial_A X= 0 \quad \bar\partial_AY= 0 \quad \bar\partial_A\varphi = 0  \nonumber \\
\varphi \cdot X = 0 \quad \varphi \cdot Y = 0 \quad X \cdot Y = 0 \nonumber \\
\sigma \cdot \varphi= 0 \quad \sigma \cdot X= 0 \quad \sigma \cdot Y =0
\label{eq:vortex}
\end{gather}
where it is understood that $\sigma$, $\varphi$ and $\bar\partial_A$ act in the appropriate representation. In addition, there are fermion zero modes in the background of such configurations. These must combine to form 1d $\cN=(0,2)$ supermultiplets whose structure will be elucidated in subsequent sections.

\subsection{The Vortex Moduli Space}
\label{sec:moduli space}

We now consider the moduli space of solutions to the generalised vortex equations~\eqref{eq:vortex} for the class of supersymmetric theories introduced in section~\ref{sec:background}. Recall that we consider quiver gauge theories with $G = \prod_{I=1}^L U(N_I)$. 

First, solutions of the generalised vortex equations form topologically distinct sectors labelled by the flux
\be
\m_I := \frac{1}{2\pi} \int_{\Sigma} \tr ( F_I ) \, .
\ee
We can equivalently write $\mathfrak{m}_I = c_1(\cV_I)$ where $\cV_I$ denotes the vector bundle on $\Sigma$ in the fundamental representation of $U(N_I)$. We use shorthand notation $\m := \{\m_I\}\in \mathbb{Z}^L$.

 The allowed fluxes $\m \in \mathbb{Z}^L$ generate a lattice in the Lie algebra of the abelian part of $G$. The latter can be identified with the dual of the Cartan subalgebra the Coulomb branch flavour symmetry, $\mathfrak{t}^\vee_C \cong \R^L$. The flux lattice is then naturally identified with the character lattice
\be
\Lambda^\vee_C := \mathrm{Hom}(T_C,U(1)) \, .
\ee
The homomorphism $\zeta  \mapsto e^{2\pi i \la \zeta, \m \ra}$ arises in the contribution to the path integral from the FI parameter. Through the identification~\eqref{eq:equiv-resol} the flux lattice is equivalently
\be
\Lambda_C^\vee \simeq  H_2(\cM_H,\mathbb{Z}) \, ,
\ee
which suggests that solutions of the generalised vortex equations are related to holomorphic maps $\Sigma \to \cM_H$ of degree $\m$. We will explain below in  what sense this is realised.

Second, the parameter $\tau \in \R^L$ appearing in the generalised vortex equations~\eqref{eq:vortex} arises from an exact contribution to the lagrangian. In what follows, we always choose this parameter to lie in the same connected component or chamber of the parameter space $\R^L \backslash \cup_\al H_\al$ as the physical FI parameter $\zeta$. 

The parameter $\tau$ (or more precisely the combination $e^2\tau$) controls the tension or inverse size of the vortex solutions. We therefore expect an intricate  dependence on $\tau$ and $\mathrm{Vol}(\Sigma)$ as the moduli space may jump when walls are crossed in the parameter space where additional vortices may be supported. \footnote{We will discuss the moduli space of gauge theories at finite $\tau$ and their wall-crossing phenomena in upcoming work \cite{toappear}.} In order to obtain a uniform description of the moduli space of solutions to~\eqref{eq:vortex} for all fluxes $\mathfrak{m} \in \mathbb{Z}^L$, we will send the parameter $\tau \to \infty$ within the appropriate chamber of $\R^L \backslash \cup_\al H_\al$. This corresponds to an `infinite-tension' limit where an arbitrary number of vortices can be supported.

In the infinite tension limit, the magnetic flux is concentrated at a finite set of points $P$ on $\Sigma$. Provided we restrict to $\Sigma - P$, the magnetic flux may be neglected in the first line of equation~\eqref{eq:vortex} and therefore
\be
\mu_\R -2[\varphi^\dagger , \varphi ]  = \tau  \, .
\ee
This is identical to the $D$-term equation in the $\cN=2$ supersymmetry description of the Higgs branch described in equation~\eqref{eq:hbN=2}. Under the assumptions of section~\ref{sec:assumption}, solution of the generalised vortex equations therefore have the property that for each point in $\Sigma -P$, $\sigma = \varphi = 0$ and they determine a point on $\cM_{H,\tau}$ . Together with the remaining equations in~\eqref{eq:vortex} this is sufficient to determine that $\sigma = \varphi = 0$ everywhere.

In the infinite tension limit, it is therefore sufficient to restrict our attention to the following system of equations
\begin{gather}
*F_A+ e^2 \left(\mu_\R -\tau  \right) = 0 \nonumber \\
 \bar\partial_A X= 0 \qquad \bar\partial_AY= 0 \qquad  X \cdot Y = 0 
\label{eq:vortex-2}
\end{gather}
whose solutions with a fixed degree $\m \in \mathbb{Z}^L$ describe holomorphic maps $\Sigma \to \cM_H$ away from a finite set of points on $\Sigma$. 
Let us then denote the moduli space of solutions to the generalised vortex equations~\eqref{eq:vortex-2} modulo gauge transformations by $\mathfrak{M}$. As explained above, this is a disjoint union of topologically distinct components,
\be
\mathfrak{M} = \bigcup_{\m \in \Lambda_C^\vee} \mathfrak{M}_\m \, .
\ee
We emphasise that the moduli space encompasses both boson and fermion zero modes. More precisely, the moduli space is parametrised by the vacuum expectation values of both 1d $\cN=(0,2)$ chiral multiplets and 1d $\cN=(0,2)$ Fermi multiplets. In the following section, we explain that the algebraic description of the solutions to these equations coincide with that of `quasi-maps'.

\subsection{Algebraic Description}
\label{sec:algebraic}

To understand the vortex moduli spaces $\fM_\m$ and the mathematical interpretation of the twisted index, we consider an algebraic description of the moduli space of generalised vortex equations~\eqref{eq:vortex-2} in the infinite-tension limit $\tau \to \infty$. We show that this description coincides with two variations of moduli spaces of stable quasi-maps $\Sigma \to \cM_H$ in the H-twist and C-twist respectively.

As for the Higgs branch, the algebraic description of the moduli space $\fM_\m$ is found schematically by removing the D-term vortex equation from~\eqref{eq:vortex-2} in favour of a stability condition and dividing by complex gauge transformations (under which the equation $X\cdot Y$ is invariant). A solution is then 
represented by the following holomorphic data:
\begin{itemize}
\item A holomorphic $G_\C$-bundle $E$ on $\Sigma$;
\item Holomorphic sections $X$, $Y$ of the associated holomorphic vector bundles $E_X$, $E_Y$ subject to the complex moment map constraint $\mu_\C = X \cdot Y = 0$;
\item Subject to a stability condition;
\end{itemize}
and modulo complex gauge transformations. We refer to a collection of such algebraic data as $(E,X,Y)$. This associates to each point on $\Sigma$ a point in $\mu_\C^{-1}(0) \subset T^*M$. We can therefore regard this algebraic data as a twisted holomorphic map $\Sigma \to \mu_\C^{-1}(0)$ of degree $\m$.

Let us now consider the stability condition arising from the vortex equation,
\be
*F_A+ e^2 \left(\mu_\R -\tau  \right) = 0 \, .
\label{eq:vortex-real}
\ee
The determination of the relevant stability condition depends intricately on the choice of parameter $\tau$ and has been studied extensively in particular examples~\cite{Banfield:2000,AlvarezConsulGarciaPrada:2003,BiswasRomao:2013}. 

The infinite tension limit leads to a simplification in the stability condition: away from a finite set of points on $\Sigma$ the curvature term in equation~\eqref{eq:vortex-real} can be ignored and the image of the map $\Sigma \to \mu_\C^{-1}(0)$ determined by the algebraic data must lie in the stable locus $\mu_\C^{-1}(0)^s$. This is precisely the stability condition introduced in~\cite{CIOCANFONTANINE201417,Okounkov:2015spn,kim:2016} to define quasi-maps $\Sigma \to \cM_H$. Accounting for the R-charges as in \eqref{associated bundle 1}, in the C-twist we therefore have an algebraic description of $\fM_\m$ as the moduli space of quasi-maps $\Sigma \to \cM_H$ of degree $\m\in \mathbb{Z}^L$ as considered in \cite{Okounkov:2015spn} for the special case $\Sigma=\CP^1$. In the H-twist, we have a similar algebraic description as twisted quasi-maps as described in \cite{kim:2016}.

\subsection{Virtual Tangent Bundle}
\label{sec:virtual}

We can further study this identification by computing the massless fluctuations around solutions of the generalised vortex equations~\eqref{eq:vortex}. By supersymmetry preserved on $S^1 \times \Sigma$, these fluctuations must organise into supermultiplets of 1d $\cN=(0,2)$ supersymmetry. We will demonstrate that the massless fluctuations reproduce the structure virtual tangent bundles or perfect obstruction theories for $\fM_\m$ considered in~\cite{CIOCANFONTANINE201417,Okounkov:2015spn,kim:2016}.

Let us fix a point on the moduli space represented by the algebraic data $(E,X,Y)$. Then each of the three-dimensional chiral multiplets $\phi =  X, Y, \varphi $ generates a pair of 1d $\cN=(0,2)$ supermultiplet fluctuations at this point:
\begin{itemize}
\item Chiral multiplets: $(\delta \phi, \psi_\phi) \in H^0(E_\phi)$.
\item Fermi multiplets: $(\eta_\phi) \in H^1(E_\phi)$.
\end{itemize}
In addition, the three-dimensional vectormultiplet contributes a chiral multiplet fluctuation $(\delta \bar A, \bar \Lambda) \in H^1(E_V)$, where $E_V$ is the holomorphic vector bundle associated with the adjoint representation, corresponding to deformations of the holomorphic vector bundle $E$ via the derivative operator $\bar\partial_A$, and a Fermi multiplet $\lambda \in H^0(E_V)$ corresponding to infinitesimal holomorphic gauge transformations.

Not all of these fluctuations remain massless. First, let us fix the holomorphic vector bundle $E$ and consider fluctuations of the hypermultiplets $(X,Y)$. For the scalar fluctuations $(\delta X, \delta Y)$, linearisation of the complex moment map equation $X \cdot Y = 0$ generates the complex of vector spaces
\be
H^0( E_V) \overset{\alpha^0}{\longrightarrow} H^0(E_X \oplus E_Y) \overset{\beta^0}{\longrightarrow} H^1(E_\varphi)^* \ ,
\label{eq:complex_0}
\ee
where the map $\al^0 : \delta g \mapsto (\delta g \cdot X,\delta g \cdot Y)$ is an infinitesimal complex gauge transformation and $\beta^0 : (\delta X, \delta Y) \to X \cdot \delta Y+ \delta X \cdot Y$ is the differential of the complex moment map. The massless fluctuations of the complex scalars lie in $\mathrm{Ker}(\beta^0)/\Im(\al^0)$. We note that under our assumptions $\al^0$ is injective.

The same result must hold for the fermion components $(\psi_X,\psi_Y)$ of the chiral multiplets by 1d $\cN=(0,2)$ supersymmetry but it is illuminating to check this explicitly. This can be understood from the Yukawa couplings with the Fermi multiplet fluctuations $\lambda \in H^0(E_V)$ and $\eta_\varphi \in H^1(E_\varphi)$. First, there is
\be\label{yukawa 1}
\int_\Sigma *  \la \lambda ,  \psi_X \cdot  X^\dagger \ra +  \int_\Sigma * \la \lambda ,   Y^\dagger \cdot  \psi_Y \ra \, .
\ee
Here we denote by $\la \cdot , \cdot \ra$ the pairing between the lie algebra $\mathfrak{g}$ and its dual $\mathfrak{g}^*$. Other contractions are implicit. Let us suppose that the fermion fluctuations take the form $(\psi_X,\psi_Y) = (\ep \cdot X ,  \ep \cdot Y)$ for some fermion $\ep \in H^0(E_V)$, meaning they lie in the image of $\alpha^0$. Then the above contributions are proportional to
\be\label{massive epsilon}
\int_\Sigma   * \la \mu_\R\left( X,Y \right) ,  \lambda \epsilon \ra .
\ee
By the stability condition, the real moment map cannot vanish identically on $\Sigma$ and therefore this coupling generates a mass for the fermions $\epsilon$ and $\lambda$. We conclude that the fermion fluctuations $(\psi_{X},\psi_Y)$ in the image of $\alpha^0$ become massive. 
Second, the superpotential~\eqref{eq:superpotential} generates the Yukawa couplings
\be\label{yukawa 2}
\int_{\Sigma}  \langle \, \eta_{\varphi} \, , \, X \cdot \psi_Y + \psi_X \cdot Y  \, \rangle \ .
\ee
It is clear that if the fermion fluctuations satisfy $X \cdot \psi_Y + \psi_X \cdot Y = 0$, then the sum of these couplings vanish and these fluctuations are massless. Otherwise they pair up with $\eta_\varphi$ to become massive. We therefore conclude that the remaining massless fluctuations $(\psi_X, \psi_Y)$ lie in $\mathrm{Ker}(\beta^0) / \Im(\al^0) $. In addition there are Fermi multiplet fluctuations $\bar \eta_\varphi$ in the cokernel of $\beta^0$. In summary, there are massless 1d $\cN=(0,2)$ fluctuations given by the cohomology of the complex~\eqref{eq:complex_0}.

Let us now return to consider fluctuations of the holomorphic bundle $E$ via the derivative operator $\bar\partial_A$. Deformations of the holomorphic vector bundle $E$ correspond to elements in $H^1(E)$. However, these deformations must be such that $(X,Y)$ remain holomorphic sections, meaning they lie in the kernel of the map
\bea
\alpha^1 : H^1(E_V) \longrightarrow H^1(E_X \oplus E_Y)\ ,
\eea
where $\al^1 : \delta \bar A \to (\delta \bar A \cdot X , \delta \bar A \cdot Y )$. The same condition must hold for the fermion component of the chiral multiplet $(\delta \bar A, \bar\Lambda)$, but it is again illuminating to show this directly. This follows by noting that the Yukawa couplings
\be
\int_{\Sigma} \, \la  \bar\Lambda , X \bar\eta_X \ra + \int_{\Sigma} \la \bar\Lambda  ,  Y    \bar\eta_Y \ra
\ee
vanish when the fermion $\bar \Lambda$ lies in the kernel of $\al^1$.

Finally, let us consider the chiral multiplet fluctuations $(\delta \varphi, \psi_\varphi) \in H^0(E_\varphi)$. The complex scalar fluctuations must obey $\delta \varphi  \cdot X = 0$ and $\delta \varphi \cdot Y   = 0$, which means that they lie in the cokernel of the map
\be
\beta^1 : H^1(E_X \oplus E_Y) \to H^0(E_\varphi)^*\, ,
\ee
where $\beta^1 : (A,B) \to X \cdot B + A \cdot Y$. Under our assumptions, $\varphi$ vanishes identically on solutions to the generalised vortex equations and therefore $\beta^1$ is surjective. Once again, the same condition must hold for the fermion components of the supermultiplet. This time we consider the remaining Yukawa couplings
\be
\int_{\Sigma} \, \langle \, \psi_\varphi \, , \, X \cdot \eta_Y + \eta_X \cdot Y \, \rangle  \, ,
\ee
which shows that the combination of fermion fluctuations $X \cdot \eta_Y + \eta_X \cdot Y$ that are not kernel of $\beta^1$ pair up with the fluctuations $\psi_\varphi$ and become massive.

In summary, the massless fluctuations around a point on the moduli space $\fM$ of quasi-maps $\Sigma \to \cM_H$ represented by algebraic data $(E,X,Y)$ are encoded in the cohomology of the following pair of complexes
\bea
\label{virtual tangent space 3}
& H^0( E) \overset{\alpha^0}{\longrightarrow} H^0( E_X \oplus E_Y) \overset{\beta^0}{\longrightarrow} H^1(E_\varphi)^* \\
& H^1( E) \overset{\alpha^1}{\longrightarrow} H^1( E_X \oplus E_Y) \overset{\beta^1}{\longrightarrow} H^0(E_\varphi)^* \ .
\eea
This can be promoted to a complex of $G_H \times U(1)_t$ equivariant sheaves on the moduli space $\mathfrak{M}$ using the universal construction on $\mathfrak{M} \times \Sigma$. The starting point is the universal $G$-bundle $\cP \to \mathfrak{M} \times \Sigma$.  We then have
\be\label{virtual tangent space 4}
R\pi^\bullet( \cP ) \overset{\alpha}{\longrightarrow} R\pi^\bullet( \cP_X \oplus \cP_Y) \overset{\beta}{\longrightarrow} R\pi^\bullet (\cP_\varphi)^* \ ,
\ee
where $\pi : \mathfrak{M} \times \Sigma \to \mathfrak{M}$ is the other projection and the associated vector bundles $\cP_X, \cP_Y, \cP_\varphi$ are defined as before in \eqref{associated bundle 1} using the pullback $\cK = f^* K_\Sigma$ where $f : \mathfrak{M} \times \Sigma \to \Sigma$ is the projection. Note that this mirrors the structure of the complex whose cohomology computes the tangent bundle to $\cM_H$ outlined in section~\ref{sec:hb}. In the remainder of the paper, we will mainly refer to $T^{\text{vir}}$ as the equivariant K-theory class of the complex \eqref{virtual tangent space 4}.

This construction coincides with the perfect obstruction theory constructed in \cite{Okounkov:2015spn} for $\Sigma = \mathbb{CP}^1$ in the C-twist \cite{kim:2016} and in H-twist on a general curve $\Sigma$ of genus $g$. The two obstruction theories have remarkably different features. The obstruction theories for the H-twist is symmetric, meaning that there is an isomorphism between the complex in degree 0 in \eqref{virtual tangent space 4} and the dual of the complex in degree 1. This implies that the virtual dimension of the moduli space is zero. In the C-twist the obstruction fails to be symmetric unless the curve is elliptic, so that the canonical bundle is trivial. A Hirzebruch-Riemann-Roch computation shows that
\be\label{virtual dimension}
\mathrm{dim}_\mathrm{vir} \left( \fM_\m \right)=
\begin{cases}
0  &\text{H-twist} \\
\mathrm{dim}\left(\mathcal{M}_H\right)(1-g) & \text{C-twist} \, .
\end{cases}
\ee
The difference between the two twists will be particularly manifest when we will attempt to give an interpretation of the twisted indices.

\subsection{Mass Parameters and Fixed Loci}

The moduli spaces $\fM_\m$ introduced above are in general expected to be non-compact. The presence of massless non-compact fluctuations would render the computation of the twisted index on $S^1 \times \Sigma$ ill-defined. To remedy this, we introduce real mass parameters for flavour symmetries that, as for the Higgs branch in section~\eqref{sec:hb}, will cut down the moduli space to the fixed locus of this flavour symmetry. 

The mass parameter for the $U(1)_t$ symmetry associated to the breaking to $\cN=2$ supersymmetry is enough to ensure the twisted index on $S^1 \times \Sigma$ is well-defined and identify its mathematical interpretation. Further introduction of mass parameters for $G_H$ will make the twisted index explicitly computable in our localisation scheme.

\subsubsection{$U(1)_t$ Mass Parameters}
\label{subsec: t mass}

Let us introduce the mass parameter $m_t$ for $U(1)_t$. The effect of this deformation is to replace $\sigma \to \sigma+ m_t$ in the generalised vortex equations~\eqref{eq:vortex}, where $m_t$ acts with the appropriate weight according to table~\ref{tab:charges}. 
The remaining moduli space of solutions is the fixed locus of the $U(1)_t$ action on $\fM_\m$. 

First recall from section~\ref{sec:masses} that turning on the mass parameter $m_t$ restricts the Higgs branch to a compact holomorphic Lagrangian known as the compact core $\cM_H^{U(1)_t} = \cL_H$. This is characterised by a holomorphic Lagrangian splitting $T^*M = L \oplus L^*$ such that he hypermultiplet fields in $L^*$ vanish on the fixed locus and $\cL_H = L /\!/_\zeta \, G$.

Similarly, solutions of the generalised vortex equations invariant under $U(1)_t$ correspond to configurations where the hypermultiplet fields in $L^*$ vanish and correspond algebraically to twisted quasi-maps $\Sigma \to \cL_H$ to the compact core. We denote the fixed locus of the moduli space by $\fM^{U(1)_t}_\m = \mathfrak{L}_\m$. Upon restriction to the fixed locus, the virtual tangent bundle splits into two pieces
\be\label{virtual tangent bundle splits}
H^\bullet(E) \overset{\alpha}{\longrightarrow} H^\bullet( E_L )  \qquad H^\bullet (E_{L^*}) \overset{\beta}{\longrightarrow} H^{1-\bullet} (E_\varphi)^* \ .
\ee
transforming with weight $0$ and $+1$ respectively under $U(1)_t$. They can be identified with the virtual tangent bundle to $\fL_\m$ and the virtual normal bundle respectively. At the level of K-theory classes we have
\be
\left. T^{\mathrm{vir}} \right|_{\fL} = T_{\fL}^{\vir} + t N\ ,
\ee
where $t = e^{2\pi i m_t}$ is the equivariant parameter for the $U(1)_t$ symmetry.

In the $H$-twist, the tangent and normal fluctuations at the fixed locus are related by Serre duality
\be
H^\bullet (E_L) = H^{1-\bullet} (E_{L^*})^*\ ,~~ H^\bullet (E) = H^{1-\bullet}(E_{\varphi})^*
\ee
and $N^{\text{vir}}= -\left(T_\fL^{\text{vir}}\right)^\vee$ as K-theory classes. Although it is not necessary in our computations, when the Higgs branch is a cotangent bundle we expect that the extended moduli space including fermionic fluctuations is a shifted cotangent bundle $T^*[-1]\fL$. In the C-twist, the virtual normal bundle $N^\text{vir}$ can be identified with the class of the complex
\be\label{C-twist normal 3}
H^\bullet (E\otimes K_\Sigma)^* \longrightarrow H^\bullet(E_L\otimes K_\Sigma)^* 
\ee
by an application of Serre duality.


\subsubsection{$G_H$ Mass Parameters}
\label{subsec: GH mass}
Let us now introduce real mass parameters $m \in \mathfrak{t}_H$ and consider localisation with respect to $T_H\subset G_H$. Under our assumption that fixed points $\{v_I\}$ of $\cM_H$ are isolated, the fixed locus in $\fM$ corresponds to a union of $\fM_I$, where the gauge group $G$ is broken to its maximal torus
\be
G\rightarrow U(1)^{\text{rk}(G)}\ .
\ee
Then the associated degree $\m$ vector bundle $E$ decomposes into the sum of line bundles
\be\label{sum of line bundles}
E = L_1 \oplus \cdots \oplus L_{\text{rk}(G)}\ ,
\ee
where deg$(L_i) = \m_i$. The rk$(G)$-vector $\underline{\m} = (\m_1,\cdots, \m_{\text{rk}(G)})$ is valued in the co-character lattice $\Lambda_G$ of the gauge group $G$, and satisfies the relation $\text{tr}(\underline{\m}) = \m$. This implies that each fixed locus $\fM_I$ can be further decomposed into
\be
\fM_I = \bigcup_{\underline{\m}\in \Lambda_{G}} \fM_{\underline{\m},I}\ .
\ee

Furthermore, integrating the abelianised vortex equations over $\Sigma$, one can show that there are exactly rk$(G)$ non-vanishing chiral multiplet fields $Z_a$, $a=1,\ldots,\text{rk}(G)$ at each fixed locus $I$. This corresponds to the isolated vacua $v_I$ in \eqref{isolated vacua 2}. Then each component of the fixed locus $\fM_{\underline{\m},I}$ parametrises the holomorphic line bundle $L_a$ together with non-vanishing holomorphic section $Z_a$, which can be identified with
the rk$(G)$-fold product of symmetric products of a curve $\Sigma$
\be\label{fixed loci sec3}
\fM_{\underline{\m},I} = \prod_{a=1}^{\text{rk}(G)} \mathrm{Sym}^{\m_a+r(g-1)}\Sigma\ .
\ee
This is a compact smooth K\"ahler manifold of complex dimension $\m+ \text{rk}(G)r(g-1)$.

Now the massless fluctuations transform in the tangent bundle to the fixed loci $T\fM_{\underline{\m},I}$ and the remaining fluctuations are massive. This corresponds to a decomposition of the virtual tangent bundle 
\be
T^{\mathrm{vir}}|_{\fM_{\underline{\m},I}} = T\fM_{\underline{\m},I} + N_{\underline{\m},I} \ ,
\ee
where the virtual normal bundle $N_{\underline{\m},I}$ encodes the fluctuations that have become massive upon turning on the mass parameter. These two contributions are known as the `fixed' and `moving' parts and are characterised as those transforming with trivial weight and non-trivial weight under the $T_H \times U(1)_t$ transformation generated by the mass parameters $m_H$, $m_t$.

\subsection{Evaluating the Partition Function}
\label{sec:evaluating}
The path integral of the twisted index computes the generating function of the equivariant virtual Euler characteristic of the moduli spaces $\fM_\m$. This is defined by the following integral
\be\label{definition of virtual euler}
\chi(\fM, \hat\cO_{\text{vir}}) = \sum_{\m \in \Lambda_C^\vee}(-q)^{\m}\int_{\fM_\m} \hat A(T_{\text{vir}})\ ,
\ee
where $\hat A(T_{\text{vir}})$ is the A-roof genus of the virtual tangent bundle. This quantity has been extensively studied in \cite{Nekrasov:2014nea,Okounkov:2015spn} in the context of the enumerative geometry of curves in Calabi-Yau five-folds. The analogous construction for the four-dimensional Vafa-Witten invariants has been recently studied in \cite{Thomas:2018lvm}. 

Due to the non-compactness of $\fM_\m$, this formula should be evaluated with a proper virtual localisation theorem. 
Let us first consider the localisation with respect to the $U(1)_t$ action, which leads to the expression
\bea
\chi(\fM, \hat\cO_{\text{vir}}) & = \sum_{\m \in \Lambda_C^\vee}(-q)^{\m}\int_{[\fL_\m]}\frac{ \hat A\left(T^{\text{vir}}_{\fL_\m} \right) }{\text{ch}\left(\widehat\wedge^\bullet N_\m^{\vee}\right)} \\
& = \sum_{\m \in \Lambda_C^\vee}(-q)^{\m}\int_{[\fL_\m]}  \hat A\left(T^{\text{vir}}_{\fL_\m} \right)  \text{ch}\left(\widehat S^\bullet N_\m^{\vee}\right) \, .
\eea
Here we introduced the ``symmetrised" symmetric and exterior algebras
\be\label{normalised}
\widehat S^\bullet V := (\det V)^{1/2} \otimes S^\bullet V\ ,~~\widehat \wedge^\bullet V : =  (\det V)^{-1/2}\otimes \wedge^\bullet V\ ,
\ee
where 
\be
S^\bullet V = \bigoplus_{i\geq 0} S^iV\ ,~~\wedge^\bullet V = \bigoplus_{i\geq 0}(-1)^i \wedge^iV\ 
\ee
are the symmetric and exterior algebra of $V$. In the H-twist, the identification of the virtual normal bundle with $N = - (T^{\text{vir}}_\fL)^\vee$ means we can also interpret the twisted index as a symmetrised virtual $\chi_y$-genus with $y = -t$.



These integrals can be explicitly evaluated by a further localisation with respect to $T_H\subset G_H$. In turning on the real mass parameters $m$, we have seen that the solutions of the BPS equations are restricted to the fixed locus $\fM^T$, which is a disjoint union of smooth compact fixed loci. Let us denote the inclusion by $\sigma_{\underline{\m},I} : \fM_{\underline{\m},I} \hookrightarrow \fM$. Then the integral decomposes as a sum of contributions from the distinct components of the fixed locus
\bea\label{fixed point localisation}
\chi(\fM, \hat\cO_{\text{vir}}) 
& = \sum_{\underline{\m} \in \Lambda_G}(-q)^{\m}\sum_I\int_{\fM_{\underline{\m},I}}\frac{ \hat A\left(T\fM_{\underline{\m},I} \right) }{\text{ch}\left(\widehat\wedge^\bullet N_{\underline{\m},I}^\vee\right)}  \\
& = \sum_{\underline{\m} \in \Lambda_G}(-q)^{\m} \sum_{I}\int_{\fM_{\underline{\m},I}} \hat A \left(T\fM_{\underline{\m},I} \right)~ \text{ch}(\widehat S^\bullet N_{\underline{\m},I}^\vee)\ .
\eea

We note the individual contributions from the components of the fixed locus may be interpreted as the index of the Dirac operator on the smooth space $\fM_{\underline{\m},I}$ twisted by a complex of holomorphic vector bundles represented by $\widehat S^\bullet N_{\underline{\m},I}^\vee$. This is an expected form of the partition function of a finite-dimensional $\cN=(0,2)$ supersymmetric quantum mechanics with target space $\fM_{\underline{\m},I}$.

As discussed in the previous section, under our assumptions, the fixed loci are smooth products of symmetric products and these integrals can be evaluated explicitly. We will explore an extensive set of examples in section \ref{sec:examples}.

\subsection{Relation to Contour Integral Formulae}
\label{sec:contour}

One main focus of this paper is to provide a concrete 
geometric interpretation for the twisted indices of 3d $\cN = 4$ theories on $S^1 \times \Sigma$. For the class of theories we consider in this paper, the twisted index is defined as
\be\label{twisted index}
I[g] = \text{Tr}_{\cH}~(-1)^F e^{2\pi i \zeta \cdot J_C}e^{2\pi i m\cdot J_{H}}t^{J_t} = \sum_{\m \in \Lambda_{C}^\vee}\text{Tr}_{\cH_\m}~(-1)^F (-q)^\m  e^{2\pi i m\cdot J_{H}} t^{J_t}\ ,
\ee
where $\cH$ is the Hilbert space of states on $\Sigma$. This can be decomposed into topological sectors labelled by $\m \in \Lambda_C^\vee$. We defined $q = e^{2\pi i \zeta}$ and multiplied by $(-1)^\m$ for each topological sector for future convenience. $m$ is the real mass for the Higgs branch symmetry $G_H$.

The twisted indices of general $\cN=2$ gauge theories have been studied extensively in \cite{Nekrasov:2014xaa,Benini:2015noa,Benini:2016hjo,Closset:2016arn} from various perspectives. Below we briefly summarise the result obtained from UV Coulomb branch localisation \cite{Benini:2015noa,Benini:2016hjo,Closset:2016arn}. Using the $\delta,\t\delta$-exactness of the Yang-Mills action \eqref{n=2 ym} and the matter Lagrangian \eqref{n=2 chiral}, the path integral can be localised to the solutions to the BPS equations by taking the limit $e\rightarrow 0$ and $g\rightarrow 0$ in a careful way. In addition, we will add a $\delta,\t\delta$-exact term \eqref{Q exact 1} to the action as \eqref{localizing action 2} and take the limit $t,g\rightarrow 0$ with $t/g\rightarrow 0$ instead, in order to land on a particular bosonic moduli space considered in section \ref{sec:lagrangians}. This does not modify the procedure of the localisation computation, except for the contribution from the asymptotic boundary of the classical Coulomb branch which we briefly explain below and in appendix \ref{JK appendix}.

After carefully integrating out the fermionic zero modes, the localised path integral can be written as a $r$-dimensional residue integral over the classical coulomb branch parametrised by $u= i\beta(\sigma + ia_0)$ valued in a complexified maximal torus of $G$:
\be\label{index JK}
I[g] = \frac{(2\pi i)^{\text{rk}(G)}}{|W_{G}|}\sum_{\underline\m \in \Lambda_{G}}
\sum_{u_* = \{u_i\}} \underset{u=u_*}{\text{JK-Res}}(Q_{u_*}(u),\eta)~ (-q)^\m ~Z_{\text{1-loop}}^{\text{vector}}(u) Z_{\text{1-loop}}^{\text{hyper}}(u,m) H(u,m)^g~d^{\text{rk}(G)}u \ .
\ee
The summation is over the GNO quantized flux $\underline\m$ valued in the co-character lattice of the gauge group $\Lambda_{G}$, and $W_G$ is the Weyl group of $G$. For $G = U(k)$, the lattice elements in \eqref{index JK} and \eqref{twisted index} are related by tr$(\underline\m) = \m \in \mathbb{Z}$.
The one-loop determinants evaluated at the BPS locus are
\be\label{vector 1-loop}
Z_{\text{1-loop}}^{\text{vector}}(u) = \left(2i\sin \pi m_t\right)^{(g-1)(2r-1)\text{rk}(G)} \prod_{\alpha \in \Delta} \frac{(2i\sin \pi\alpha(u))^{\alpha(\underline\m)+(g-1)(2r-1)}}{(2i\sin \pi (\alpha (u)-m_t))^{\alpha(\underline\m) - (g-1)(2r-1)}}
\ee
and
\be\label{hyper 1-loop}
Z_{\text{1-loop}}^{\text{hyper}}(u,m) = \prod_i\prod_{\rho \in M} \frac{\left(2i\sin\pi  \left(-\rho(u)- m_i+\frac{m_t}{2}\right)\right)^{\rho(\underline\m) -(g-1)(r-1)}}{\left(2i\sin\pi  \left(\rho(u)+ m_i+\frac{m_t}{2}\right)\right)^{\rho(\underline\m) + (g-1)(r-1)}}\ ,
\ee
where $\Delta$ is the set of all roots of $\mathfrak g$ and $\rho$ is the weights in a complex representation $M$ of $G$. Note that we adopt a symmetric quantization for the one-loop computation, because this will be in accordance with computation of the virtual Euler characteristic constructed from the normalised symmetric and exterior algebra \eqref{normalised}. The last term in \eqref{index JK} can be obtained from integrating out the gaugino zero modes $\Lambda_1, \Lambda_{\bar 1}$:
\be
H(u,m) = \underset{ab}{\text{det}} \left[H^{\text{vector}}_{ab}(u)+ H^{\text{hyper}}_{ab}(u,m)\right]\ ,
\ee
where
\be
H^{\text{vector}}_{ab}(u) = \sum_{\alpha \in { \Delta}} \alpha^a\alpha^b\frac{\cos \pi(\alpha(u)-m_t)}{2i\sin \pi (\alpha(u)-m_t)}
\ee
and
\be
H_{ab}^{\text{hyper}}(u,m) =\sum_i\sum_{\rho\in M} \rho^a\rho^b \left(\frac{\cos \pi (\rho(u)+m_i + m_t/2)}{2i\sin \pi (\rho(u)+m_i + m_t/2)} + \frac{\cos \pi (-\rho(u)-m_i + m_t/2)}{2i\sin \pi (-\rho(u)-m_i + m_t/2)} \right)\ .
\ee

The integrand of \eqref{index JK} has four types of singular hyperplanes in the domain of the $u$-integral, where each of the hyperplane $H_Q$ is assigned a charge vector $Q\in {\frak t}^*$:
\begin{itemize}
\item There exist potential singularities where a chiral multiplet becomes massless:
\be \label{hyper sing}
H_{\pm\rho} = \left\{ u \in t_{\mathbb{C}} ~\Big|~ \pm \rho(u)\pm m + \frac{m_t}{2} = 0\right\}
\ee
The order of the pole is $\pm \rho(\underline{\m}) + (g-1)(r-1)+g$.
\item For each $\alpha \in \Delta$, there exist potential singularities at
\be\label{adjoint chiral sing}
H_{\alpha}^\Phi =  \left\{ u \in t_{\mathbb{C}}  ~\Big|~\alpha(u)-m_t = 0\right\}\ ,
\ee
where the adjoint chiral multiplets in the $\cN=4$ vector multiplets become massless.
\item For $g>1$, we have additional singularities at
\be\label{W-boson sing}
H_{\alpha}^W =  \left\{ u \in t_{\mathbb{C}}  ~\Big|~\alpha(u) = 0\right\}\ ,
\ee
where the W-boson becomes massless.
This singularity corresponds to the boundary of the Weyl chamber, where the non-abelian gauge symmetry enhances.
\item Finally, the integrand can have a potential singularity at
\be\label{asymptotic hyperplane}
H_{Q_{a,\pm}} = \left\{ u \in t_{\mathbb{C}}  ~\Big|~u_a\rightarrow \pm i\infty\right\}\ .
\ee
The behaviour of the integrand at infinity is governed by the charge of the monopole operators $T^\pm$ under the gauge and global symmetries in the theory \cite{Benini:2015noa,Benini:2016hjo,Closset:2016arn}, whose explicit form is not needed for this paper.
\end{itemize}

The integral is given by a sum of the $\text{rk}(G)$-dimensional residue integral over the poles defined by the intersections of $\text{rk}(G)$ singular hyperplanes. The JK-integral \cite{JK1995,1999math......3178B,2004InMat.158..453S} is defined by the property
\be\label{JK definition}
\underset{u=u_*}{\text{JK-Res}}(Q_{u_*}(u),\eta)\left[ \frac{d^{\text{rk}(G)}u}{Q_1(u)\cdots Q_{\text{rk}(G)}(u)}\right] = \left[\begin{array}{cc}\displaystyle\frac{1}{|\text{det}(Q_1,\cdots ,Q_{\text{rk}(G)})|} ~~& \text{if }\eta\in \text{Cone}(Q_1,\cdots, Q_{\text{rk}(G)}) \\
0 & \text{else}\end{array}\right.
\ee
where $\text{Cone}(Q_1,\cdots, Q_r)$ is the positive cone spanned by the charge vectors $(Q_1,\cdots, Q_r)$. 
This definition includes the charge vector $Q_{\pm}$ from the hyperplanes at asymptotic infinities \eqref{asymptotic hyperplane}. The charges $Q_{a,\pm}$ of these hyperplanes can be defined by examining the integral of the auxiliary field $\hat D$ in the large $u$ region. For $G=U(1)$, we show in appendix \ref{JK appendix} that the natural choice is \footnote{The definition of the charge of the pole at infinity $Q_{a\pm}$ is different from that of \cite{Benini:2016hjo,Closset:2016arn}. This is because the localising action we used in this paper, \eqref{localizing action 2}, modifies the integral of the auxiliary field $\hat D$ in the large $|u^a|$ region as discussed in appendix \ref{JK appendix}.}
\be
Q_\pm = \frac{2\pi \m}{e^2}-\text{vol}(\Sigma)\tau\ ,
\ee
for each GNO flux sector $\m \in \mathbb{Z}$, where $\tau$ is the parameter we introduced in \eqref{Q exact 1}.

Each residue integral defined by \eqref{JK definition} depends on the auxiliary parameter $\eta\in {\frak t}^*$. In this paper, we will choose
\be\label{choice of eta}
\eta = -\frac{2\pi \m}{e^2}+ \text{vol}(\Sigma) \tau :=\eta_0\ ,
\ee
so that the residue integral \eqref{index JK} does not pick up the poles involving the hyperplane at asymptotic boundaries. For $G=U(1)$ theory, one can show that the residue integral does not depend on the choice of $\eta$, due to the residue theorem.
As discussed in section \ref{sec:moduli space}, we will work on the chamber where $\tau$ is sufficiently large, so that we have $\eta>0$ for all values of $\m \in \mathbb{Z}$. We claim that this procedure can be generalised to the non-abelian gauge theories considered in this paper. Note that the Jeffrey-Kirwan residue integral operation in \eqref{JK definition} is ill-defined for the poles which intersect with the W-boson singularities \eqref{W-boson sing}. These singularities need to be properly resolved, and following \cite{Benini:2016hjo,Closset:2016arn}, we will exclude the residues from these poles in the final formula. \footnote{For non-abelian $G$ with $g>0$, the final result may depend on the choice of $\eta$, once we exclude the poles from the W-boson singularities. For the theories we consider in this paper, however, we will show in section \ref{subsec:SQCD} that the uniform choice $\eta = \eta_0 >0$ with the residues from the W-boson singularities excluded reproduces the correct integral representation of the Euler characteristics of the moduli spaces in the $\tau\rightarrow \infty$ chamber.}

Integrating the D-term equation \eqref{eq:vortex-2} over $\Sigma$, we obtain
\be\label{real moment map}
\int_{\Sigma}*~\mu_{\mathbb{R}} = \eta_0\ .
\ee
From this relation we can check that the poles that passes the JK condition with the choice \eqref{choice of eta} are in one-to-one correspondence with the fixed loci of the moduli space \eqref{fixed loci sec3}. In particular, the real moment map condition \eqref{real moment map} can be explicitly written as
\be
\sum_i Q_i^a \int_\Sigma * |\phi_i^a|^2 = \eta_0^a\ ,~~~ \forall a = 1,\cdots, \text{rk}(G)\ ,
\ee
where the index $i$ runs over all the chiral multiplets in the theory. This equation implies that the rk$(G)$-vector $\eta_0$ is in the positive cone of the charge vectors $\{Q_i^a\}$, which is precisely the condition that selects the charge vectors in the JK-residue integral \eqref{JK definition}.

Furthermore, the poles that involve the hyperplanes of type \eqref{adjoint chiral sing} do not contribute to the integral as the residues of such poles always vanish due to the order of zeros in the numerator \eqref{hyper 1-loop}. Therefore, for the class of theories we consider, the non-trivial contributions are from the residue integrals which consist of the first type of hyperplanes \eqref{hyper sing} only.
They correspond to the fixed loci
\be
\sum_I \prod_{a=1}^{\text{rk}(G)}\text{Sym}^{\m_{I_a}+r(g-1)}\Sigma\ ,
\ee
parameterised by the sum of the line bundles \eqref{sum of line bundles} and non-vanishing sections thereof.   
This discussion gives a geometric interpretation of the contour expressions, which we extensively study with various examples in section \ref{sec:examples}. In particular, using the intersection theory of the symmetric product of a curve $\Sigma$ studied in \cite{macdonald1962symmetric,thaddeus1994stable}, the contour integral expressions of the twisted indices can be converted to the equivariant integrals computing the virtual Euler characteristics discussed in section \ref{sec:evaluating}. This provides a powerful way to compute enumerative invariants of moduli spaces of quasi-maps.

\section{The Limit $t \to 1$}
\label{sec:t1limit}

As discussed in section \ref{sec:twisting}, compactifying 3d $\cN=4$ theories on a Riemann surface preserves 1d $\cN=(2,2)$ or $\cN=(0,4)$ supersymmetry in the H- and C-twist respectively.
So far, we have considered a localisation scheme which preserves a $\cN=(0,2)$ subalgebra only. Once we turn off the $U(1)_t$ mass parameter, we can add various exact terms to the localising action with respect to the supercharges that do not commute with the $U(1)_t$ symmetry. This further constrains the BPS moduli space and the twisted indices in the limit $t\rightarrow 1$ are expected to provide a geometric invariant for a reduced moduli space.

As we will see, the localisation scheme which preserves four supercharges turns out to be most powerful in the C-twist, where we can reduce the bosonic BPS moduli space to the Higgs branch itself, and the twisted indices can be interpreted as the Rozansky-Witten invariants \cite{Rozansky:1996bq} of the Higgs branch $\cM_H$. From the 3d mirror symmetry that exchanges the C- and H-twist, these considerations imply remarkable statements relating invariants of very differently-looking spaces, which we elaborate on in section \ref{sec:mirror-symmetry}.

The notation for the fields and the supersymmetry algebra used in this section are summarized in appendix \ref{app:susy algebra}.

\subsection{C-twist}

Let us start from the C-twist. In addition to the localizing action \eqref{localizing action 1} with the term \eqref{localizing action 2}, we can write down additional $Q$-exact terms using the four supercharges in the $\cN=(0,4)$ algebra:
\be
\frac{1}{t^2_C}\cL_{C,\text{vector}} =  \t \delta_C^2  \left( \t \lambda_1 V^{\dagger} \right)  +  \delta_C^2 \left( \lambda_1  V^{\dagger} \right) + \t \delta_C^1 \left( \t \lambda_2 V^{\dagger} \right)  + \delta_C^1\left(  \lambda_2 V^{\dagger} \right)\ ,
\ee
where
\be
V= \frac{1}{4 t^2_C}\left( \t \delta_C^2 \t \lambda_1 - \delta_C^2 \lambda_1 \right) - \frac{1}{4t_C^2}\left(\t \delta_C^1 \t  \lambda_2 - \delta_C^1 \lambda_2 \right)\ .
\ee
The bosonic part of this action is a total square
\be
\frac{1}{t^2_C} \|*F_A -2[\varphi^\dagger , \varphi]\|^2\ .
\ee
If we take the limit $t_C\rightarrow 0$, the field configuration of the vector multiplet localises to the intersection of \eqref{eq:vortex} and
\be
*F_A -2[\varphi^\dagger , \varphi]=0\ .
\ee
For the hypermultiplet, we can add
\be
\frac{1}{s_C^2} \cL_{C,\text{hyper}}[V,X] = \frac{1}{s_C^2} \left( -\t \delta_C^{ B} ( \psi_z \delta_{C,B} \t \psi_{\bar z} + \psi_{\bar z} \delta_{C,B} \t \psi_{ z}  ) - \delta_H^{B} (\psi_z \t \delta_{C,B} \t \psi_{z} + \psi_{\bar z} \t \delta_{C,B} \t \psi_{\bar z} )  \right) \, .
\ee
The bosonic part of this action is \be
\frac{1}{s_C^2} \cL_{C,\text{hyper}}^{\text{bosonic}} =
4D_{ 1} X_B D_{\bar 1} \t X^B+ 4 D_{\bar 1} X_B {D_1} \t X^B +  \varphi X_B \t  X^B \varphi^\dagger +  \varphi^\dagger X_B \t X^B \varphi \ .
\ee
Taking $s_C\rightarrow 0$, the path integral localizes to the equations
\be
D_1 X_A= D_{\bar 1}  X_A = \varphi \cdot X_B = \varphi^\dagger\cdot X_B =0 \ ,
\ee
which in particular implies that $X^A$'s are covariantly constant on $\Sigma$.

Combining these results, we can define the bosonic C-twisted $\cN=4$ moduli space $\cM_{\cN=4}$ to be the space of field configurations $(A, \varphi, X_B)$ satisfying the following set of equations:
\bea\label{C twist N=4}
&*F_A -2[\varphi^\dagger , \varphi] = 0\ ,\\
& \bar\partial_A\varphi =  0\ ,\\
& d_{A} X^B=0\ ,\\
&\varphi \cdot X_B = \varphi^\dagger\cdot X_B = 0 \ ,\\
&\mu_{\mathbb{R}} - \tau =0\ ,~\mu_{\mathbb{C}} =0\ .
\eea
Note that the equations for the vector multiplet fields $(A, \varphi)$ alone define the \emph{Hitchin moduli space} \cite{hitchin1987self} associated with the gauge group $G$. 
For the class of theories that we are interested in, the BPS equations \eqref{C twist N=4} imply $\varphi = 0$. 
Furthermore, the real moment map condition together with the condition that the sections $X_A$ are covariantly constant implies that the vector bundle $E$ must be trivial. Therefore the bosonic moduli space reduces to the Higgs branch $\mathcal{M}_H$ itself. Let us now look at the various contribution in the virtual tangent bundle. The first complex (the deformation space) in \eqref{virtual tangent space 3} reduces to 
\be
\mathfrak{g}_\C \overset{\alpha}{\longrightarrow} M \oplus M^* \overset{\beta}{\longrightarrow} \mathfrak{g}_\C^* \ ,
\ee
which defines the tangent space of $\cM_H$. Similarly, the second complex becomes
\be
\mathbb{C}^g \otimes\left[ \mathfrak{g}_\C \overset{\alpha}{\longrightarrow} M \oplus M^* \overset{\beta}{\longrightarrow} \mathfrak{g}_\C^* \right]\ .
\ee
This can be identified with the $g$ copies of the tangent bundle $T\cM_H$. In the limit $t\rightarrow 1$, the virtual Euler characteristic gets contributions from the zero-flux sector only and is therefore independent of $q$. In particular, we recover the holomorphic Euler characteristic of $\cM_H$ valued in $\left( \widehat{\wedge}^{\bullet} T^* \mathcal{M}_H \right)^g$,
\be \label{C-twist N=4 geometric}
\chi (\cM, \hat \cO_{\text{vir}})\big|_{t\rightarrow 1} = \chi \left( \mathcal{M}_H ,  \left( \widehat{\wedge}^{\bullet} T^* \mathcal{M}_H \right)^g \right)\ ,
\ee
which is the Rozansky-Witten invaraint on $\Sigma\times S^1$ associated with the Higgs branch $\cM_H$. Notice that the in this limit the virtual dimension \eqref{virtual dimension} is manifest. This relation between the twisted indices and the Rozansky-Witten invariants has been also studied in \cite{Gukov:2016gkn}.

\subsection{H-twist}

Similarly for the H-twist, we can write down additional $Q$-exact terms using the four supercharges in $\cN=(2,2)$ algebra. We choose
\be
\frac{1}{t_H^2} \cL_{H,\text{vector},1} =  \delta_H^1 \left( \lambda_{\dot 2} V^\dagger \right) + \delta_H^{\dot 2} \left( \t \lambda_1 V^\dagger \right)\ ,
\ee
where
\be
V= \frac{1}{4 t_H^2}\left(\delta_H^{\dot 1} \lambda_2 - \delta_H^{\dot 1} \t \lambda_1 \right) \, .
\ee
As in the C-twist, the bosonic part is a sum of squares, but now takes the form
\be
\|  *F_A  +  iD\|^2 ,
\ee
where $D$ is the auxiliary field for the $\cN=2$ vector multiplet. Solving the equation of motion for the $D$-term, and taking the limit $t_H\rightarrow 0$ gives rise to the condition \be\label{solution1}
* F_A  + e^2 (\mu_{\mathbb{R}}-\tau) =0 \, .
\ee
Therefore the H-twisted moduli space for the vector multiplet on $\Sigma$ can be viewed as the intersection of solutions to \eqref{eq:vortex} and \eqref{solution1}, which can be written as 
\footnote{Notice that there are interesting additional terms that could be added to the action. For example, 
\be
\frac{1}{t_H^2}  \cL_{H,\text{vector},2} = \frac{1}{4 t_H^2} \delta_H^{\dot 2} \left( \t \Lambda_{\bar 1 , \dot2}  V^\dagger \right) 
\ee
where 
\be
V=  \delta_H^{\dot 2} \t \Lambda_{\bar 1 , \dot 2} \, , 
\ee
whose bosonic part is
\be
\| D_{1} \varphi \|^2 \, .
\ee
This forces $\varphi$ and $\varphi^\dagger$ to be covariantly constant on $\Sigma$, not just covariantly holomorphic. We could also add terms coming from the hypermultiplet, giving $ \varphi^\dagger \cdot  X_A = \varphi\cdot X_A^\dagger = 0$ .}
\bea\label{N=4 H-twist}
& *F_A + e^2 (\mu_{\mathbb{R}}-\tau)  = 0\ , \\
& \bar\partial_A X_A = 0\ ,\\
& \bar\partial_A \varphi = [\varphi^\dagger , \varphi] = 0\ ,\\
& \varphi \cdot X_A = 0\ ,\\
& \mu_{\mathbb{C}}=0\ .
\eea

\noindent Since $\varphi$ vanishes on the moduli space under our assumptions, the bosonic moduli space remains the same as in the $\cN=2$ case. However, since $\varphi$ decouples from the D-term equation, the derivation of the stability condition simplifies.

As mentioned, the additional supercharges do not commute with $U(1)_t$ and therefore we consider the limit $t\rightarrow 1$ of the twisted index. In this limit, the virtual $\chi_t$-genus greatly simplifies to the generating function of the integral of the Euler class of the fixed loci $\fL_\m$ of the $U(1)_t$ action
\be\label{virtual euler section 4}
\chi({\fM,\hat\cO_{\text{vir}}})\big|_{t\rightarrow 1}  = \sum_{\m\in \Lambda_C^\vee} (-q)^\m (-1)^{\text{dim}_{\text{vir}}(\fL_\m)}\int_{[\fL_\m]} e\left(T^{\text{vir}}_{\fL_\m}\right)\ .
\ee
For the class of the theories we consider, the localisation with respect to the Higgs branch flavour symmetry $G_H$ provides an alternative expression for the index in the $t\rightarrow 1$ limit. Since the fixed loci $\fM_{\underline{\m},I}$ with respect to $T_H\subset G_H$ are smooth and compact, the expression \eqref{virtual euler section 4} can be explicitly evaluated by a computation of the sum of the Euler characteristic of the fixed loci:
\be\label{H-twist t->1}
\chi({\fM,\hat\cO_{\text{vir}}})\big|_{t\rightarrow 1} = \sum_{\underline{\m} \in \Lambda_G} (-q)^\m \sum_{I} (-1)^{\text{dim}_{\mathbb{C}}(\fM_{\underline{\m},I})}\int_{\fM_{\underline{\m},I}} e(\fM_{\underline{\m},I})\ ,
\ee

As discussed in the paper \cite{Hori:2014tda}, the supersymmetric ground states in the effective quantum mechanics that preserve $\cN=(2,2)$ supersymmetries are singlet under the flavour symmetry $G_H$. This agrees with the result \eqref{H-twist t->1}, which is independent of the equivariant parameters $m$.

\section{Examples}
\label{sec:examples}

In this section, we apply the strategy outlined above to some concrete examples. We explicitly prove that the virtual Euler characteristics of the appropriate moduli spaces of quasi-maps, computed via equivariant localisation \eqref{fixed point localisation}, reproduces the contour integral formulae of the twisted indices derived in \cite{Closset:2016arn,Benini:2015noa} and summarised in \ref{sec:contour}. For each example, we also discuss and verify interpretations that become available in the $t\rightarrow1$ limit, where $\mathcal{N}=4$ supersymmetry is restored, as anticipated in the previous section.

\subsection{Free hypermultiplets} \label{key section}

We start the study of our examples by briefly collecting some facts about the free hypermultiplet, since they are going to be useful in view of mirror symmetry. In $\cN=2$ language, the hypermultiplet corresponds to two chiral multiplets $\Phi_X$ and $\Phi_Y$, which have a $U(1)_H$ flavour symmetry, and which are charged as follows:
\begin{center}
\begin{tabular}{c|c|c}
& $U(1)_H$ & $U(1)_t$ \\
\hline
$X$ & $+1$ & $\frac12$\\
$Y$ & $-1$ & $\frac12$\\
\end{tabular}
\end{center}
For an arbitrary R-charge $r$, the index reads
\be\label{freehyper}
I = \left( \frac{\left(t^{1/2} x\right)^{1/2}}{1-t^{1/2}x} \right)^{\m_H + \m_t + (r-1)(g-1)}\left( \frac{\left(t^{1/2}/x\right)^{1/2}}{1-t^{1/2}/x} \right)^{-\m_H + \m_t + (r-1)(g-1)} \, ,
\ee
where $\m_H$ and $\m_t$ are the degrees of the line bundles $L_H$ and $L_t$, and $x$ and $t$ are the fugacities for $U(1)_H$ and $U(1)_t$ respectively. The two factors in \eqref{freehyper} correspond to the indices of $\Phi_X$ and $\Phi_Y$. The contribution of each $\cN=2$ chiral multiplet can be understood from the point of view of the 1d $(0,2)$ multiplets mentioned in \eqref{N=2chiralfields}. In fact, it is the index of a 1d quantum mechanics with $h^0(\Sigma, K_{\Sigma}^{r/2}\otimes L_H \otimes L_t)$ chirals and $h^1(\Sigma, K_\Sigma^{r/2}\otimes L_H \otimes L_t)$ fermi multiplets, whose difference is controlled by the Riemann-Roch theorem.

\subsection{SQED[1]}

Let us first consider a $U(1)$ gauge theory with a hypermultiplet $(X, Y^\dagger)$ having the following charges \footnote{Strictly speaking this theory falls short of the class we have previously defined in section \ref{sec:assumption}. However, the resolved Higgs branch is well-defined (it is a point) and the computations are still possible. This example contains the basic building blocks needed in more elaborate examples.
}
\be
\begin{array}{c|cc|cc}
&~ U(1)_{G}~& ~U(1)_t ~&~ U(1)_{H} ~&~ U(1)_{C} \\
\hline
X & 1 & \frac12 & \frac12 & 0 \\
Y & -1& \frac12 & \frac12 & 0 \\
\varphi & 0 & -1 & 0 & 1 
\end{array} \, ,
\ee
The $\mathcal{N}=2$ BPS equations become
\bea\label{moduli 2}
& * F_A + e^2 ( X X^\dagger - Y^\dagger  Y - \tau) = 0\ , \\
& \bar\partial_{A} X = \bar\partial_{A} Y =0\ ,~ X\cdot Y=0\ ,\\
&\bar\partial_{A}\varphi=0\ ,~ \varphi\cdot X =  \varphi \cdot Y=0 \ .
\eea
The moduli space of solutions to the above equations is a disjoint union of topological components
\be
{\frak M}= \bigcup_{\m \in \mathbb{Z}} {\frak M}_\m\ ,
\ee
indexed by the degree of the holomorphic line bundles $L$ associated to the connection $A$. $X$ and $Y$ are holomorphic sections of $L\otimes K^{r/2}$ and $L^{-1}\otimes K^{r/2}$ respectively. Integrating the D-term equation over $\Sigma$, we can check that $X$ is non-vanishing provided
\be\label{chamber plus}
\tau> \frac{2\pi \m}{e^2 \text{vol}(\Sigma)}\ .
\ee
Note that this condition is equivalent to the choice $\eta_0>0$ in the twisted index computation \eqref{choice of eta}.  
Since $X$ is a holomorphic section of a line bundle, the number of zeros of $X$ on $\Sigma$ is finite and equal to the degree of $L \otimes K^{r/2}$. The remaining BPS equations imply that $Y=\varphi=0$. Therefore the moduli space in this chamber is
\be
\fM_{\m}^+ = \big\{(A, X)~|~ * F_A + e^2 ( X X^\dagger  - \tau) = 0\ ,~  \bar\partial_{A} X =  0\big\}~/U(1)_{G}\  \, .
\ee
This space defines the moduli space of abelian vortices. The points in $\fM^+_\m$ can be parametrised by the zeros of $X$, or equivalently by divisors on $\Sigma$, which can be viewed as points in the $\m+r(g-1)$-th symmetric product of $\Sigma$. Importantly, given a divisor $D \in \Sigma$, the line bundle $L$ can then be recovered as $\mathcal{O}(D)$.  Introducing the notation
\be
\Sigma_n := \text{Sym}^n \Sigma \, ,
\ee
we have
\be\label{plus chamber}
\fM_{\m}^+ = \Sigma_{n_+}\ ,~~ n_+ = \m + r(g-1)\ .
\ee

The chamber $\tau <\frac{2\pi \m}{e^2\text{vol}(\Sigma)}$ can be treated in a similar way. The bosonic moduli space is constructed from non-vanishing sections $Y$ (when they exist) and their corresponding line bundles $L$, whereas $X$ is set to zero. In this chamber the bosonic moduli space becomes
\be
\fM_{\m}^{-} = \Sigma_{n_-}\ ,~~ n_- = -\m + r(g-1)\ .
\ee
For concreteness, we will work in the chamber \eqref{chamber plus} for all the flux sectors $\m \in \mathbb{Z}$ by formally sending $\tau \rightarrow \infty$, and omit the superscript $+$ from $\fM_\m^+$.

\paragraph {H-twist} In order to compute the index using virtual localisation, we need to study the virtual tangent space to $\fM_\m$. In the H-twist, the physical fluctuations around the bosonic moduli space are given by 
\bea
(\delta X, \psi_X) \in H^0(L\otimes K_\Sigma^{1/2})\ ,&~~(\eta_X,F_X) \in H^1(L \otimes K_\Sigma^{1/2})\ , \\
(\delta Y, \psi_Y) \in H^0( L^{-1}\otimes K_\Sigma^{1/2})\ ,&~~(\eta_Y,F_Y) \in H^1(L^{-1} \otimes K_\Sigma^{1/2})\ ,\\
(\varphi, \psi_{\varphi}) \in H^0( {\cal O})\ ,&~~(\eta_\varphi, F_\varphi) \in H^1({\cal O})
\eea
The virtual tangent space restricted to a point $D$ of the moduli space \eqref{virtual tangent space 3} corresponds therefore to the cohomology of the following two complexes:
\bea
&H^0({\cal O}) \overset{\alpha^0}{\longrightarrow} H^0( (\cO (D) \oplus \cO (D)^{-1})\otimes K_\Sigma^{1/2})\overset{\beta^0}{ \longrightarrow }H^1({\cal O})^* \ , \\
& H^1({\cal O}) \overset{\alpha^1}{\longrightarrow} H^1((\cO (D) \oplus \cO (D)^{-1})\otimes K_\Sigma^{1/2}) \overset{\beta^1}{\longrightarrow} H^0({\cal O})^* \ ,
\eea
where the map $\alpha$ is defined as multiplication by $(X,-Y)$, while $\beta$ is defined by taking an inner product with $(Y,X)$. Since $Y$ vanishes identically on the moduli space $\fM_\m$, these complexes split into two pieces each: 
\bea\label{sqed1 virtual}
&H^0({\cal O}) \overset{\alpha^0}{\longrightarrow} H^0( \cO (D)\otimes K_\Sigma^{1/2})\ ,~~ H^0(\cO (D)^{-1} \otimes K_\Sigma^{1/2})\overset{\beta^0}{ \longrightarrow }H^1({\cal O})^* \ , \\
& H^1({\cal O}) \overset{\alpha^1}{\longrightarrow} H^1( \cO (D)\otimes K_\Sigma^{1/2})\ ,~~ H^1(\cO (D)^{-1} \otimes K_\Sigma^{1/2})\overset{\beta^1}{ \longrightarrow }H^0({\cal O})^*\ . 
\eea
Let us first consider the cohomology of the two complexes on the left hand side.
The maps $\alpha^{0}$ and $\alpha^{1}$ are injective and surjective respectively and therefore the cohomology can be written as
\be\label{symmetrictangent}
T_D \fM_\m  = \text{ker}(\alpha^1) \oplus H^0(\mathcal{O}(D) \otimes K_\Sigma^{1/2})/\text{im}(\alpha^0)\ ,
\ee
which corresponds to the tangent space of the symmetric product at the point $D$. It follows that some of the massless fermionic fluctuations at the point $D$ encoded by the complexes span the tangent space to the bosonic moduli space, as expected \cite{Bullimore:2018yyb}. By Serre duality, it is then easy to see that the combination of the two complexes on the right of \eqref{sqed1 virtual} define the contangent space $T^*\fM_\m$ over the moduli space. Thus the virtual tangent space restricted on $\fM_\m$ is given by
\be
T_{\text{vir}}\big|_{\fM_\m} = T\fM_\m - T^* \fM_\m\ ,
\ee
where the second factor has weight $t$ under the $U(1)_t$ action. Hence, we can identify the virtual Euler characteristic with the holomorphic Euler characteristic valued in the exterior powers of the tangent bundle, 
which can be identified with the $\chi_t$-genus of the moduli space $\fM_\m$. This can be computed from the ordinary index theorem:
\be
\chi(\fM_\m, \hat \cO_{\text{vir}})  =\chi_t (\fM_{\m}) = \int_{\fM_{\m}} \hat A(T\fM_{\m}) ~ \text{ch}(\widehat\wedge^\bullet t~ T \fM_{\m}) \ ,
\ee
where $\widehat S^\bullet$ and $\widehat \wedge^\bullet$ are the normalised symmetric and exterior product defined in \eqref{normalised}.~\footnote{In standard notation for the Hirzebruch $\chi_y$-genus, this is $t^{-\text{dim}( \fM_\m)/2} \chi_{-t}(\fM_\m)$.}

 In order to relate this expression to the twisted index computation, we have to introduce classes over symmetric products as well as some useful identities. First of all, we introduce standard generators of the cohomology ring of the symmetric product $\Sigma_n$ following \cite{macdonald1962symmetric}: 
\be
\xi_i, \xi_i' \in H^1(\Sigma_n, \mathbb{Z})\ ,~~\eta \in H^2 (\Sigma_n, \mathbb{Z})\ .
\ee 
We also define the combination
\be
\sigma_i = \xi_i\xi'_i\ ,~~i=1,\cdots, g\ ~~\text{and}~~\sum_{i=1}^g \sigma_i = \sigma\ .
\ee
The generators $\xi_i$ and $\xi_i'$ anticommute with each other and commute with $\eta$ \footnote{There is also one last ring relation: 
\be
\xi_{i_1} \xi_{i_2} \ldots \xi_{i_a} \xi'_{j_1} \xi'_{j_2} \ldots \xi'_{j_b} \left( \xi_{k_1}\xi'_{k_1}-\eta \right)\ldots \left( \xi_{k_c}\xi'_{k_c}-\eta \right) \eta^q = 0
\label{ringrelations}
\ee
provided
\be
a+b+2c+q = n+1 \, .
\ee}.
The Chern class of the tangent bundle $T\Sigma_n$ is computed in \cite{macdonald1962symmetric}:
\be\label{chern class}
c(T\Sigma_n) = (1+\eta)^{n-2g+1}\prod_{i=1}^g (1+\eta - \sigma_i)\ ,
\ee
from which we obtain the Todd class:
\be
\text{td}(T\Sigma_n) = \left(\frac{\eta}{1-e^{-\eta}}\right)^{n-2g+1} \prod_{i=1}^g \frac{\eta - \sigma_i}{1-e^{-\eta+\sigma_i}}\ .
\ee
This formula can be simplifed by means of the following useful identity due to Don Zagier \cite{thaddeus1994stable}. For any power series $h(\eta)$ on $\Sigma_n$, we have the identity
\bea\label{zagier 1}
h(\eta)^{n-2g+1}\prod_{i=1}^g h(\eta-\sigma_i) &= h(\eta)^{n-g+1}\prod_{i=1}^g \left(1-\sigma_i\frac{h'(\eta)}{h(\eta)}\right)\\
&= h(\eta)^{n-g+1} \exp \left(-\sigma \frac{h'(\eta)}{h(\eta)}\right)\ ,
\eea
which follow from $\sigma_i^2 = 0$. 
If we choose $h(\eta) = \frac{\eta}{1-e^{-\eta}}$, we get
\be\label{td sym}
\text{td}(T\Sigma_n) = \left(\frac{\eta}{1-e^{-\eta}}\right)^{n-g+1}\exp \left(\frac{\sigma}{e^\eta-1}-\frac{\sigma}{\eta}\right)\ .
\ee
The $\hat A$ genus of the tangent bundle can be obtained from the todd class \eqref{td sym}:
\bea
\hat A (T\fM_{\m}) &= e^{-c_1(T\fM_\m)/2}~ \text{td}(T\fM_{\m}) \\
& = \left(\frac{\eta e^{-\eta/2}}{1-e^{-\eta}}\right)^{n-g+1}\exp \left(\frac{\sigma (e^\eta +1)}{2(e^\eta-1)}-\frac{\sigma}{\eta}\right)\ ,
\eea
with $n=n_+=\m+g-1$. Finally, the Chern character of the exterior powers of the tangent bundle can be obtained from  \eqref{chern class}. We find
\bea\label{ch y sym}
\text{ch}(\widehat\wedge^\bullet t~ T \fM_{\m})  =& \left(e^{\pi i m_t}-e^{-\pi i m_t}\right)^{g-1} \left(e^{-\eta/2+\pi i m_t}-e^{\eta/2-\pi i m_t}\right)^{n-2g+1} \\
&\prod_{i=1}^g \left(e^{\pi i m_t-(\eta-\sigma_i)/2}- e^{-\pi i m_t+(\eta - \sigma_i)/2}\right)\ ,
\eea
where $t=e^{2\pi i m_t}$.
Again using the identity \eqref{zagier 1}, we can simplify the expression to
\bea
\text{ch}(\widehat\wedge^\bullet t ~T \fM_{\m})  =&\left(e^{\pi i m_t}-e^{-\pi i m_t}\right)^{g-1} \left(e^{-\eta/2+\pi im_t}-e^{\eta/2-\pi i m_t}\right)^{n-g+1} \\ 
& \exp\left(-\frac{\sigma(1+ e^{-\eta+2\pi i m_t})}{2(1- e^{-\eta+2\pi i m_t})}\right)\ .
\eea
Combining all these expressions, we now have
\bea
\chi(\fM_\m, \hat\cO_{\text{vir}})= \left(e^{\pi i m_t}-e^{-\pi i m_t}\right)^{g-1} \int_{\Sigma_n} & \left(\frac{\eta \left(e^{-(\eta/2-\pi i m_t)}-e^{(\eta/2-\pi i m_t)}\right)}{e^{\eta/2}-e^{-\eta/2}}\right)^{n-g+1}\\
& \exp\left(\frac{\sigma (e^\eta +1)}{2(e^\eta-1)}-\frac{\sigma}{\eta}-\frac{\sigma(1+ e^{-\eta+2\pi i m_t})}{2(1- e^{-\eta+2\pi i m_t})}\right)\ .
\eea

The integral can be converted into the residue integral using the following identity, also due to Don Zagier \cite{thaddeus1994stable}. For any power series $A(\eta)$ and $B(\eta)$, one can show that
\be\label{zagier 2}
\int_{\Sigma_n} A(\eta)e^{\sigma B(\eta)}= \underset{u=0}{\text{res}}~ du~\frac{A(u)(1+u B(u))^g}{u^{n+1}}\ .
\ee
Note that this formula holds also for $n=0$ where $\Sigma_n= \text{pt}$. Using this identity, we find
\bea
\chi_t (\fM_{\m}) =  &~2\pi i \left(e^{\pi i m_t}-e^{-\pi i m_t}\right)^{g-1} \underset{u = 0}{\text{res}} \left(\frac{ e^{-\pi i (u-m_t)}-e^{\pi i (u-m_t)}}{e^{\pi i u}-e^{-\pi i u}}\right)^{n-g+1} \\ &\cdot \left(\frac{ e^{2\pi i u} +1}{2(e^{2\pi i u}-1)}-\frac{1+ e^{2\pi i(-u+m_t)}}{2(1- e^{2\pi i(-u+m_t)})}\right)^g\ .
\eea
This exactly reproduces the integral formula of the twisted index in the chamber $\tau > \frac{2\pi \m}{e^2 \text{vol}(\Sigma)}$. One can check that the residue is non-zero in the region
\be
-g + 1 \leq \m \leq g-1\ .
\ee
This is consistent with the geometric observation that $\Sigma_n$ becomes a holomorphic fibration over the Jacobian with fiber $\mathbb{CP}^{\m-1}$ when $\m>g-1$ \cite{macdonald1962symmetric}. In fact, the cohomology of $\Sigma_n$ factorizes into $H^\bullet (\Sigma_n) = H^\bullet (\mathbb{CP}^{\m-1}) \otimes H^\bullet \left(\text{Jac}[g]\right)$ and therefore the index vanishes in this region since $\chi_t\left(\text{Jac}[g]\right)=0$. Multiplying by the weight $(-q)^{\m}$ for each flux sector and summing over $\m$, we have 
\be
\chi_t(\fM) = \sum_{\m\in \mathbb{Z}} (-q)^{\m}\chi_t (\fM_{\m})  = (-q)^{-g+1} \left[(1-q t^{-1/2})(1-q t^{1/2})\right]^{g-1}\ ,
\ee
which agrees with the generating function of the $\chi_t$ genus of the symmetric product of $\Sigma$ computed in \cite{macdonald1962symmetric}, up to an overall sign. Notice that as dictated by mirror symmetry, this also agrees with the index of the C-twist of the free hypermultiplet in the absence of background fluxes, see \eqref{freehyper}.

In the limit $t\rightarrow 1$, because of the relation $\hat A(TM) \text{ch}(\widehat\wedge^\bullet TM) = (-1)^{\text{dim}_{\mathbb{C}}M}e(M)$ the virtual Euler characteristic becomes
\be\label{euler symmetric 5}
\chi_t(\fM)\big|_{t\rightarrow 1} =(-1)^{g-1} \sum_{\m\in \mathbb{Z}} q^\m \int_{\fM_\m} e(\fM_\m) = (-1)^{g-1}q^{-g+1}(1-q)^{2(g-1)}\ .
\ee 
This reproduces the generating function of the Euler characteristic of the symmetric product of $\Sigma$.

\paragraph{C-twist} In the case of the C-twist, the underlying moduli space is  $\mathfrak{M}_{\m}=\Sigma_\m$. The fluctuations of the various fields on $\fM_\m$ can be written as follow: 
\bea
(\delta X, \psi_{X}) \in H^0(L)\ ,&~~(\eta_{X},F_X) \in H^1(L)\ , \\
(\delta Y, \psi_{Y}) \in H^0(L^{-1})\ ,&~~(\eta_{Y}, F_Y )\in H^1(L^{-1} )\ ,\\
(\varphi, \psi_{\varphi}) \in H^0(K_\Sigma)\ ,&~~ (\eta_\varphi, F_\varphi) \in H^1(K_\Sigma)
\eea
where deg$(L)=\m$. In this case, the virtual tangent bundle at a point \eqref{virtual tangent space 3} coincides with\footnote{We omit the details about the maps, which we have already spelled out for the H-twist.}
\bea
& H^0({\cal O}) \longrightarrow H^0((L\oplus L^{-1}) )\longrightarrow H^1(K_\Sigma)^*  \\
& H^1({\cal O}) \longrightarrow H^1((L\oplus L^{-1})) \longrightarrow H^0(K_\Sigma)^*  \ .
\eea
$Y$ vanishes identically in the chamber \eqref{plus chamber} and the complex split in various pieces. Note furthermore that $H^0(L^{-1})$ is empty when $\m>0$. Let us first assume $\m>0$. Then the virtual tangent bundle restricted to the bosonic moduli space can be written as
\be \label{virtual 2}
T_{\text{vir}}\big|_{\fM_\m} = T\fM_\m + N_\m\ ,
\ee
where $T\fM_\m$ is again the tangent space of the underlying moduli space defined by the complexes
\be
H^0({\cal O}) \longrightarrow H^0(L) \ ,~~ H^1({\cal O}) \longrightarrow H^1(L) \ .
\ee
The second component $N_\m$ is the contribution from the normal bundle, which can be obtained from the cohomology of the remaining complex
\be
H^0(L^{-1})\rightarrow H^1(K_\Sigma)^*\ ,~~H^1(L^{-1})\rightarrow H^0(K_\Sigma)^*\ ,
\ee
which defines a smooth vector bundle whose class is 
\be\label{NM contribution}
[N_\m] = -[H^\bullet(L\otimes K_\Sigma)^*] + [H^\bullet(K_\Sigma)^*] \ .
\ee
Therefore, for the C-twist, the virtual Euler characteristic computes the holomorphic Euler characteristic valued in $\widehat S^\bullet N_\m^\vee$:
\be\label{glg}
\chi(\fM_\m, \hat \cO_{\text{vir}}) = \int_{\fM_\m} \hat A(T\fM_{\m}) \wedge \text{ch}(\widehat S^\bullet  N_\m^\vee)\ .
\ee
Note that this can be extended to $\m=0$, where the moduli space is a point and the virtual tangent space is trivial.

The characteristic classes of the normal bundle $N_\m$ can be most easily computed by introducing the universal divisor
\be
\Delta \subset \Sigma \times \text{Sym}^\m \Sigma
\label{universal divisor}
\ee 
of degree $\m$.
This is defined by the property that if we restrict to an effective divisor $D$ on $\Sigma \simeq \Sigma\times \{D\}$, we have
\be
\Delta|_{\Sigma \times \{D\}} = D \times \{D\}\ ,
\ee
which implies
\be
\cO (\Delta)|_{\Sigma \times \{D\}} = \cO (D)\ .
\ee
Let us denote by $\pi$ and $f$ the projection onto each factors:
\be
\xymatrix{ & \Sigma \times \ar[dl]_{\pi}\text{Sym}^\m\Sigma \ar[dr]^{f} & \\
           \text{Sym}^\m\Sigma &  &\Sigma }
\ee
It is useful to note that 
\bea
&R^0\pi_* \left(\cO(\Delta)\otimes f^* M\right)|_D = H^0 (\Sigma,\cO(D)\otimes M) \, , \\
&R^1\pi_* \left(\cO(\Delta)\otimes f^* M\right)|_D = H^1 (\Sigma,\cO(D)\otimes M) \, ,
\eea
for any line bundle $M$ on $\Sigma$, where $R^\bullet \pi_*$ is the derived pushforward. For simplicity, we will denote it by $\pi_*$. In particular, we can write the class of the vector bundle $N_\m$ in \eqref{NM contribution} as 
\be\label{normal sqed1}
[N_\m] = -[\pi_* \left(\cO(\Delta)\otimes f^* K_\Sigma\right)^*] + [H^\bullet (K_\Sigma)^*]\ .
\ee
The Grothendieck-Riemann-Roch formula tells us
\be
\text{td}(T\mathfrak{M}_{\m}) ~  \text{ch}\left( \pi_* \left(\cO(\Delta)\otimes f^* K_\Sigma\right)\right)
= \pi_* \left[\text{td}(\Sigma \times \Sigma_\m)~  \text{ch}\left(\cO(\Delta)\otimes f^* K_\Sigma\right)\right]\ .
\ee
Using $\pi_*\text{td}(\Sigma \times \Sigma_\m) = \text{td} (\Sigma_\m) \wedge\pi_*\text{td}(\Sigma )$, we find
\be\label{chern class of normal}
\text{ch}\left( \pi_* \left(\cO(\Delta)\otimes f^* K_\Sigma\right) \right)= \pi_* \left[\text{td} (\Sigma ) \wedge \text{ch}\left( \cO(\Delta)\otimes f^* K_\Sigma\right)\right]\ .
\ee
The cohomology class of $\Delta$ on the product $\Sigma \times  \Sigma_\m$ is computed in \cite{arbarello1985geometry} using the K\"unneth decomposition. We can write a class $\delta \in H^2(\Sigma \times\Sigma_\m,\mathbb{Z} )$ as
\be
\delta = \delta ^{2,0} + \delta^{1,1} + \delta^{0,2}\ ,
\ee
where $\delta^{i,j}$ is an element of $H^i(\Sigma) \otimes H^{j}(\Sigma_\m)$. The result is
\be
\delta = \m \eta_{\Sigma} + \gamma + \eta\ ,
\ee
where $\eta_\Sigma$ is the K\"ahler class on $\Sigma$, and $\gamma$ is an element of $H^1(\Sigma) \otimes H^1(\text{Sym}^\m\Sigma)$. One can check that they satisfy $\eta_{\Sigma}^2 = \eta_{\Sigma}\gamma = \gamma^3=0$ and $\gamma^2 = -2\eta_{\Sigma}\sigma$. Using these identities, we find
\be
\text{ch}(\cO(\Delta)) = e^\eta + \m \eta_{\Sigma} e^{\eta} - \eta_{\Sigma} \sigma e^{\eta} + \gamma e^{\eta}\ .
\ee
The remaining factors in \eqref{chern class of normal} can be easily computed:
\be
\text{td}(\Sigma) = 1 + (1-g)\eta_{\Sigma}\ ,~~~\text{ch}(f^* K_\Sigma) = 1+ 2(g-1)\eta_{\Sigma}\ .
\ee
Combining all these expressions, we find
\be\label{chern character normal}
\text{ch}\left( \pi_* \left(\cO(\Delta)\otimes f^* K_\Sigma\right)\right) = (\m-\sigma + g-1)t^{-1}e^{\eta}\ ,
\ee
where  $t=e^{2\pi i m_t}$. From this expression, we obtain the Chern class of this bundle. Using $\sigma_i^2=0$, we can rewrite \eqref{chern character normal} as 
\be
\text{ch}\left( \pi_* \left(\cO(\Delta)\otimes f^* K_\Sigma\right) \right) =\left[ (\m-1)t^{-1}e^{\eta} + \sum_{i=1}^g t^{-1}e^{\eta -\sigma_i}\right], 
\ee
which implies 
\be
c\left( \pi_* \left(\cO(\Delta)\otimes f^* K_\Sigma\right)\right)= (1+\eta-2\pi i m_t)^{\m-1}\prod_{i=1}^g (1+\eta-2\pi i m_t -\sigma_i) .
\ee
Applying the identity \eqref{zagier 1}, we arrive at the expression
\be
\text{ch}(\widehat S^\bullet  N_\m^\vee) =(e^{\pi i m_t}-e^{-\pi i m_t})^{1-g}(e^{-(\eta/2 - \pi i m_t)}-e^{(\eta/2 - \pi i m_t)/2})^{\m + (g-1)} \exp\left[-\frac{\sigma (e^{\eta-m_t}-1)}{2(e^{\eta-m_t}-1)}\right]\ .
\ee
Multiplying all the contributions, the holomorphic Euler characteristic can now be written as
\bea
\chi(\fM_\m, \hat\cO_{\text{vir}})
=(e^{\pi i m_t}-e^{-\pi i m_t})^{1-g} \int_{\cM_{\m}} &\left(\frac{\eta e^{-\eta/2}}{1-e^{-\eta}}\right)^{\m-g+1}(e^{-(\eta/2 - \pi i m_t)}-e^{(\eta/2 - \pi i m_t)})^{\m + (g-1)} \\
&\wedge\exp\left[-\frac{\sigma}{\eta}+\frac{\sigma(e^\eta+1)}{2(e^\eta -1)} -\frac{\sigma (e^{\eta-2\pi i m_t}+1)}{2(e^{\eta-2\pi i m_t}-1)}\right]\ .
\eea
Using the identity \eqref{zagier 2}, we can convert this formula into the residue integration
\bea
\chi(\fM_\m, \hat \cO_{\text{vir}}) 
= 2\pi i (e^{\pi i m_t}-e^{-\pi i m_t})^{1-g}\underset{u=0}{\text{res}} &\frac{\left(e^{\pi i (-u + m_t)}-e^{\pi i (u-m_t)}\right)^{\m+g-1}}{\left(e^{\pi i u}-e^{-\pi i u}\right)^{\m-g+1}} \\ &\cdot \left(\frac{e^{2\pi i u}+1}{2(e^{2\pi i u} -1)} -\frac{ e^{2\pi i (u-m_t)}+1}{2(e^{2\pi i (u-m_t)}-1)}\right)^g\ ,
\eea
which exactly reproduces the twisted index computation. One can check that
\be\label{euler1}
\chi(\fM_\m, \hat \cO_{\text{vir}}) = \left\{\begin{array}{cc} 1\ ,~~~ & \m=0\\ 0\ , ~~~& \m\neq 0 \end{array}\right.
\ee
Notice that this result is also compatible with the $t \rightarrow 1$ limit as described below \eqref{C-twist N=4 geometric}, as well as with the result of the H-twist of the free hypermultiplet in the absence of background fluxes \eqref{freehyper}, in accordance with mirror symmetry.

\subsection{SQED[$N$]}

Let us now generalise the previous discussion to a $U(1)$ gauge theory with $N$ fundamental hypermultiplets. These theories have non-trivial Higgs-branch flavour symmetry, and they satisfy the conditions spelled out in \ref{sec:assumption} provided $N\geq 2$. We assume the following charge assignment:
\be
\begin{array}{c|ccc|cc} &~ U(1)_{G}~ &~ U(1)_t ~&~ SU(N)_H~ &~U(1)_{H}~ &~ U(1)_C\\
\hline
X & 1 &  \frac12 & \overline{\tiny\yng(1)}& \frac12 & 0\\
Y & -1 & \frac12 &\tiny\yng(1) & \frac12 &0\\
\varphi & 0 & -1 & 0 & 0 & 1
\end{array} .
\ee
\subsubsection{$\cN=2$ moduli space}
The BPS moduli space $\fM$ that preserves $\cN=2$ supersymmetry is given by triples $(A,X,Y)$ which satisfy following equations:
\be\label{n=2sqed}
*F_A + e^2\left(  X X^\dagger -  Y^\dagger Y -\tau\right) = 0\ ,~~ \bar\partial_{A} X_i = \bar\partial_{A} Y_i=0\ ,~ \sum_{i=1}^N X_i Y_i = 0\ ,
\ee
modulo $U(1)$ gauge transformations. As in SQED[1], the moduli space of solutions decomposes into topological sectors
\be\label{moduli 32}
\fM=\bigcup_{\m\in\mathbb{Z}} \fM_\m\ ,
\ee
where $\m$ is the degree of the gauge bundle. We will work in the infinite-tension limit $\tau \rightarrow +\infty$, so that the moduli space is uniformly described with non-vanishing $X$. As explained in section \ref{sec:algebraic}, the algebraic description of the moduli space coincides the space of stable quasi-maps into the Higgs branch $T^*\mathbb{CP}^{N-1}$ (C-twist) or twisted stable quasi-maps (H-twist).

In order to compute the index, we consider the action of the Higgs branch flavour symmetry $G_H=PSU(N)$ and apply the localisation principle to the diagonal subgroup 
\be\label{flavour action sqed}
t_H = \text{diag}\left(a_1\ ,\cdots,a_N \right) \in T_H\ ,~~\prod_{i=1}^Na_i = 1\ .
\ee
The variables $a_i$'s are chosen to be completely generic, so that we have $a_{i}\neq a_j$ for any pair $i,j = 1,\cdots, N$.
The subgroup acts on the moduli space as
\be
t_{H}:\left(d_A, \{X_i, Y_i\}\right) \rightarrow \left(d_A, \{a_i X_i, a^{-1}_iY_i\}\right)\ .
\ee
In addition to the $T_H$ action, we can also consider the action of $U(1)_t$ which acts on $X$ and $Y$ as multiplication by $t^{1/2} = e^{\pi i m_t}$. The fixed loci of \eqref{flavour action sqed} is determined up to the action of the gauge symmetry. In our limit, the fixed locus is a disjoint union of $N$ components
\be
\fM_\m^{\text{fixed}} = \bigcup_{i=1}^N\fM_{\m}^{(i)}\ ,
\ee 
which are defined by setting all the bosonic fields to zero except for one of the $X_i$'s:
\be\label{fixed sqed}
\fM_{\m}^{(i)} = \big\{(A,X_i)| *F_A + e^2  ( X_i  X^{\dagger}_i - \tau)=0\ ,~\bar\partial_A X_i =0\big\}/~U(1)_{G}\ .
\ee
Note that $\fM_\m^{(i)}$ can be again identified with a symmetric power of the curve $\Sigma$
\be\label{fixed 2wd}
\fM_{\m}^{(i)} = \Sigma_n\ ,~~n=\m+r(g-1)\ ,
\ee
and that for each fixed locus there exists an inclusion
\be\label{embedding1}
\sigma_{i}: \fM_{\m}^{(i)} \rightarrow \fM_\m\ .
\ee

From now on, we understand the moduli space algebraically and work with its virtual tangent space. The virtual tangent space at a generic point on $\fM_\m$ is given by the cohomology of the following complexes:
\bea \label{sqedn complex}
&H^0( {\cal O}) \overset{\alpha^0}{\longrightarrow }
H^0 ( M_X \oplus  M_Y) \overset{\beta^0}{\longrightarrow } H^1 (K^{1-r}_\Sigma)^*\ ,\\
&H^1( \cO) \overset{\alpha^1}{\longrightarrow }
H^1 (M_X \oplus  M_Y) \overset{\beta^1}{\longrightarrow } H^0 (K^{1-r}_\Sigma)^*\ .
\eea
Here we defined 
\be
M_X = \bigoplus_{i=1}^NL \otimes K_\Sigma^{r/2} ~~\text{ and }~~M_Y = \bigoplus_{i=1}^N L^{-1} \otimes K_\Sigma^{r/2}\ ,
\ee
where each summand has weight $a_i t^{1/2}$ and $a_i^{-1}t^{1/2}$ respectively under the action of $T_H \times U(1)_t$.
We recall that the maps are defined by
\bea
\alpha &: \epsilon \mapsto (\epsilon X_1, \cdots, \epsilon X_N, -\epsilon Y_1,\cdots, -\epsilon Y_N ) \\
\beta & : (A_1,\cdots, A_N, B_1, \cdots, B_N) \mapsto \sum_{i=1}^N A_i Y_i + B_i X_i\ .
\eea
We notice that if we restrict to points in a component of the fixed locus $ \fM_\m^{(i)}$, the complexes split into various pieces. From the first line of \eqref{sqedn complex}, we have 
\bea\label{llk1}
H^0({\cal O}) \longrightarrow H^0(L \otimes K_\Sigma^{r/2})\ ,~~&H^0( L^{-1}\otimes K_\Sigma^{r/2}) \longrightarrow H^1(K^{1-r})^*\ ,\\
& H^0\left(\bigoplus_{j\neq i}^N (L\oplus L^{-1})\otimes K_\Sigma^{r/2}\right)\longrightarrow 0\ .
\eea
From the second line, we obtain similar complexes with the degree shifted by one:
\bea\label{llk2}
H^1({\cal O}) \longrightarrow H^1(L\otimes K_\Sigma^{r/2})\ ,~~&H^1(L^{-1}\otimes K_\Sigma^{r/2}) \longrightarrow H^0(K_\Sigma^{1-r})^* \longrightarrow 0\ ,\\
& H^1\left(\bigoplus_{j\neq i}^N (L\oplus L^{-1})\otimes K_\Sigma^{r/2}\right)\longrightarrow 0\ .
\eea
As explained around \eqref{symmetrictangent}, we can identify the first complex of \eqref{llk1} and \eqref{llk2} as the tangent space of the fixed locus $\fM_{\m}^{(i)}$. Therefore the virtual tangent space restricted to the fixed locus can be written as
\be
T_{\text{vir}}\big|_{\fM_\m^{(i)}} = T\fM_{\m}^{(i)} + N_\m^{(i)} \ .
\ee
The second piece corresponds to the contributions from the virtual normal bundle $N_\m^{(i)}$, which have a non-zero weight under the action of $T_H\times U(1)_t$. The class of the virtual normal bundle is
\be\label{normal 12}
[N_\m^{(i)}] = [H^\bullet (L^{-1}\otimes K_\Sigma^{r/2})] - [H^\bullet (K_\Sigma^{1-r})^*] + \left[H^\bullet\left(\bigoplus_{j\neq i}^N (L\oplus L^{-1})\otimes K_\Sigma^{r/2}\right) \right]\ .
\ee
The first two terms
\be
 [H^\bullet (L^{-1}\otimes K_\Sigma^{r/2})] - [H^\bullet (K_\Sigma^{1-r})^*]:=\widetilde N_\m^{(i)} \ ,
\ee
are the contributions from the multiplet $Y^i$ and the vector multiplet, whose Chern classes were computed in the last section. 
The last summand of \eqref{normal 12} contains contributions from the hypermultiplets $(X_j, Y_j)$ with $j\neq i$, which have non-zero weights $(a_{ji}, a_{ji}^{-1}t)$ under the action $t_H\times U(1)_t$. We will denote the contributions from these fields as 
$N_\m^{(i), X_j}:=[H^\bullet (L\otimes K^{r/2}_\Sigma)]$ and $N_\m^{(i), Y_j}:=[H^\bullet (L^{-1}\otimes K^{r/2}_\Sigma)]$ for $j\neq i$. Now the equivariant virtual Euler characteristic can be written as
\bea
~\chi(\fM_\m, \hat \cO_{\text{vir}})  &=  \sum_{i=1}^N\int_{\fM^{(i)}_\m} \hat A(T\fM^{(i)}_\m)~  \text{ch} ( \widehat S^\bullet N_{\m}^{(i)\vee} )\\
&= \sum_{i=1}^N\int_{\fM^{(i)}_\m} \hat A(T\fM^{(i)}_\m) ~ \text{ch}(\widehat S^\bullet   \widetilde N_{\m}^{(i)\vee}) ~ \prod_{j\neq i}^N \left[\text{ch}(\widehat S^\bullet N_{\m}^{(i),X_j\vee}) ~ \text{ch}(\widehat S^\bullet N_{\m}^{(i),Y_j\vee})\right]\ .
\eea
The Chern class of $[N_{\m}^{(i),X_j}]$ and $[N_{\m}^{(i),Y_j}]$ for $j\neq i$ can also be computed from the universal construction discussed in the last section. We can derive
\be\label{chern normal sqed}
c\left(N_{\m}^{(i),X_j},m_{ji}\right) = (1+\eta + 2\pi i m_{ji})^{n-2g+1} \prod_{a=1}^g (1+\eta +2\pi i  m_{ji} -\sigma_a)\ .
\ee
where we defined $a_i = e^{2\pi i m_i}$ and $m_{ij}:=m_i-m_j$.
Using the identity \eqref{zagier 1}, we can write the Chern characteristics of the symmetric powers as
\be\label{eq:SQEDNclasses1}
\text{ch}\left(\widehat S^\bullet N_{\m}^{(i),X_j\vee}\right)= \left(\frac{e^{(-\eta/2+\pi i m_{ji})}}{1-e^{-(\eta-2\pi i m_{ji})}}\right)^{\m+(r-1)(g-1)}\exp\left[ \frac{\sigma( e^{\eta -2\pi i m_{ji}}+1)}{2(e^{\eta-2\pi i m_{ji}}-1)}\right]\ .
\ee
From the multiplet $Y_j~(j\neq i)$, 
\bea\label{eq:SQEDNclasses2}
\text{ch}\left(\widehat S^\bullet N_{\m}^{(i),Y_j\vee}\right) = &\left(e^{(-\eta/2+\pi i m_{ji}+\pi i m_t)} - e^{\eta/2 - \pi i m_{ji}-\pi i m_t}\right)^{\m-(r-1)(g-1)} \\
&\cdot \exp\left[ \frac{\sigma( e^{-\eta +2\pi i (m_{ji}+m_t)}+1)}{2(e^{-\eta+2\pi i (m_{ji}+m_t)}-1)}\right]\ .
\eea
The contribution from $\text{ch}\left(\widehat S^\bullet\widetilde N_{\m}^{(i)}\right)$ is the same as the normal bundle contribution studied in the last example. We have
\bea\label{eq:SQEDNclasses3}
\text{ch}\left(\widehat S^\bullet   \widetilde N_{\m}^{(i)\vee}\right)  = &\left(e^{\pi i m_t}-e^{-\pi i m_t}\right)^{(2r-1)(g-1)}\left(e^{\pi i m_t-\eta/2}-e^{-\pi i m_t+\eta/2}\right)^{\m-(r-1)(g-1)}\\
&\cdot \exp\left[ \frac{\sigma( e^{-\eta+2\pi i m_t}+1)}{2(e^{-\eta+2\pi i m_t}-1)}\right]\ .
\eea
Converting this expression into the residue integral using \eqref{zagier 2}, we find that the equivariant virtual Euler characteristic can be written as
\bea\label{SQED integral}
\chi(\fM_\m, \hat\cO_{\text{vir}}) 
=&~2\pi i \left(t^{1/2}-t^{-1/2}\right)^{(2r-1)(g-1)} \\
&\cdot \sum_{i=1}^N \underset{u=m_i}{\text{res}}du \prod_{j=1}^N\frac{\left(e^{\pi i (-u+m_j +m_t)}-e^{\pi i (u-m_j -m_t)}\right)^{\m-(g-1)(r-1)}}{\left(e^{\pi i (u-m_j)}-e^{\pi i (-u+m_j)}\right)^{\m+(g-1)(r-1)}} \\
&\cdot\left[\sum_{j=1}^N\left(\frac{1+e^{2\pi i (-u+m_j)}}{2(1-e^{2\pi i (-u+m_j)})}+\frac{1+e^{2\pi i (u-m_j-m_t)}}{2(1-e^{2\pi i (u-m_j-m_t)})}\right)\right]^g\ .
\eea
This again reproduces the integral representation of the twisted index computation.

\subsubsection{The $t \rightarrow 1$ limit}

\paragraph{H-twist}

For the H-twist, the expression \eqref{SQED integral} (with $r=1$) can be understood as the virtual $\chi_t$ genus of the $U(1)_t$-fixed locus 
\be
\fL =\bigoplus_{\m\in \mathbb{Z}}\fL_\m\ ,
\ee
where $\fL_\m$ can be identified as a space of degree $\m$ twisted quasi-maps to the compact core $\mathbb{CP}^{N-1}$, the base of Higgs branch $\cM_H$. This space is parametrised by the solution $(A,{X_i})$ to the equations
\be
*F_A + e^2\left( X X^\dagger - \tau\right)=0\ ,~\bar\partial_A X =0\ ,
\ee
modulo $U(1)$ gauge tranformations.

The H-twisted $\cN=4$ moduli space for this theory is identical to that of the $\cN=2$ moduli space defined in \eqref{moduli 32}. In the limit $t\rightarrow 1$, we recover the expression for the integral of the virtual Euler class of the fixed locus $\fL$ inside the moduli space. This quantify can be directly computed using the alternative localisation scheme with respect to the $T_H \subset G_H$ action. Then the index can be written as a sum of the Euler characteristics of the smooth compact fixed loci $\fM_\m^{(i)}$ defined in \eqref{fixed sqed}. We have
\bea\label{sqedn h}
\sum_{\m\in\mathbb{Z}} (-q)^{\m}\chi(\fM_\m, \hat\cO_{\text{vir}}) \big|_{t\rightarrow 1}&= (-1)^{g-1}\sum_{\m\in \mathbb{Z}}q^{\m}~\sum_{i=1}^N\int_{\fM^{(i)}_{\m}} e(\fM^{(i)}_{\m}) \\
&= (-1)^{g-1}N(q^{1/2}-q^{-1/2})^{2(g-1)}\ ,
\eea
which correctly reproduces the generating function for the Euler characteristic of the $N$ copies of the Sym$^\m\Sigma$. Note that the residue integration at each $i$ is independent of the equivariant parameters $\{a_i\}$, which agrees with the fact that the Hilbert space of the effective quantum mechanics is the de Rham cohomology \cite{Hori:2014tda}.

\paragraph{C-twist}
 
For the C-twist, imposing $\cN=4$ BPS equations trivialises the line bundle $L$, and the moduli space parametrises the solutions $(\{X_i, Y_i\})$ to the equation
\be
X X^\dagger - Y^\dagger Y = \tau\ ,~~ \sum_{i=1}^N X_iY_i=0\ ,
\ee
for constant $X_i$ and $Y_i$, modulo $U(1)$ gauge transformation. This is the resolution of the Higgs branch $\cM_{H}=T^*(\mathbb{CP}^{N-1})$ inside the $\cN=2$ moduli space $\fM$.

The $t\rightarrow 1$ limit with $r=0$ of the result \eqref{SQED integral} can be understood as the Rozansky-Witten invariant computing the holomorphic Euler characteristic of $\cM_H$ valued in the vector bundle $\left(\widehat\wedge^\bullet T^*\cM_H\right)^g$:

\be
\chi(\fM, \hat \cO_{\text{vir}}) \big|_{t\rightarrow 1}=\int_{\cM_H} \hat A\left(T\cM_H\right) \wedge\text{ch}\left[\left(\widehat\wedge^\bullet T^*\cM_{H}\right)^{\otimes g}\right]\ .
\ee
$\cM_H$ is non-compact and this expression can be evaluated from the equivariant localisation with respect to the $T_H\subset G_H$ action. Let us consider the action of $g_H$ defined in \eqref{flavour action sqed}. When $\tau>0$, the fixed loci are $N$ isolated points, where $i$-th fixed point is defined by $X_i\neq 0$ and all the other bosonic fields are identically zero. From the fixed point formula we arrive at the expression
\bea
\chi(\fM, \hat\cO_{\text{vir}}) (y\rightarrow 1)&=\sum_{i=1}^N \prod_{j\neq i}\left(\frac{e^{-\pi i m_{ij}}}{1-e^{-2\pi im_{ij}}}\right)^2 \prod_{j\neq i}\left( e^{\pi i m_{ij}} - e^{-\pi i m_{ij}}\right)^{2g}\\
&= \sum_{i=1}^N \prod_{j\neq i} \left( e^{- \pi i m_{ij}} - e^{\pi i m_{ij}}\right)^{2(g-1)}\ .
\eea

\subsection{SQCD[$N_c, N_f$]}\label{subsec:SQCD}

We can generalise our previous analysis to non-abelian gauge groups, provided that the fixed loci of the moduli space are products of symmetric products of the curve $\Sigma$. In this section we present the simplest example, which is SQCD[$N_c, N_f$] where $N_f\geq 2N_c$, as discussed in section \ref{sec:assumption}. The fields of the theory are charged as follows:
\be
\begin{array}{c|ccc|cc} &~ U(N_c)_{G}~ &~ U(1)_t ~&~ SU(N)_H~ &~U(1)_{H}~ &~ U(1)_C\\
\hline
X & \tiny\yng(1) &  \frac12 &\overline{\tiny\yng(1)} & \frac12 & 0\\
Y & \overline{\tiny\yng(1)} & \frac12 & \tiny\yng(1)& \frac12 &0\\
\varphi & \text{adj} & -1 & \mathbf{1} & 0 & 1
\end{array}
\ee

\subsubsection{$\mathcal{N}=2$ moduli space}

The $\mathcal{N}=2$ BPS equations for SQCD$[N_c,N_f]$ read
\bea\label{n=2sqcd}
&*F_A + e^2\left(X X^\dagger - Y^\dagger Y  -2 [\varphi^\dagger, \varphi] - \tau  \right)= 0 \ \\
&\bar\partial_{A} X = \bar\partial_{A} Y=D_{\bar z}\varphi=0\ \\
&\varphi \cdot X =\varphi\cdot  Y =  X \cdot Y = 0\ .~
\eea
The moduli space of solutions to BPS equations modulo gauge transformations can be decomposed into topological sectors labelled by the degree of the holomorphic bundle in the fundamental representation associated to the gauge bundle $P$:
\be
\fM = \bigcup_{\m \in \Lambda_{C}^\vee} \fM_{\m}\ .
\ee
We again consider the infinite-tension limit $\tau \rightarrow +\infty$. It follows from the discussion in section \ref{sec:moduli space} that by using a Hitchin-Kobayashi correspondence, the moduli space has the follwing algebraic description for every $\m$. A point of the moduli space in the component $\fM_\m$ is given by
\begin{itemize}
\item{A holomorphic $GL(N_c,\mathbb{C})$-bundle $E$ of degree $\m$;}
\item{Holomorphic sections $(X,Y)$ of associated bundles $E_X$ and $E_Y$, corresponding to $N_f$  copies of the fundamental and anti-fundamental representation respectively;}
\item{Subject to the complex moment map condition $Y \cdot X = 0$;}
\item{Subject to the stability condition that $X$ has generically maximal rank on $\Sigma$;}
\item{Modulo complexified gauge transformations.}
\end{itemize}
This can be thought of as the space of stable quasi-maps into the Higgs branch $\cM_H = T^*G(N_c,N_f)$ (C-twist) or twisted stable quasi maps (H-twist).

Let us consider the fixed points of a maximal torus $T_H$ of the flavour symmetry, which locally acts as
\be\label{GH action}
X \mapsto X t_H,\ ,~Y\mapsto t_H Y\ ,~ \ \bar \partial_{A_{\mathbb{C}}} \mapsto \bar \partial_{A_{\mathbb{C}}}\ ,
\ee
for $t_H$ represented as a diagonal $N_f \times N_f$ matrix, and with $A_\mathbb{C}$ the connection on the holomorphic bundle $E_\mathbb{C}$. The fixed points are solutions to the equations
\be\label{gauge transf}
g_\mathbb{C}X = X t_H \ ,~Yg_\mathbb{C} =t_H Y  \ ,~g_{\mathbb{C}}^{-1}d_{A_{\mathbb{C}}} g_{\mathbb{C}} = d_{A_{\mathbb{C}}} \ ,
\ee
for an element of the complex gauge transformation $g_{\mathbb{C}} \in G_{\mathbb{C}}$. Given the stability condition on $X$, $g_\mathbb{C}$ must act non-trivially. From the last equation of \eqref{gauge transf}, $E_\mathbb{C}$ decomposes at fixed points as a direct sum of line bundles
\be \label{eq:SQCDdecomposition}
E_{\mathbb{C}}=L_1 \oplus \cdots \oplus L_{N_c} \, .
\ee
Let us denote $\m_a = \text{deg}(L_a)$, which satisfy
\be
\m = \sum_a \m_a\in H_2(\mathcal{M}_H,\mathbb{Z}) \cong \mathbb{Z} \ .
\ee
The associated bundles $E_X$ and $E_Y$ decompose accordingly
\bea
E_X &\cong \left(L_1\oplus \cdots \oplus L_{N_c} \right)^{\oplus N_f} \otimes K^{r/2}\\
E_Y &\cong \left( L_1^{-1} \oplus \cdots \oplus L_{N_c}^{-1}\right)^{\oplus N_f}\otimes K^{r/2}  \, .
\eea
For later convenience, we also note that on any fixed locus $E_V$ and $E_\Phi$ decompose as \footnote{In fact, the complexified Lie algebra decomposes under the adjoint action as
\be 
\mathfrak{g}_{\mathbb{C}} = \mathfrak{t}_{\mathbb{C}} \oplus \bigoplus_{\alpha \in \mathfrak{g}_{\mathbb{C}}} {\mathfrak{g}_{\mathbb{C}}}_{\alpha}  \ ,
\ee
where the summands can be identified with diagonal matrices (over which the adjoint action of $\mathfrak{t}_{\mathbb{C}} \cong (a_1 , \cdots , a_N)$ is trivial) and matrices with one single off-diagonal entry $e_{ij}$ (the action corresponding to $x_{ij}\mapsto a_{i}a_j^{-1} x_{ij}$).}
\be
E_{V} \cong \left( \mathcal{O}^{N_c} \oplus \bigoplus_{a \neq b} L_a \otimes L_b^{-1}  \right)~~\text{ and }~~
E_{\Phi} \cong \left( \mathcal{O}^{N_c} \oplus \bigoplus_{a \neq b} L_a \otimes L_b^{-1}  \right) \otimes K^{1-r}_{\Sigma}\, .
\ee
Fixed points are labelled by $N_c$-subsets $I= \{i_1, \cdots, i_{N_c} \}\subset \{1, \cdots , N_f\}$ so that the only non-vanishing sections are
\be
X^a_{i_a} \neq 0   \, ,
\ee
and fixed loci reduce to disjoint unions of  $N_c$ copies of symmetric product products
\be
\fM_{\m}^T = \bigsqcup_{\substack{(\m_1, \cdots , \m_{N_c} )\\ \sum_a \m_a = \m } } \left( \prod_{a=1}^{N_c} \mathrm{Sym}^{\m_a+r(g-1)} \Sigma \right) \, , 
\ee
where as usual $r=1,0$ for H- and C-twist respectively. 

The breaking of the gauge bundle into a sum of abelian contributions makes the generalisation from SQED[$N$] to SQCD[$N_c,N_f$] rather straightforward, and we will therefore be brief, mainly working at the level of K-theory classes. We will work on the component $\underline \m = (\m_1,\cdots,\m_{N_c})$ of the fixed locus $I$, which we denote $\fM_{I, \underline \m}$.

Over $\fM_{I , \underline \m}$, the virtual tangent bundle decomposes into the following contributions
\bea\label{virtual sqcd}
\left[T_{\text{vir}}|_{\fM_{I , \underline \m }} \right] =&\sum_{a=1}^{N_c} \left( -\left[ H^{\bullet}  \left( \mathcal{O} \right) \right] + \left[ H^{\bullet}  \left(  L^{i_a}_a \otimes K_{\Sigma}^{r/2} \right) \right] \right) \\
&+\sum_{a=1}^{N_c}\sum_{\substack{j=1\\  j \neq i_a}}^{N_f}\left[  \underbrace{H^{\bullet}  \left(  L^{j}_a \otimes K_{\Sigma}^{r/2} \right)}_{:= NX^a_j} \right] +\sum_{a=1}^{N_c}\sum_{j=1}^{N_f}\left[ \underbrace{H^{\bullet}  \left( \left(L^j_a\right)^{-1}  \otimes K_{\Sigma}^{r/2} \right)}_{:=NY^a_j} \right]  \\
&-\sum_{a=1}^{N_c} \left[ \underbrace{H^{\bullet}  \left( K_{\Sigma}^r \right)}_{:=N\Phi_{aa}} \right] - \sum_{a\neq b}  \left[ \underbrace{H^{\bullet}  \left(L_a^{-1} \otimes L_b \otimes K^r_{\Sigma}\right) }_{:=N\Phi_{ab}} \right]
- \sum_{a\neq b}\left[\underbrace{H^\bullet (L_a \otimes L_b^{-1})}_{:=NV_{ab}}\right] \, .
\eea
The first line includes all contributions tangent to the fixed locus (fixed part), whereas all other contributions are normal (moving part). In order to express these contributions in terms of characteristic classes over the fixed locus $\prod_{a=1}^{N_c}\Sigma_{n_a}$, let us first define the generators of cohomology class as follow:
\be
\eta_a \in H^2(\Sigma_{n_a},\mathbb{Z})~~\text{and}~~
\sigma^{ab} = \sum_{i=1}^g \xi_i^a {\xi_i'}^b\ ,~~\xi_i^a,{\xi'}_i^a \in H^1(\Sigma_{n_a},\mathbb{Z})\ .
\ee
Then from the fixed part, we obtain the tangent bundle over the fixed locus which contributes
\be\label{SQCDclass0}
\hat{A}\left(T{\fM}_{I , \underline \m} \right) = \prod_{a=1}^{N_c} \left(\frac{\eta_a e^{-\eta_a/2}}{1-e^{-\eta_a}}\right)^{\m_a+(r-1)(g-1)}\exp \left(\frac{\sigma^{aa} (e^{\eta_a} +1)}{2(e^{\eta_a}-1)}-\frac{\sigma^{aa}}{\eta_a}\right) \ . 
\ee
The contributions from the moving part $N_{I,\underline{\m}}$ can be summarised as
\be\label{eq:SQCDnormal}
\prod_{a=1}^{N_c}\left[\prod_{\substack{j=1\\ j \neq i_a}}^{N_f}\text{ch}(\widehat S^\bullet N{X^{a\vee}_j})\prod_{j=1}^{N_f}\text{ch}(\widehat S^\bullet NY^{a\vee}_j)\prod_{b=1}^{N_c}\text{ch}(\widehat\wedge^\bullet N\Phi_{ab}^\vee)\right] \prod_{a\neq b} \text{ch}(\widehat \wedge^\bullet NV_{ab}^\vee)\ .
\ee
The arguments \eqref{eq:SQEDNclasses1}-\eqref{eq:SQEDNclasses3} can be straightforwardly generalised to obtain the contribution from the hypermultiplets:
\be\label{eq:SQCDNclasses1}
\text{ch}(\widehat S^\bullet NX^{a\vee}_j)= \left(\frac{e^{-\eta_a/2+\pi i m_{ji_a}}}{1-e^{-\eta_a+2\pi i m_{ji_a}}}\right)^{\m_a+(r-1)(g-1)}\exp\left[ \frac{\sigma^{aa}( e^{\eta_a -2\pi i m_{ji_a}}+1)}{2(e^{\eta_a-2\pi i m_{ji_a}}-1)}\right]
\ee
and
\bea\label{eq:SQCDNclasses2}
\text{ch}(\widehat S^\bullet NY^{a\vee}_j) = &\left(e^{-\eta_a/2+\pi i( m_{ji_a}+m_t)} - e^{\eta_a/2 - \pi i (m_{ji_a}+m_t)}\right)^{\m_a-(r-1)(g-1)} \\
&\cdot \exp\left[ \frac{\sigma^{aa}( e^{-\eta_a +2\pi i (m_{ji_a}+m_t)}+1)}{2(e^{-\eta_a+2\pi i (m_{ji_a}+m_t)}-1)}\right] \, .
\eea
The contribution from the multiplet in the adjoint representation, $NV_{ab}$ and $N\Phi_{ab}$ can be written as classes on $\text{Sym}^{\m_a + r(g-1)}\Sigma\times \text{Sym}^{\m_b + r(g-1)}\Sigma$. We computed the Characteristic classes of these contributions in appendix \ref{app:class sym}. To summarise, the vector multiplet contribution is 
\bea\label{normal adjoint}
\prod_{a\neq b}\text{ch}(\widehat\wedge^\bullet NV_{ab}^\vee) =&\prod_{a\neq b}\left(e^{(-\eta_a + \eta_b)/2 -\pi i (m_{i_a}-m_{i_b})} -e^{(\eta_a - \eta_b)/2+\pi i (m_{i_a}-m_{i_b})} \right)^{-\m_a + \m_b + 1-g}\\
& \exp\left[(\sigma^{aa}+\sigma^{bb}-\sigma^{ab}-\sigma^{ba})\frac{( e^{-\eta_a + \eta_b -2\pi i (m_{i_a}-m_{i_b})}+1)}{2( e^{-\eta_a + \eta_b -2\pi i (m_{i_a}-m_{i_b})  }-1)} \right]\ .
\eea
Note that the exponential terms in \eqref{normal adjoint} with positive and negative root $\alpha$ cancel each other out, and we are left with a simple expression
\be
\prod_{a\neq b}\text{ch}(\widehat\wedge^\bullet NV_{ab}^\vee) =   (-1)^{\sum_{\alpha>0}\alpha(\underline{\m})}\prod_{a\neq b}\left(e^{(-\eta_a + \eta_b )/2-\pi i (m_{i_a}-m_{i_b})} -e^{(\eta_a - \eta_b) +\pi i (m_{i_a}-m_{i_b})} \right)^{1-g}\ .
\ee
Contribution from the adjoint chiral $N\Phi_{ab}$ can be similarly written as
\bea\label{SQCDclass3}
&\prod_{a,b=1}^{N_c}\text{ch}(\widehat\wedge ^\bullet N\Phi_{ab}^\vee) \\
=&\prod_{a,b=1}^{N_c} \left(e^{(\eta_a- \eta_b) +\pi i (m_{i_a}-m_{i_b})+ \pi i m_t} - e^{(-\eta_a+\eta_b )/2-\pi i (m_{i_a}-m_{i_b}) -\pi i m_t}\right)^{\m_a - \m_b -(1-2r)(g-1)} \\
& \exp \left[(\sigma^{aa}+\sigma^{bb}-\sigma^{ab}-\sigma^{ba})\frac{e^{\eta_a + \eta_b +2\pi i (m_{i_a}-m_{i_b})+2\pi i m_t } - 1}{2(e^{\eta_a - \eta_b +2\pi i (m_{i_a}-m_{i_b})+2\pi i m_t } - 1)}\right]
\eea

We can now compute the equivariant virtual Euler characteristic, which by \eqref{fixed point localisation} can be written as
\bea\label{SQCD9}
~&\chi(\fM,\hat\cO_{\text{vir}}) \\
=&\sum_{\m\in \mathbb{Z}}(-q)^\m\sum_{\substack{(\m_1, \cdots , \m_{N_c} )\\ \sum_a \m_a = \m }}\sum_{I\subset\{1,\cdots, N_f\}}\int_{\fM_{I,\underline{\m}}} \hat A (T\fM_{I,\underline{\m}}) \text{ch}(\widehat S^\bullet N^\vee_{I,\underline{\m}}) \, .
\eea
where by \eqref{eq:SQCDnormal}, the integral can be expanded as
\be\label{SQCD10}
\int_{\fM_{I,\underline{\m}}} \hat A (T\fM_{I,\underline{\m}})\prod_{a\in I}\left[\prod_{\substack{j=1\\ j \neq i_a}}^{N_f}\text{ch}(\widehat S^\bullet N{X^{a\vee}_j})\prod_{j=1}^{N_f}\text{ch}(\widehat S^\bullet NY^{a\vee}_j)\prod_{b=1}^{N_c}\text{ch}(\widehat \wedge^\bullet N\Phi_{ab}^\vee)\right] \prod_{a\neq b} \text{ch}(\widehat \wedge^\bullet NV_{ab}^\vee) \, .
\ee
Combining the result from \eqref{SQCDclass0}-\eqref{SQCDclass3}, these contributions are equal to
\be
\int_{\fM_{I,\underline{\m}}} \left(\prod_{a\in I} \eta_a^{\m_a + (r-1)(g-1)}\right) A_I(\eta_1,\cdots, \eta_{N_c}) \exp\left[\sum_{a,b}^{N_c}\sigma^{ab}B_{I,ab}(\eta_1,\cdots, \eta_{N_c})\right] \, .
\ee
where
\bea
~&A_I(\eta_1,\cdots, \eta_{N_c}) \\
= & \prod_{a\in I} \prod_{j=1}^{N_f} \left(\frac{e^{-\eta_a/2 + \pi i m_{ji_a}}}{1- e^{-\eta_a + 2\pi i m_{ji_a}}}\right)^{\m_a + (r-1)(g-1)}  \left(e^{-\eta_a/2 + \pi i (m_{ji_a} + m_t)} - e^{(\eta_a/2 - \pi i (m_{ji_a} + m_t)}\right)^{\m_a -(r-1)(g-1)}\\
&\prod_{\substack{a,b\in I\\ a\neq b}} \left(e^{(-\eta_a + \eta_b)/2-\pi i (m_{i_a}-m_{i_b})} - e^{(\eta_a - \eta_b)/2+\pi i (m_{i_a}-m_{i_b})} \right)^{-\m_a + \m_b + 1-g} \\
&\prod_{a,b \in I} \left(e^{(\eta_a - \eta_b)/2+\pi i (m_{i_a}-m_{i_b}) + \pi im_t} - e^{(-\eta_a + \eta_b )/2-\pi i (m_{i_a}-m_{i_b}) -\pi i m_t}\right)^{\m_a -\m_b - (1-2r)(g-1)} 
\eea
and
\be
B_{I,ab}(\eta_1,\cdots, \eta_{N_c}) = H_{I,ab}(\eta_1,\cdots, \eta_{N_c}) - \delta_{ab}\eta_a^{-1}\ ,
\ee
where $H_{ab}$ is given by the expression
\bea
H_{I,ab} =&~ \delta_{ab}\left[\sum_{j=1}^{N_f} \frac{1+e^{-\eta_a + 2\pi i m_{ji_a}}}{2(1-e^{-\eta_a + 2\pi i m_{ji_a}})} + \sum_{c\neq a}^{N_c} \frac{1+e^{\eta_a -\eta_c+2\pi i (m_{i_a}-m_{i_c})+2\pi i m_t}}{2(1-e^{(\eta_a-\eta_c+2\pi i (m_{i_a}-m_{i_c})+2\pi i m_t)})}\right.\\
&~~~~~~+\left.\sum_{j=1}^{N_f}\frac{1+ e^{\eta_a - 2\pi i (m_{ji_a}+m_t)}}{2(1-e^{\eta_a -2\pi i(m_{ji_a}+m_t)})} + \sum_{c\neq a}^{N_c} \frac{1+ e^{\eta_c - \eta_a +2\pi i (m_{i_c}-m_{i_a})+ 2\pi i m_t}}{2(1-e^{\eta_c -\eta_a+2\pi i (m_{i_c}-m_{i_a}) +2\pi i m_t})}\right]\\
&+ (1-\delta_{ab})\left[\frac{1+e^{\eta_a -\eta_b+2\pi i (m_{i_a}-m_{i_b}) + 2\pi i m_t}}{2(1-e^{\eta_a - \eta_b+2\pi i (m_{i_a}-m_{i_b}) + 2\pi i m_t})} + \frac{1+e^{\eta_b - \eta_a-2\pi i (m_{i_a}-m_{i_b}) + 2\pi i m_t}}{2(1-e^{\eta_b -\eta_a-2\pi i (m_{i_a}-m_{i_b}) + 2\pi i m_t})}\right]\ .
\eea
The last expression in \eqref{SQCD9} can be converted to a product of residue integrals as in the abelian examples. We show in appendix \ref{app:class sym} that the identity \eqref{zagier 2} over  $\Sigma_n$ can be generalised to integrals over $\prod_{i=1}^N \Sigma_n$: For any power series $A(\eta_1,\cdots, \eta_{N_c})$ and $B_{ab}(\eta_1,\cdots, \eta_{N_c})$ on $\prod_{i=1}^{N_c} \Sigma_{n_i}$, we have
\bea
~&\int_{\prod_{i=1}^{N_c} \Sigma_{n_i}} A(\eta_1,\cdots, \eta_{N_c}) \exp\left[\sum_{a,b=1}^{N_c} \sigma^{ab}B_{ab}(\eta_1,\cdots, \eta_{N_c})\right] \\
=&\underset{u_1=0}{\text{res}}\cdots  \underset{u_{N_c}=0}{\text{res}}~ \frac{A(u_1,\cdots u_{N_c})}{u_1^{n_1+1}\cdots u_{N_c}^{n_{N_c}+1}}~ \left[\underset{ab}{\text{det} }\left(\delta_{ab}+ u_a B_{ab}(u_1,\cdots, u_{N_c})\right)\right]^g\ .
\label{generaliseddonZagier}
\eea
Then the integral \eqref{SQCD9} becomes
\bea
~&\chi(\fM,\hat\cO_{\text{vir}}) \\
&=\sum_{\underline{\m}\in \mathbb{Z}^{N_c}}(-q)^{\sum_{a=1}^{N_c}\m_a} \sum_{I\subset\{1,\cdots, N_f\}}\underset{u_1=0}{\text{res}}\cdots  \underset{u_{N_c}=0}{\text{res}} A_I(u_1,\cdots, u_{N_c})\left[\underset{ab}{\text{det}}~ H_{I,ab}(u_{1},\cdots, u_{N_c})\right]^g\ .
\eea
By a redefinition of the integration variables $u_a \rightarrow u_a -m_{i_a} + m_t/2$ for each summand labelled by $I$, the integral can be rewritten as
\bea\label{index SQCD final}
~&\chi(\fM,\hat\cO_{\text{vir}}) \\
&=(2\pi i )^{N_c}\sum_{\underline{\m}\in \mathbb{Z}^{N_c}}(-q)^{\sum_{a=1}^{N_c}\m_a} \sum_{I\subset\{1,\cdots, N_f\}}\left(\prod_{a=1}^{N_c}\underset{u_a = m_{i_a}-m_t/2}{\text{res}}\right) \mathbf{A}(u_1,\cdots, u_{N_c})\left[\underset{ab}{\text{det}}~ \mathbf{H}_{ab}(u_{1},\cdots, u_{N_c})\right]^g\ .
\eea
where $\mathbf{A}$ and $\mathbf{H}$ is
\bea
\mathbf{A}(u_1,\cdots, u_{N_c}) = 
& \prod_{a=1}^{N_c} \prod_{j=1}^{N_f} \frac{(e^{\pi i (-u_a + m_{j} + m_t/2)} - e^{\pi i (u_a - m_{j} - m_t/2)})^{\m_a - (r-1)(g-1)}}{\left(e^{\pi i (u_a - m_j + m_t/2)}- e^{\pi i (-u + m_j - m_t/2)}\right)^{\m_a + (r-1)(g-1)}}\\
&\prod_{\substack{a,b=1\\ a\neq b}}^{N_c} \left(e^{\pi i (-u_a + u_b)} - e^{\pi i (u_a - u_b)} \right)^{-\m_a + \m_b + 1-g} \\
&\prod_{a,b =1}^{N_c} \left(e^{\pi i (u_a - u_b + m_t)} - e^{\pi i (-u_a + u_b -m_t)}\right)^{\m_a -\m_b - (1-2r)(g-1)} 
\eea
and
\bea
\mathbf{H}_{ab}(u_{1},\cdots, u_{N_c}) =&~ \delta_{ab}\left[\sum_{j=1}^{N_f} \frac{1+e^{2\pi i (-u_a + m_{j}+m_t/2)}}{2(1-e^{2\pi i (-u_a + m_{j}+m_t/2)})} + \sum_{c\neq a}^{N_c} \frac{1+e^{2\pi i (u_a -u_c+m_t)}}{2(1-e^{2\pi i (u_a-u_c+m_t)})}\right.\\
&~~~~~~+\left.\sum_{j=1}^{N_f}\frac{1+ e^{2\pi i (u_a - m_{j}+m_t/2)}}{2(1-e^{2\pi i (u_a -m_{j}+m_t/2)})} + \sum_{c\neq a}^{N_c} \frac{1+ e^{2\pi i (u_c - u_a + m_t)}}{2(1-e^{2\pi i (u_c -u_a +m_t)})}\right]\\
&+ (1-\delta_{ab})\left[\frac{1+e^{2\pi i (u_a -u_b + m_t)}}{2(1-e^{2\pi i (u_a - u_b + m_t)})} + \frac{1+e^{2\pi i (u_b - u_a + m_t)}}{2(1-e^{2\pi i (u_b -u_a + m_t)})}\right]\ .
\eea
Finally, it is straightforward to show that the residue integral together with the choice of fixed point is the equivalent to the Jeffrey-Kirwan residue integral of the integrand with the choice $\eta>0$:
\be
\sum_{I\subset\{1,\cdots, N_f\}}\left(\prod_{a=1}^{N_c}\underset{u_a = m_{i_a}-m_t/2}{\text{res}}\right) = \frac{1}{N!}\sum_{u_* = \{u_i\}} \underset{u=u_*}{\text{JK-Res}}(Q_{u_*}(u),\eta>0)\ .
\ee
Therefore we again proved that the equivariant virtual Euler characteristic of the moduli space reproduces the twisted indices computation. This procedure can be generalised to the class of the theories defined in section \ref{sec:assumption}.  
\subsubsection{The limit $t\rightarrow 1$}

\paragraph{H-twist} 
The $\cN=4$ BPS equation for the H-twist is given by
\bea
& *F_A + e^2\left(X X^\dagger - Y^\dagger Y - \tau \right) = 0\ ,\\
& \bar\partial_{A}X = \bar\partial_{A} Y = \bar\partial_{A} \varphi = [\varphi^\dagger, \varphi]=0\ ,\\
& X\cdot \varphi = \varphi \cdot Y = X\cdot Y =0\ ,
\eea
modulo $U(N_c)$ gauge transformation. If we consider localisation with respect to the $U(1)_t$ action, the fixed locus $\fM^{U(1)_t}$ is parametrised by the solutions $(A,X)$ to the equations
\be
*F_A + e^2\left(X X^\dagger - \tau \right) = 0\ ,~~\bar\partial_{A}X =0\ ,
\ee
modulo $U(N_c)$ gauge transformations, which can be identified as the space of twisted quasi-maps to the compact core inside the Higgs branch $\cM_H = T^*G(N_c, N_f)$, which we denote by $\fL$. Similarly to the SQED example, the index \eqref{index SQCD final} with $r=1$ can also be thought of as the virtual $\chi_t$ genus
\be
\chi(\fM, \hat\cO_{\text{vir}})|_{r=1 }  = \chi_t^{\text{vir}} (\fL)\ .
\ee
In the limit $t\rightarrow 1$, the index reduces to the generating function of the integral of the virtual Euler class. This can be evaluated using the virtual localisation with respect to the $T_H$ action \eqref{GH action}:
\be
\chi(\fM, \hat\cO_{\text{vir}})\big|_{\substack{r=1 \\ t\rightarrow 1}}  = (-1)^{N_c(g-1)} \sum_{\underline{\m} \in \mathbb{Z}^{N_c}}q^{\m}\sum_{{\substack{(\m_1, \cdots , \m_{N_c} )\\ \sum_a \m_a = \m } }}\sum_{I} \int_{\fM_{\underline{\m},I}} e(T\fM_{\underline{\m},I})\ .
\ee
Summing over $\m$, we have
\be
\chi(\fM, \hat\cO_{\text{vir}})\big|_{\substack{r=1 \\ t\rightarrow 1}}  =(-1)^{N_c(g-1)}{N_f\choose N_c} \left(q^{1/2}-q^{-1/2}\right)^{2N_c(g-1)}\ ,
\ee
using the generating function of the Euler numbers for a symmetric product \eqref{euler symmetric 5}.

\paragraph{C-twist} 
As we studied in section \ref{sec:t1limit}, imposing the $\cN=4$ BPS equations trivialises the vector bundle $E$, and the moduli space reduces to the resolution of the Higgs branch $\cM_{H} = T^*G(N_c, N_f)$. The index in the $t\rightarrow 1$ limit then computes the equivariant Rozansky-Witten invariants of the target $\cM_{H}$. This can be directly computed from the $\m=0$ sector of the expression \eqref{SQCD9}, taking the $t\rightarrow 1$ limit and extracting the constant term in the power series expansion of the characteristic classes. This procedure gives
\bea
\chi(\fM, \hat\cO_{\text{vir}})\big|_{\substack{r=0 \\ t\rightarrow 1}} 
=& \sum_{I} \prod_{\substack{i\in I\\ j\in I^\vee}}
\left(e^{-\pi i m_{ij}}-e^{\pi i m_{ij}}\right)^{2(g-1)}\ ,
\eea
where $I^\vee$ is the complement of the index set $I$ in $ \{1,\cdots, N_f\}$.

\section{Mirror Symmetry}
\label{sec:mirror-symmetry}
\subsection{Symplectic duality for twisted stable quasi-maps}

A distinctive feature of the class of the theories we consider in this paper is mirror symmetry. We can find a pair of UV theories $\cT$ and $\cT^\vee$, which are dual under exchange of the following pairs of objects and parameters:
\bea\label{mirror map}
\text{H-twist}~ &\leftrightarrow~ \text{C-twist} \\
\cM_H ~&\leftrightarrow ~\cM_C\\
G_H~ &\leftrightarrow~ G_C\\
\{m_i\}~ &\leftrightarrow~ \{\zeta_i\} \\
t ~&\leftrightarrow~ t^{-1} \, .
\eea
The duality holds when the theory flows to the deep infrared, which is in general described by taking the limit $e^2\rightarrow \infty$. On the other hand, the description of the moduli space of solutions to \eqref{eq:vortex} depends on the chamber specified by a dimensionless parameter
\be
s:=\frac{\tau e^2 \text{vol}(\Sigma)}{2\pi}\ .
\ee
Therefore, flowing to the deep infrared corresponds to studying the moduli space in the limit $|s|\rightarrow \infty$. This can be alternatively described by taking the infinite tension limit, $|\tau|\rightarrow \infty$ (with $e^2$ finite), which we have studied so far. It follows that the twisted indices computed in this chamber are expected to exhibit the duality exchanging the parameter as in \eqref{mirror map}.

Given the interpretation of the twisted indices we have offered in this paper, mirror symmetry implies an extremely non-trivial relation between two generating functions of enumerative invariants of twisted stable quasi-maps into a conical symplectic resolution of the Higgs branch $\cM_H$. In fact, mirror symmetry implies two relations
\be
I_H(\zeta,m,t)[\cT] = I_C(m,\zeta,t^{-1})[\cT^\vee]\ ,
\ee
and
\be
I_C(\zeta,m,t)[\cT] = I_H(m,\zeta,t^{-1})[\cT^\vee] \ .
\ee
In particular, this exchanges the equivariant parameters $m$ and the degree counting parameters $\zeta$ of two generating functions. We may call this \emph{symplectic duality for stable quasi-maps}.

The simplest example is the theory SQED[2], which is a self-dual theory ${\cal T} = {\cal T}^\vee$. The generating functions for the first few genera are explicitly computed in \cite{Closset:2016arn}. For example, the generating function for the H-twist with $g=2$ is
\be\label{HTSU2}
I_H(q,a,t)\big|_{g=2} = -\frac{(1+t)\left[t(a+a^{-1}-2)(q+q^{-1}-2)+4(1-t)^2\right]}{t^{1/2}(t-a)(t-a^{-1})}\ ,
\ee  
where $a= e^{2\pi i {(m_1-m_2)}}$ and $q = e^{2\pi i \zeta}$. This can now be interpreted as the generating function for the equivariant virtual $\chi_t$ genus of $\fL$, where $\fL$ can be identified with the space of twisted stable quasi-maps into $\mathbb{P}^1$. 
On the other hand, for the C-twist, we have
\be
I_C(q,a,t)\big|_{g=2} = -\frac{(1+t)[t(a+a^{-1}-2)(q+q^{-1}-2)+4(1-t)^2]}{t^{1/2}(t-q)(t-q^{-1})}\ .
\ee
This corresponds to the generating function of the virtual Euler characteristic for stable quasi-maps. It agrees with \eqref{HTSU2} by exchanging $q\leftrightarrow a$ and $t\leftrightarrow t^{-1}$ as expected.

\subsection{Mirror symmetry for the $\cN=4$ index}

As studied in section \ref{sec:t1limit}, the twisted indices
drastically simplify in the limit $t\rightarrow 1$.
The H-twisted indices in this limit can be identified with a sum over the integrals of the Euler class of the fixed loci
\be\label{mirror H}
I_H(q):=I_H(q,a,t)|_{t\rightarrow 1}= \sum_{\underline{\m} \in \Lambda_G^\vee}(-q)^\m\sum_{I} (-1)^{\text{dim}_{\mathbb{C}}\fM_{\underline{\m},I}} \int_{M_{\underline{\m},I}} e\left(\fM_{\underline{\m},I}\right)\ ,
\ee
which is independent of the equivariant parameters $a$.
On the other hand, $C$-twisted indices receive contributions from the degree zero sector only and therefore independent of $q$ in the limit $t\rightarrow 1$. The index in this limit computes the Rozansky-Witten invariants 
\be\label{mirror C}
I_C(a):=I_C(q,a,t)|_{t\rightarrow 1}  =\int_{\cM_H} \hat A\left(T\cM_H\right)~\text{ch}\left[\left(\widehat\wedge^\bullet T^*\cM_{H}\right)^{\otimes g}\right]\ ,
\ee
where $\cM_H$ is a resolution of the Higgs branch.

\begin{figure}[h]
\centering
\includegraphics[height=1.25cm]{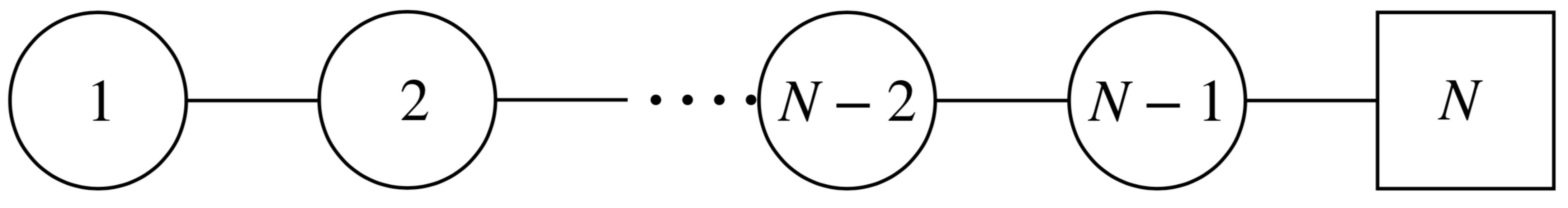}
\caption{$T[SU(N)]$ theory}
\label{fig:TSUN}
\end{figure}

Symplectic duality in this limit then takes a simple form 
\be
I_H(q)[{\mathcal{T}}] = I_C(a)[\cT^\vee]\ ,
\ee
with the identification $q=a$. Below we will explicitly show this identity for $T[SU(N)]$ theories. This class of theories plays an important role in the S-duality of the half-BPS boundary conditions in $\cN=4$ Yang-Mills theory \cite{Gaiotto:2008ak}. These can be represented as the quiver in Figure \ref{fig:TSUN}.
The twisted indices of $T[SU(N)]$ quiver depends on the FI-parameters $\zeta_{i=1,\cdots, N-1}$ for each factor of the gauge group $U(1)\times \cdots \times U(N-1)$ and the mass parameters $m_{i=1,\cdots, N}$ for the $PSU(N)$ flavour symmetry, which satisfies $\sum_{i=1}^{N}m_i=0$. This class of theories are known to be self dual under exchanging $\epsilon_i\leftrightarrow m_i$ for all $i$, where $\zeta_i = \epsilon_i - \epsilon_{i+1}$.

\paragraph{H-twist} 
The moduli space of $T[SU(N)]$ theory can be decomposed into the topological sectors weighted by the FI-paramaters $q_i = e^{2\pi i \zeta_i}$:
\be
\cM = \bigcup_{\substack{(\m_1,\cdots, \m_{N-1})\\
\in \mathbb{Z}^{N-1}}}q_1^{\m_1}q_2^{\m_2}\cdots q_{N-1}^{\m_{N-1}} ~\cM_{\m_1,\cdots, \m_{N-1}}\ .
\ee
For each factor of the gauge group labelled by $a=1,\cdots, N-1$, we denote by
\be
\m_a = \frac{1}{2\pi}\text{tr}\int_{\Sigma}F_a \in \mathbb{Z}\ ,
\ee
the degree of the vector bundle $E_a$ of rank $a$, associated with the $U(a)$ gauge bundle $P_a$.
Let us denote $X_a^{a+1}, Y_{a+1}^{a}$ by the bi-fundamental fields between $a$-th and $a+1$-th nodes, where $X_a^{a+1}$ can be regarded as a $a \times (a+1)$ matrix whose components are $X_{a(k_a)}^{a+1(k_{a+1})}$, and similarly for the $Y$. Then the $\cN=4$ moduli space for the H-twist is given by the space of solutions $(A_1,\cdots A_{N-1}, X,Y)$ to the equations
\bea
& *F_a + e^2\left({X_a^{a+1}} {X_a^{a+1}}^\dagger - {Y_{a+1}^{a}}^\dagger{Y_{a+1}^{a}} - { X_{a-1}^{a}}^\dagger{X_{a-1}^{a}} +  {Y_a^{a-1}} {Y_a^{a-1}}^\dagger -\tau_a \right) =0\ ,\\
& \bar\partial_{A} X_a^{a+1} =  \bar\partial_{A}  Y_{a+1}^{a} = 0\ ,~~ X_a^{a+1} Y_{a+1}^{a}=0\ ,~~\text{for } a=1,\ldots, N-1\ ,
\eea
modulo $U(1)\times \cdots \times U(N-1)$ gauge transformation.

Let us consider the chamber where all $\tau_a$'s are sufficiently large. As in the previous examples, we perform the equivariant localisation with respect to the action of the flavour symmetry $g_H\in PSU(N_f)$, by turning on the mass parameters $m_i$. If we keep these parameters generic, on the fixed locus, each factor of the gauge group is abelianised
\be
U(a) \rightarrow U(1)^a\ ,~~\text{for } a=1,\cdots, N-1\ ,
\ee
and  accordingly the vector bundle $E_a$ is decomposed into the sum of the line bundle
\be
E_a = L_{a(1)} \oplus \cdots \oplus L_{a(a)}\ ,
\ee
on the fixed loci, where deg$(L_{a(k)}) = \m_{a(k)}$ with $\m_a = \sum_{k=1}^a \m_{a(k)}$. Then the moduli space reduces to a disjoint union of $N!$ fixed loci labelled by a set of the index sets
\be
\{I_1,\cdots, I_{N-1}\}\ ,~\text{ where }I_a\subset \{1,\cdots, a+1\}\ ,~~|I_a|=a\ .
\ee
If we denote the element of the index set $I_a$ by $i_{a(k_a)}$ with $k_a=1,\cdots, a$, then on the fixed locus $\{I_1,\cdots, I_{N-1}\}$, the only non-vanishing bosonic fields in the chiral multiplets are
\be
X^{a+1(i_{a(k_a)})}_{a(k_a)} \neq 0\ ,~~\text{ for }~a=1,\cdots, N-1\ .
\ee
To simplify the notation, let us make a choice of fixed locus defined by $I_a = \{1,\cdots, a\}$ for all $a$, where the only non-vanishing bosonic fields in the chiral multiplets are $X_{a(k_a)}^{a+1(k_a)}$ for all $k_a=1,\cdots, a$. All other fixed points can be obtained by an action of the $PSU(N)$ flavour symmetry. The fixed locus is described by the equations
\bea
&* F_{a(i_a)} + e^2 \left(|X_{a(i_a)}^{a+1(i_a)}|^2 - |X_{a-1(i_a)}^{a(i_a)}|^2-\tau_a\right) =0\ ,\\
&  \bar\partial_{A}  X_{a(i_a)}^{a+1(i_a)} =0\ ,~~\text{ for }~i_a = 1,\ldots a.
\eea
modulo $\prod_{a=1}^{N-1}U(1)^{a}$ gauge symmetries, where we defined $X_{a-1(a)}^{a(a)}=0$. Here $X_{a(i_a)}^{a+1(i_a)}$ is a holomorphic section of the line bundle $L_{a(i_a)}\otimes L^{-1}_{a+1(i_a)}\otimes K^{1/2}$.
Therefore the fixed locus can be described as the $1+\cdots+ N-1$ copies of the symmetric product:
\be
\fM_{I_a}=\prod_{a=1}^{N-1}\prod_{i_a=1}^a \Sigma_{\m_{a(i_a)} - \m_{a+1(i_a)}+g-1}\ .
\ee

Then the contribution of this fixed locus to the virtual Euler characteristic in the limit $t\rightarrow 1$ \eqref{mirror H} becomes
\be\label{tsun 1}
\sum_{\substack{\m_{a(k_a)}\in\mathbb{Z}\\ \text{for } k_a = 1,\cdots, a-1 \\a=1,\cdots, N-1} } q_1^{\m_{1(1)}}q_2^{\m_{2(1)}+\m_{2(2)}}\cdots q_{N-1}^{\sum_{i=1}^{N-1}\m_{N-1(i)}} \int_{\fM_{I_a}}\prod_{a=1}^{N-1} \prod_{i_a=1}^a e\left(\Sigma_{\m_{a(i_a)} - \m_{a+1(i_a)}+g-1}\right)\ .
\ee
It is convenient to change the summation variable as
\be
\m_{a(k_a)} \rightarrow \m_{a(k_a)} + \sum_{r=a+1}^{N-1} \m_{r(k_a)}\ ,~~\text{ for all }~a,k_a
\ee
the expression \eqref{tsun 1} then becomes
\be
\sum_{\substack{\m_{a(k_a)}\in\mathbb{Z}\\ \text{for } k_a = 1,\cdots, a-1 \\a=1,\cdots, N-1} }\prod_{a=1}^{N-1} \left(q_1\cdots q_a\right)^{\m_{a(1)}}\left(q_2\cdots q_a\right)^{\m_{a(2)}}\cdots \left(q_a\right)^{\m_{a(a)}}\int_{\fM_{I_a}} \prod_{a=1}^{N-1}\prod_{i_a=1}^a e\left(\Sigma_{\m_{a(i_a)+g-1}}\right)\ .
\ee
Let us redefine
\be
q_i = e^{2\pi i (\epsilon_i - \epsilon_{i+1})}\ ,~~\text{ with }~\sum_{i=1}^N \epsilon_i = 0\ ,
\ee
and sum over $1+\cdots+N-1$ copies of integers $\m_{a(i_a)}$. Using the relation \eqref{euler symmetric 5}, we find a simple expression
\be
\prod_{i\neq j}^{N}\left(e^{\pi i (\epsilon_i - \epsilon_j)} -e^{\pi i (-\epsilon_i + \epsilon_j)}\right)^{(g-1)}
\ee
for the fixed locus $\fM_{I_a}$. Since the result does not depend on the equivariant parameters $\{m_i\}$, the contribution from $N!$ fixed locus are the same. Therefore we conclude
\be\label{tsunh final}
I_H(\{\epsilon_i\}) = (-1)^{N(N-1)(g-1)/2}N!\prod_{i< j}^{N}\left(e^{\pi i (\epsilon_i - \epsilon_j)} -e^{\pi i (-\epsilon_i + \epsilon_j)}\right)^{2(g-1)}\ .
\ee
Note that the result shows the structure of the full $SU(N)$ Coulomb branch symmetry $G_C$ enhanced from the UV topological symmetry $U(1)^{N-1}$.

\paragraph{C-twist}

Once we impose the $\cN=4$ BPS equation for the C-twist, the associated vector bundle $E_1 \oplus \cdots \oplus E_{N-1}$ trivialises. In the large $\tau$ limit, the moduli space reduces to the resolved Higgs branch $\cM_{H,\tau}$, which can be identified as a cotangent space of a flag variety.

Similarly to the H-twist, the fixed loci of the $g_H$ action are given by the 
choice of the index set $\{I_1,\cdots, I_{N-1}\}$, where $I_a = \{i_{a(1)},\cdots, i_{a(N)}\} \subset \{1,\cdots, a+1\}$, for all $a$. Each fixed locus is an isolated point, characterised by the non-vanishing bi-fundamental chiral fields
\be
X_{a(k_a)}^{a+1(i_{a(k_a)})} \neq 0\ .
\ee
The C-twisted index in the limit $t\rightarrow 1$ gets contribution from the $\m=0$ sector only. It is straightforward to compute \eqref{mirror C} equivariantly at each fixed points, which gives the expression
\bea
I_C(\{m_i\}) = \sum_{\{(I_1,\cdots, I_{N-1})\}} &\prod_{\substack{i\in I_{N-1}\\j\in I^\vee_{N-1}}} \left(e^{-\pi im_{ij}/2} - e^{\pi i m_{ij}/2}\right)^{2(g-1)} \prod_{\substack{i\in I_{N-2}\\j\in I^\vee_{N-2}}}\left(e^{-\pi im_{ij}/2} - e^{\pi i m_{ij}/2}\right)^{2(g-1)} \\
& \ldots \prod_{\substack{i\in I_{1}\\j\in I^\vee_{1}}}\left(e^{-\pi im_{ij}/2} - e^{\pi i m_{ij}/2}\right)^{2(g-1)}\ ,
\eea
where the summation is over $N!$ choices of the fixed locus. $I^{\vee}_{a-1}$ is defined as the complement of $I_{a-1}$ inside the index set $\{1,\cdots, a\}$. Note that each term in the summation is invariant under the Weyl group $W_{G_H}$ of the flavour symmetry and therefore the contributions from all the fixed loci are identical. The expression simplifies to
\be\label{tsunc final}
I_C(\{m_i\}) = N! \prod_{i<j}\left(e^{-\pi im_{ij}/2} - e^{\pi i m_{ij}/2}\right)^{2(g-1)}\ .
\ee

Comparing two expressions \eqref{tsunh final} and \eqref{tsunc final}, we find an agreement
\be
I_{C}(\{\epsilon_i\})[T[SU(N)]] = I_{H}(\{m_i\})[T[SU(N)]]\ 
\ee
up to an overall sign, under the identification of the parameters $\epsilon_i = m_i\ ,\forall i$. This agrees with the self-dual property of $T[SU(N)]$ theories.

\section{Conclusions and future directions}

In this paper, we have demonstrated that the twisted indices of a large class of 3d $\mathcal{N}=4$ theories on $S^1 \times \Sigma$ have a fascinating interpretation in enumerative geometry as generating functions for equivariant Euler characteristics of moduli spaces of twisted quasi-maps to the Higgs branch. Other interesting interpretations become available in the strict $\mathcal{N}=4$ limit $t \to 1$, and mirror symmetry implies surprising relations between these generating functions. We conclude with directions for further research.

First, it should be possible to extend our computations to a larger class of gauge theories labelled by a compact gauge group $G$ and quaternionic representation $Q = T^*M$, for which the contour integral for the twisted index applies. In this case, we expect to recover quasi-maps to the Higgs branch `stack' and that it will be useful to consider virtual localisation with respect to $G$ to reduce the problem to integrals of characteristic classes on the Picard stack. 
We expect this to make contact with the work~\cite{Teleman:2009}.
It would further be interesting to extend our work to theories with $\mathcal{N}=2$ supersymmetry. This will in general involve additional considerations regarding topological vacua.

A second interesting extension would be to introduce background vector bundles for the flavour symmetries $G_H$ and $G_C$ on $\Sigma$. This is relatively straightforward for $G_H$, but in the case of $G_C$ these are expected to induce vector bundles on the moduli spaces of solutions themselves \cite{Bullimore:2018yyb}, and it would be of great mathematical appeal to uncover generalisations of this phenomenon. 
It is also possible to study the inclusion of line operators wrapping $S^1$, which are similar to the above. 

Finally, it is possible to fully embrace the Hamiltonian point of view and consider the theories on $\mathbb{R}\times \Sigma$ as quantum mechanics on $\mathbb{R}$. The space of supersymmetric ground states of these quantum mechanics would provide a categorification of the enumerative invariants we have studied here. First steps in this program were taken in~\cite{Bullimore:2018yyb}. In the limit $t \to 1$, we expect a close connection to conformal blocks for the vertex operator algebras introduced in \cite{Gaiotto:2016wcv,Costello:2018fnz} including in the presence of line operators \cite{Costello:2018swh}. This suggests the existence of a vast generalisation of the Verlinde formula and its extension to Higgs bundles \cite{Andersen:2016hoj}.

\acknowledgments
We would like to thank Stefano Cremonesi and Tudor Dimofte for fruitful discussions.
The work of H.K. is supported by ERC Consolidator Grant 682608 ``Higgs bundles: Supersymmetric Gauge Theories and Geometry (HIGGSBNDL). A.F. is supported by the SNF Doc.Mobility fellowship P1SKP2\_181340 ``Twisted Hilbert Spaces of 3D Supersymmetric Gauge Theories.''

\appendix

\section{Supersymmetric Algebra}
\label{app:susy algebra}

In this section, we summarize the supersymmetry algebra of twisted $\cN=4$ theories. We start by fixing our notation for the $\cN=4$ vector multiplet and the $\cN=4$ chiral multiplet, and the respective decomposition into $\cN=2$ multiplets. We then address the H- and C- twist algebras in turns. \footnote{We use the convention
\be
\epsilon^{12} = -\epsilon_{12}=\epsilon^{\dot1 \dot2} = -\epsilon_{\dot1\dot2}=1
\ee
where ${(\sigma^\mu)_\alpha}^\beta$ are standard Pauli matrices.}

The $\cN=4$ vector multiplet consists of the fields
\be
V_{\cN=4} = \left(A_\mu, \lambda_{\alpha}^{A\dot B}, \varphi^{\dot A \dot B}, D^{AB}\right)\ .
\ee
The fields are subject to the reality condition as follows:
\be
D_{AB} = \left(D^{AB}\right)^\dagger\ ,~ \varphi_{\dot A\dot B} = -\left(\varphi^{\dot A \dot B}\right)^\dagger\ .
\ee
The multiplet decomposes into an $\cN=2$ vector multiplet $V = (A_\mu, \sigma,\lambda,\bar\lambda, \Lambda_1,\bar\Lambda_{\bar 1},D)$ and an $\cN=2$ chiral multiplet $\Phi_\varphi=(\varphi, \psi_{\varphi}, \eta_{\varphi}, F_{\varphi})$ in the adjoint representation with the identification 
\bea
&\lambda = \frac{1}{2} \lambda_2^{\dot 2 \dot 2} \ , \bar\lambda = \  \frac{1}{2} \lambda_1^{\dot 1 \dot 1} \ , \ \Lambda_1 =   \lambda_1^{\dot 2 \dot 2} \ ,  \bar\Lambda_{\bar 1} = \lambda_2^{\dot 1 \dot 1} \, ,\\
& \sigma =  \varphi_{\dot 1 \dot 2} \ , \ D =   D_{12} \, , \\
&\varphi^\dagger= -\frac12 \varphi_{\dot 2 \dot 2}\ , \  \varphi = \frac12 \varphi_{\dot 1 \dot 1} \, , \bar\psi_{\varphi} = \lambda_{1}^{\dot 2 \dot 1} , \ \psi_{\varphi} = \lambda_{1}^{\dot 1 \dot 2} \, ,\\
&\bar\eta_{\varphi} = \lambda_{2}^{\dot 1 \dot 2}\, ,  \eta_{ \varphi} = \lambda_{2}^{\dot 2 \dot 1} \, , \ F^\dagger_{\varphi} = -D_{22}, \ F_{\varphi} = D_{11} \, .
\eea

The $\mathcal{N}=4$ hypermultiplet consists of fields
\be
H_{\cN=4} =  \left( X_A, {X^A}^\dagger , \psi_{\alpha \dot A} , \bar \psi_{\alpha \dot A} \right)
\ee
In terms of $\cN = 2$ fields, we identify 
\be
(X, Y^\dagger) = (X_1,X_2)  , \ (Y, X^\dagger) = ({X^2}^\dagger, {X^1}^\dagger) \, ,
\ee
\subsection{H-twist}

The four supercharges preserved under the H-twist can be written as
\be
\zeta_1^{1 \dot A} := \zeta_H^{\dot A}\ ,~~\zeta_2^{2 \dot A} := \t\zeta_H^{\dot A}\ .
\ee
Note that the $\cN=2$ subalgebra is generated by $\zeta_H^{\dot 1}$ and $\t\zeta_H^{\dot 2}$. For convenience, we redifine the vector multiplet fermions as
\be
\lambda_{1}^{1 \dot A} = \lambda^{\dot A}\ ,~\lambda_{2}^{2 \dot A} = \bar\lambda^{\dot A}\ ,~\lambda_{1,1\dot A} = \bar \Lambda_{\bar 1,\dot A} \ ,~ \lambda_{2,2\dot A} = \Lambda_{ 1,\dot A}\ .
\ee
The supersymmetry transformation of the vector multiplet is given by
\bea
\delta A_0 &= \frac{i}{2} \t  \zeta_H^{\dot A} \lambda_{\dot A} - \frac{i}{2} \zeta_H^{\dot A} \bar \lambda_{\dot A} \ , \\
\delta A_1 &= i\t \zeta_H^{\dot A}\Lambda_{\dot A, 1}\ ,\\
\delta A_{\bar 1} &= -i\zeta_H^{\dot A}\bar\Lambda_{\dot A,\bar 1}\ ,\\
\delta \varphi_{\dot A \dot B} &= \frac{1}{2} \left( \zeta_{H,\dot A}\bar\lambda_{\dot B} + \t\zeta_{H,\dot A}\lambda_{\dot B} + \zeta_{H,\dot B}\bar\lambda_{\dot A} + \t\zeta_{H,\dot B}\lambda_{\dot A} \right) \ ,\\
\delta D_{11} &= -i \t \zeta_H^{\dot A} \left( 2 D_1 \bar \lambda_{\dot A} - D_0  \bar \Lambda_{\bar 1 , \dot A} \right) - i\t \zeta_H^{\dot B}\left[\bar \Lambda_{\bar 1}^{\dot C}, \varphi_{\dot B \dot C} \right] \ , \\
\delta D_{22}& = i\zeta_H^{\dot A} \left(2 D_{\bar 1} \lambda_{\dot{A}}-D_0 \Lambda_{ 1, \dot{A}} \right) - i \zeta_H^{\dot B}\left[ \Lambda_{ 1}^{\dot C}, \varphi_{\dot B \dot C} \right] \ ,\\
\delta D &=  \frac{i}{2} \zeta_H^{\dot B} \left(2 D_{\bar 1}\bar \Lambda_{\bar 1,\dot B} + D_0 \bar \lambda_{\dot B} +\left[ \bar  \lambda^{\dot C}, \varphi_{\dot B \dot C}\right]  \right) + \frac{i}{2} \t \zeta_H^{\dot B} \left( 2 D_{ 1} \Lambda_{1,\dot B} + D_0\lambda_{\dot B} - \left[\lambda^{\dot C}, \varphi_{\dot B\dot C}\right]\right)\ ,\\
\delta \lambda_{\dot A} &=  -\left(2F_{1 \bar 1}  - D \right)  \zeta_{H, \dot A} - i \zeta_H^{\dot B} D_0  \varphi_{\dot B \dot A} +\frac{i}{2} \zeta_{H, \dot D} \left[ \varphi_{\dot A}^{\ \dot C}, \varphi_{\dot C}^{\ \dot D}  \right] \ , \\
\delta \bar\lambda_{\dot A} & = \left( 2 F_{1 \bar 1} - D  \right) \t \zeta_{H, \dot A} + i \t \zeta_{H}^{\dot B} D_0 \varphi_{\dot B \dot A}  +\frac{i}{2} \t \zeta_{H, \dot D} \left[ \varphi_{\dot A}^{\ \dot C}, \varphi_{\dot C}^{\ \dot D}  \right] \ , \\ 
\delta \Lambda_{1,\dot A} &=  2F_{0\bar 1}  \zeta_{\dot A} - 2i\zeta^{\dot B}D_{\bar{1}}\varphi_{\dot B\dot A} + D_{22} \t \zeta_{\dot A}\ ,\\
\delta \bar\Lambda_{\bar 1,\dot A} &=2F_{0 1}  \t \zeta_{\dot A} + 2i \t \zeta^{\dot B}D_{ 1 }\varphi_{\dot B\dot A} + D_{11}\zeta_{\dot A}\ .
\eea
The supersymmetry transformations of the hypermultiplet can be written as 
\bea
\left(\begin{array}{c}\delta X_1 \\ \delta X_2\end{array} \right) &= \left( \begin{array}{c} - \t \zeta_H^{\dot B}\psi_{1, \dot B} \\ - \zeta_H^{\dot B}\psi_{2,\dot B} \end{array}\right) \ ,\\
\left(\begin{array}{c}\delta  \t X^1 \\ \delta  \t X^2\end{array}\right) &= \left(\begin{array}{c}- \zeta_H^{\dot B} \bar\psi_{ 2 , \dot B} \\ \t\zeta_H^{\dot B}\bar\psi_{1, \dot B} \end{array}\right)\ ,\\ 
\left( \begin{array}{c}\delta\psi_{1, \dot A} \\\delta\psi_{2, \dot A}\end{array}\right) &= \left(\begin{array}{c} - i \zeta_{H,\dot A} D_0 X_1 - 2 i \t\zeta_{H, \dot A} D_1 X_2 -i \zeta_H^{\dot B}X_1 \varphi_{\dot B \dot A} \\  i\t\zeta_{H, \dot A} D_0 X_2 -  2i\zeta_{H, \dot A} D_{\bar 1}X_1  -  i\t\zeta_H^{\dot B} X_2 \varphi_{\dot B\dot A}\end{array}\right) \ ,\\
 \left( \begin{array}{c}\delta\bar\psi_{1,\dot A} \\\delta\bar\psi_{2,\dot A}\end{array}\right) &= \left(\begin{array}{c} i \zeta_{H, \dot A} D_0  {X^2}^\dagger  - 2i \t\zeta_{H, \dot A} D_1  {X^1}^\dagger - i \zeta_H^{\dot B} {X^2}^\dagger \varphi_{\dot B,\dot A} \\  i\t\zeta_{H, \dot A} D_0 {X^1}^\dagger + 2 i \zeta_{H,\dot A} D_{\bar 1} {X^2}^\dagger  +  i \t\zeta_H^{\dot B}  {X^1}^\dagger \varphi_{\dot B\dot A}\end{array}\right)\ .
\eea

\subsection{C-twist}
The four supercharges preserved under the C-twist can be written as
\be
\zeta_{1}^{A,\dot 1} = \zeta_{C}^A\ ,~\zeta_{2}^{A,\dot 2} = \t\zeta_{C}^A\ .
\ee
The $\cN=2$ subalgebra is generated by $\zeta_C^{1}$ and $\t\zeta_C^2$. For the vector multiplet, we define
\bea
&\varphi = \frac12\varphi_{1\dot 1}\ ,~\varphi^\dagger  = -\frac12\varphi_{2\dot 2}\ ,\\
&\lambda_1^{A\dot 1} = \lambda^A\ ,~\lambda_2^{A\dot 2} = \bar\lambda^A\ ,~
\lambda_{1, A\dot 1} =\bar\Lambda_{\bar 1, A} \ ,~ \lambda_{2,A\dot 2} = \Lambda_{ 1, A}\ .
\eea
The supersymmetry transformation of the vector multiplet is given by
\bea
\delta A_0 &= \frac{i}{2} \t\zeta_{C}^A\lambda_A -\frac{i}{2} \zeta_C^A \bar\lambda_A\ , \\
\delta A_{1} &= i\t \zeta_{C}^A \Lambda_{A,{1}}\ ,\\
\delta A_{\bar 1} &= -i\zeta_{C}^A \bar \Lambda_{A,\bar 1} \\
\delta \sigma &= \frac{1}{2}\zeta_C^A \bar\lambda_{A} - \frac{1}{2}\t\zeta_C^A \lambda_A\ ,\\
\delta \varphi & =   -\frac12 \t\zeta_C^A \bar \Lambda_{\bar 1,A}\ ,\\
\delta \varphi^\dagger & =  \frac12 \zeta_C^A\Lambda_{ 1,A}\ ,\\
\delta D^{AB} &= i\zeta_C^A \left( D_{\bar 1}\bar \Lambda_{\bar 1}^B + \frac{1}{2}D_0 \bar\lambda^B -  \frac{1}{2} \left[\bar \lambda^{B} , \sigma  \right] -   \left[\Lambda_{ 1}^B , \varphi  \right] \right) + ( A\leftrightarrow B)\ ,\\
&~~~ -i\t\zeta_C^A\left(D_{1}\Lambda_{ 1}^B + \frac{1}{2} D_0\lambda^B  -  \frac{1}{2} \left[ \lambda^{B} , \sigma  \right] - \left[ \bar \Lambda_{\bar 1}^B ,  \varphi^\dagger  \right]\right)  + (A\leftrightarrow B)\ .\\
\delta\lambda_A &= \left(-2F_{1 \bar 1} -  iD_0\sigma -2 i [ \varphi ,  \varphi^\dagger] \right) \zeta_{C,A} + 4i\t\zeta_{C,A}D_{ 1}\varphi_{ 1} - {D_{A}}^B \zeta_{C,B} \ ,\\
\delta\bar\lambda_{A} &= \left(2F_{1 \bar 1} - iD_0\sigma +2 i [\varphi  , \varphi^\dagger ] \right)  \t \zeta_{C,A} + 4i \zeta_{C,A}D_{ \bar 1} \varphi_{ \bar 1} - {D_A}^B\t\zeta_{C,B} \ ,\\
\delta \bar \Lambda_{\bar 1, A} &= \left(2F_{10}  + 2i D_1\sigma\right)\t\zeta_{C,A} -2i D_0  \varphi \t \zeta_{C,A} - 2i \zeta_{C, A} \left[\sigma ,  \varphi \right] \ ,\\
\delta \Lambda_{1, A} &= \left(2F_{\bar 1 0} - 2i D_{\bar 1} \sigma \right) \zeta_{C,A}  -2 iD_0   \varphi^\dagger \zeta_{C, A} -2 i \t \zeta_{C, A} [\sigma ,   \varphi^\dagger ]\ .
\eea
For the hypermultiplet, we define
\bea
\psi_{1}^{\dot 1} &= \chi\ ,~ \psi_2^{\dot 2} = \eta\ ,~\psi_1^{\dot 2} = \psi_{1}\ ,~\psi_{2}^{\dot 1} = -\psi_{\bar 1}\ ,\\
\bar\psi_{1}^{\dot 1} &= \bar\chi\ ,~ \bar\psi_2^{\dot 2} = \bar\eta\ ,~\bar\psi_1^{\dot 2} = \bar\psi_{1}\ ,~\bar\psi_{2}^{\dot 1} = -\bar\psi_{\bar 1}\ .
\eea
We have
\bea
\delta X_A& = \t \zeta_{C,A} \chi + \zeta_{C,A} \eta\ ,\\
\delta {X^A}^\dagger &= \t \zeta_C^A\bar\chi + \zeta_C^A\bar\eta\ ,\\
\left(\begin{array}{c} \delta \chi \\ \delta \psi_{\bar 1}\end{array}\right) &= \left(\begin{array}{c} -i\zeta_C^B (D_0 +\sigma)X_B \\ 2i \zeta_C^B D_{\bar 1} X_B -2 i \t\zeta_{C}^B X_B \varphi^\dagger\end{array}\right)\ , \\
\left(\begin{array}{c} \delta \psi_{1} \\ \delta\eta \end{array}\right) & = \left(\begin{array}{c} -2i \t\zeta_C^B  D_1 X_B + 2i\zeta_C^B X_B \varphi  \\
i\t\zeta_C^B (D_0+\sigma) X_B\end{array}\right)\ , \\
\left(\begin{array}{c} \delta \bar\chi \\ \delta\bar\psi_{\bar 1}\end{array}\right) &= \left(\begin{array}{c} i\zeta_{C,B} (D_0 - \sigma) {X^B}^\dagger \\ -2i\zeta_{C,B} D_{\bar 1} {X^B}^\dagger -2i\t\zeta_{C,B}{ X^B}^\dagger \varphi^\dagger\end{array}\right)\ , \\
\left(\begin{array}{c}\delta \bar\psi_1 \\ \delta\bar\eta \end{array}\right) &= \left(\begin{array}{c} 2i\t\zeta_{C,B} D_1{ X^B}^\dagger + 2i\zeta_{C,B}{X^B}^\dagger \varphi\\-i\t\zeta_{C,B} (D_0 - \sigma) { X^B}^\dagger \end{array}\right)\ .
\eea

\section{Residue integrals at the large $|u|$ regions}
\label{JK appendix}

In this appendix, we show that the residue integrals involving the hyperplanes of type \eqref{asymptotic hyperplane} does not contribute to the integral with the choice \eqref{choice of eta}. Let us consider the localising action we take for the vector multiplet
\be\label{exact app}
\frac{1}{t^2}\left[\frac{1}{e^2}\cL_{\text{YM}}+ \cL_{\text{H}}\right]\ ,
\ee
which is modified from the localising action used in \cite{Benini:2015noa,Benini:2016hjo,Closset:2016arn} by additional term $\cL_{\text{H}}$.
As explained in section \ref{sec:lagrangians}, we take the limit $t\rightarrow 0$ with $e$ finite so that the localisation locus for the vector multiplet is given by
\be
*F + i D = 0\ ,~~ D = -i \left(\mu_{\mathbb{R}}-2 [\varphi^\dagger, \varphi]-\tau\right)\ ,~~\sigma^a = \text{constant}\ ,
\ee
and therefore the path integral localises to the finite dimensional integral of the Cartan zero modes $u = i\beta(\sigma+a_0)\in {\frak t}_{\mathbb{C}}$. As discussed in \cite{Benini:2015noa,Benini:2016hjo,Closset:2016arn}, it is convenient to allow a constant non-BPS mode $\hat D$ such that the auxiliary field localises to the field configuration which satisfies
\be
*F + iD = i\hat D\ ,~~\text{where }\hat D \in {\mathbb R}^r\ .
\ee
Then the contour integral expression can be derived from the algebra of the zero mode multiplets $\cV = (u,\bar u, \lambda_0, \bar\lambda_0, \hat D)\ .$ See \cite{Benini:2015noa,Benini:2016hjo,Closset:2016arn} for more details. The modified $Q$-exact action \eqref{exact app} affects the $\hat D$ integrals in the large $|u|$ region. Let us consider $G=U(1)$ theory for simplicity. The boundary integral for a given $\eta$ in the neighbourhood of the hyperplane \eqref{asymptotic hyperplane} is governed by the expression 
\bea
I_{\text{asymp}}(\eta) = \sum_{\m \in \mathbb{Z}}&~q^\m~\lim_{t\rightarrow 0}~\oint_{u\rightarrow \pm i \infty} du \int_{\mathbf{R}+i\delta} \frac{d\hat D}{\hat D}\\
&g_\m(u,m,\hat D)~ \exp\left[\frac{\beta\text{vol}(\Sigma)}{2t^2 e^2}\hat D^2 - \frac{i\beta}{t^2}\left(-\frac{2\pi \m}{e^2}+ \text{vol}(\Sigma) \tau\right)\hat D \right]\ ,
\eea
where $g(u,m,\hat D)$ is the one-loop contribution with the non-zero $\hat D$ background, which reduces to the integrand of the expression \eqref{index JK} at $\hat D=0$. Here $\delta \in {\frak t}$ is introduced as a regulator of the $\hat D$ integral, which is chosen in a way that it satisfies $\eta(\delta)<0$ for a choice of  $\eta \in {\frak t}^*$ in the definition of the JK-residue integral. \cite{Hori:2014tda,Benini:2015noa} The integral can be performed by rescaling $\hat D \rightarrow t^2\hat D$ and taking the limit $t\rightarrow 0$. We find
\bea
I_{\text{asymp}}(\eta) &= \sum_{\m \in \mathbb{Z}}~q^\m\oint_{u\rightarrow \pm i \infty} du \int_{\mathbf{R}+i\delta} \frac{d\hat D}{\hat D} ~g_\m(u,m,0)~e^{-i\beta\left(-\frac{2\pi \m}{e^2}+ \text{vol}(\Sigma) \tau\right) \hat D} \\
& = -2\pi i~\text{sgn}(\eta)\sum_{\m \in \mathbb{Z}}~q^\m~\Theta\left[\eta\left(\frac{2\pi \m}{e^2}- \text{vol}(\Sigma) \tau\right)\right]\oint_{u\rightarrow \pm i\infty} du~ g_\m(u,m,0)\ . 
\eea
If we assign the charges to the pole at infinity at each flux sector $\m$ as
\be\label{charge at infty}
Q_{\pm \infty} = \frac{2\pi \m}{e^2}- \text{vol}(\Sigma) \tau\ ,
\ee
we can write
\be
I_{\text{asymp}}(\eta) = -2\pi i~\text{sgn}(\eta)\sum_{\m \in \mathbb{Z}}~q^\m~\underset{u\rightarrow \pm\infty}{\text{JK-Res}}(Q_{\pm\infty}(u),\eta)~ g_\m(u,m,0)~ du
\ee
In order to match with the geometric interpretation we will choose
\be
\eta = -\frac{2\pi \m}{e^2}+ \text{vol}(\Sigma) \tau
\ee
so that the boundary contribution always vanish.

\section{Characteristic classes on a fixed locus and their integration}
\label{app:class sym}
In this Appendix, we derive expressions for various characteristic classes of bundles on
\be
\fM^T_{(\m_1,\cdots, \m_k)} = \prod_{i=1}^k \text{Sym}^{\m_i}\Sigma\ .
\ee
and their integration, as needed in section \ref{sec:examples}. In particular, we would like to compute \eqref{normal adjoint} and prove the integration formula \eqref{generaliseddonZagier}.

We start by considering the universal divisor 
\be\label{universal divisor2}
\Delta \subset \Sigma \times \text{Sym}^\m\Sigma
\ee
of degree $\m$, which was introduced in \eqref{universal divisor}. We denote by $f$ and $\pi$ the projections onto $\Sigma$ and Sym$^\m\Sigma$ respectively. Following \cite{macdonald1962symmetric}, we denote classes on $\Sigma$ by
\be
e_1,\cdots, e_g,e_1',\cdots, e_g' \in H^1(\Sigma,\mathbb{Z})\ ,~~\eta_\Sigma \in H^2(\Sigma, \mathbb{Z})
\ee
and as explained in the main body, standard classes on Sym$^\m\Sigma := \Sigma_\m$ by
\be
\xi_i,\xi_i'\in H^1(\Sigma_\m,\mathbb{Z}),~~\eta \in H^2(\Sigma_\m,\mathbb{Z})\ .
\ee
By the K\"unneth decomposition, the class of the universal divisor can be written as
\be
[\Delta] = \m\eta_\Sigma + \gamma + \eta\ 
\ee
where $\gamma = \sum_{i=1}^g \xi_i'e_i - \xi_ie_i'$. By the ring relations of the cohomology of the symmetric product \eqref{ringrelations} and the standard relations on the curve, we have $\gamma^2 = -2\sigma \eta_{\Sigma}$. The Grothendieck-Riemann-Roch theorem then impies that for any $q\in \mathbb{Z}$
\be
\text{td}(\Sigma_\m)\text{ch}[\pi_* \cO(q\Delta)] = \pi_* [\text{td}(\Sigma\times \Sigma_\m)\text{ch}(\cO(q\Delta))]\ .
\ee
From this we obtain
\bea
\text{ch}[\pi_* \cO(q\Delta)] &= \pi_* [\text{td}(\Sigma)\text{ch}(q\Delta)]\\
&= \pi_* (1+(1-g)\eta_{\Sigma})\exp (q\m \eta_\Sigma + q\gamma + q \eta)\\
&= \pi_* (1+(1-g)\eta_{\Sigma})(1+q \m \eta_\Sigma)(1+ q\gamma - q^2 \sigma \eta_{\Sigma}) e^{q\eta} \\
&= (q\m + (1-g)-q^2 \sigma)e^{q\eta}\ .
\eea

In order to compute the expression \eqref{normal adjoint}, we first compute the Chern class of the line bundle $\pi_* (\cO (-\Delta_a)\otimes \cO (\Delta_b))$ on Sym${}^{\m_a}\Sigma\times \text{Sym}^{\m_b}\Sigma$, where for notational simplicity we omit the pullbacks by the projections on $\text{Sym}^{\m_a}\Sigma$, $\text{Sym}^{\m_b}\Sigma$. We can compute
\bea
\text{ch}[\pi_* \cO(-\Delta_a)\otimes \cO(\Delta_b)] &= \pi_* [\text{td}(\Sigma)\text{ch}(\cO(-\Delta_a)\otimes \cO(\Delta_b))] \\
&= \pi_* (1+(1-g)\eta_\Sigma) \exp[-\m_a \eta_\Sigma - \gamma_a -\eta_a+ \m_b \eta_\Sigma + \gamma_b+\eta_b]\\
&=[-\m_a+\m_b + (1-g)-(\sigma^{aa}+\sigma^{bb} -\sigma^{ab}-\sigma^{ba})]e^{-\eta_1+\eta_2}\ ,
\eea
where 
\be
\sigma_{i}^{ab} = \xi_i^a{\xi_i'}^{b}\ ,~~\sigma^{ab} = \sum_{i=1}^g \sigma_i^{ab}\ ,
\ee
which satisfy the relation $\gamma_a\gamma_b = -2\sigma^{ab}\eta_{\Sigma}$.
From this we obtain
\be
c[\pi_* (\cO (-\Delta_a)\otimes \cO (\Delta_b))] = (1-\eta_a+\eta_b)^{-\m_a+\m_b - (g-1)} \exp \left[-\frac{\sigma^{aa}+\sigma^{bb} -\sigma^{ab}-\sigma^{ba}}{1-\eta_a+\eta_b}\right]\ .
\ee
Let us define a function 
\be
h(\eta_a-\eta_b) = e^{(-\eta_a+\eta_b + t)/2} -  e^{(\eta_a-\eta_b -t)/2}\ .
\ee
Then using the relation $(\sigma_i^{ab})^2=0$, we can show that the contribution from the class $[N\cM^{\cV_{ba}}] = [H^\bullet (L_a^{-1}\otimes L_b)]$ can be written as \footnote{In the Appendix, we will omit the weights under the action of the flavour symmetry $T_H$ for simplicity, which can be reintroduced easily.}
\be
\text{ch}(\hat \wedge^\bullet N\cM^{\cV_{ba}}) = h(\eta_a-\eta_b)^{-m_a+\m_b - (g-1)}\exp\left[(\sigma^{aa}+\sigma^{bb} -\sigma^{ab}-\sigma^{ba})\frac{h'(\eta_a-\eta_b)}{h(\eta_a-\eta_b)}\right]\ .
\ee
We would now like to prove \eqref{generaliseddonZagier}, which is a generalisation of the formula by don Zagier \eqref{zagier 2} to integrals over $\fM^T =  \prod_{a=1}^k\text{Sym}^{n_a}\Sigma$. We want to show that for any function $A(\eta_1,\cdots,\eta_k)$ and $B(\eta_1,\cdots,\eta_k)$, we have
\bea
~&\int_{\fM^T} A(\eta_1,\cdots, \eta_k) \exp\left[\sum_{a,b=1}^k \sigma^{ab}B_{ab}(\eta_1,\cdots, \eta_k)\right] \\
=&\underset{u_1=0}{\text{res}}\cdots  \underset{u_k=0}{\text{res}}~ \frac{A(u_1,\cdots u_k)}{u_1^{n_1+1}\cdots u_k^{n_k+1}}~ \left[\underset{ab}{\text{det} }\left(\delta_{ab}+ u_a B_{ab}(u_1,\cdots, u_k)\right)\right]^g\ .
\label{generalisedDonZagier2}
\eea
This can be demonstrated as follows. First we notice that
\bea
\exp \left[\sum_{a=1,b=1}^{k} \sigma^{ab} B_{ab} \right] &= \prod_{i=1}^g\prod_{a=1}^k\prod_{b=1}^k \exp \left(\sigma_i^{ab}B_{ab}\right) \\
&= \prod_{i=1}^g \prod_{a=1}^k\prod_{b=1}^{k}\left(1+\sigma_i^{ab} B_{ab}\right) \, \\
&= \prod_{i=1}^g \left(\sum_{p=0}^k \sum_{\substack{\text{all }\{a_1,\cdots, a_p\} \\\subset \{1,\cdots, k\}}} \sum_{\substack{\text{all }\{b_1,\cdots, b_p\} \\\subset \{1,\cdots, k\}}} \prod_{l=1}^p \prod_{m=1}^p \sigma^{a_l b_m}_i B_{a_lb_m}\right) \, ,
\label{detderivation1}
\eea
where we used $\left( \sigma_i^{ab}\right)^2=0$ as well as the fact that $\sigma$'s with different indices commute. We can then make use of the identity \cite{thaddeus1994stable}
\be\label{integral 3}
1 = \int_{\Sigma_{n_a}} \eta_a^{n_a} \left(\prod_{i\in I}\eta_a^{-1}\sigma_i^{aa}\right)\ 
\ee
and its straightforward generalisations to products of symmetric products. They imply that the only monomials contained in \eqref{detderivation1} surviving integration are the ones for which the subsets $\{b_1,\cdots, b_p\}$ are permutations of the $\{a_1,\cdots, a_p\}$. Let us denote by $S_p$ the permutation group of $p$ elements and suppose there is an $ s \in S_p$ so that $s(a_i)=b_i$. Then 
\be
\prod_{l=1}^k \sigma_i^{a_ls(a_l)} = \text{sgn}(s) \prod_{l=1}^k \sigma_i^{a_l a_l} \, .
\ee
Therefore,
\bea
~&\int_{\fM^T} A(\eta_1,\cdots, \eta_k) \prod_{i=1}^g \left[\sum_{p=0}^k \sum_{\substack{\text{all }\{a_1,\cdots, a_p\} \\\subset \{1,\cdots, k\}}} \sum_{\substack{\text{all }\{b_1,\cdots, b_p\} \\\subset \{1,\cdots, k\}}} \prod_{l=1}^p \prod_{m=1}^p  \left( \sigma^{a_l b_m}_i  \eta_{a_l}^{-1} \right)\eta_{a_l} B_{a_lb_m}\right]  \\
&=\underset{u_1=0}{\text{res}}\cdots  \underset{u_k=0}{\text{res}}~ \frac{A(u_1,\cdots u_k)}{u_1^{n_1+1}\cdots u_k^{n_k+1}}~ \prod_{i=1}^g \left[\sum_{p=0}^k \sum_{\substack{\text{all }\{a_1,\cdots, a_p\} \\\subset \{1,\cdots, k\}}} \sum_{s \in {S_p}} \text{sgn}(s) \prod_{l=1}^p u_{a_l} B_{a_ls(a_l)}\right] \\
&=\underset{u_1=0}{\text{res}}\cdots  \underset{u_k=0}{\text{res}}~ \frac{A(u_1,\cdots u_k)}{u_1^{n_1+1}\cdots u_k^{n_k+1}}~ \left[\sum_{p=0}^k \sum_{\substack{\text{all }\{a_1,\cdots, a_p\} \\\subset \{1,\cdots, k\}}} \sum_{s \in {S_p}} \text{sgn}(s) \prod_{l=1}^p u_{a_l} B_{a_ls(a_l)}\right]^g \, ,
\eea
where in the first line we formally divided and multiplied by $\eta_{a_l}$ with respect to \eqref{detderivation1}. By means of the Leibniz expansion of the determinant, this coincides with \eqref{generalisedDonZagier2}, as required.

\pagebreak

\bibliographystyle{JHEP}
\bibliography{twisted-index_v3}

\end{document}